\documentclass[pra,aps,longbibliography,showpacs,groupedaddress,superscriptaddress,twocolumn,toc=flat]{revtex4-1}

\usepackage{amsfonts,amsmath,amssymb,stmaryrd}

\usepackage[utf8]{inputenc}
\usepackage[table]{xcolor}
\usepackage{bbold, bm}
\usepackage{graphicx}
\usepackage{array,multirow}
\usepackage{boldline}
\usepackage{tabularx}
\usepackage{tikz}
\usepackage{placeins}
\usepackage{bm}
\usepackage{mathrsfs}

\usepackage{ctable}
\usepackage{epsfig}
\usepackage[english, status=final]{fixme}
\fxsetup{marginface=\tiny}
\fxusetheme{color}    

\usepackage[colorlinks=true,linkcolor=blue,urlcolor=blue,citecolor=blue]{hyperref}

\renewcommand{\P}{\hat{P}}

\newcommand{\Gj}{\hat{G}_{\bm{j}}}
\newcommand{\W}{\hat{W}}
\newcommand{\Wj}{\hat{W}_{\bm{j}}}

\newcommand{\bra}[1]{\langle#1|}
\newcommand{\ket}[1]{|#1\rangle}

\renewcommand{\j}{\bm{j}}
\renewcommand{\ij}{{\langle \bm{i}, \bm{j} \rangle}}
\renewcommand{\H}{\hat{\mathcal{H}}}

\renewcommand{\a}{\hat{a}}
\newcommand{\aj}{\hat{a}_{\bm{j}}}

\newcommand{\ad}{\hat{a}^\dagger}
\newcommand{\adj}{\hat{a}^\dagger_{\bm{j}}}

\newcommand{\n}{\hat{n}}
\newcommand{\nj}{\hat{n}_{\bm{j}}}

\renewcommand{\W}{\hat{W}}

\newcommand{\taux}[2]{\hat{\tau}^x_{\langle{#1},{#2}\rangle}}
\newcommand{\tauxij}{\hat{\tau}^x_{ \ij }}

\newcommand{\tauzij}{\hat{\tau}^z_{ \ij }}

\newcommand{\Ztwo}{$\mathbb{Z}_2$}

\AtBeginDocument{%
    \newwrite\bibnotes
    \def\bibnotesext{Notes.bib}
    \immediate\openout\bibnotes=\jobname\bibnotesext
    \immediate\write\bibnotes{@CONTROL{REVTEX41Control}}
    \immediate\write\bibnotes{@CONTROL{%
    apsrev41Control,author="08",editor="1",pages="1",title="0",year="1"}}
     \if@filesw
     \immediate\write\@auxout{\string\citation{apsrev41Control}}%
    \fi
}%

\makeatletter
\def\maketitle{
\@author@finish
\title@column\titleblock@produce
\suppressfloats[t]}
\makeatother

\begin{document}

\title{Realistic scheme for quantum simulation of $\mathbb{Z}_2$ lattice gauge theories \\ with dynamical matter in $(2+1)$D}

\author{Lukas Homeier}
\email{lukas.homeier@physik.uni-muenchen.de}
\affiliation{Department of Physics and Arnold Sommerfeld Center for Theoretical Physics (ASC), Ludwig-Maximilians-Universit\"at M\"unchen, Theresienstr. 37, M\"unchen D-80333, Germany}
\affiliation{Munich Center for Quantum Science and Technology (MCQST), Schellingstr. 4, D-80799 M\"unchen, Germany}
\affiliation{Department of Physics, Harvard University, Cambridge, MA 02138, USA}

\author{Annabelle Bohrdt}
\address{ITAMP, Harvard-Smithsonian Center for Astrophysics, Cambridge, MA 02138, USA}
\affiliation{Department of Physics, Harvard University, Cambridge, MA 02138, USA}

\author{Simon Linsel}
\affiliation{Department of Physics and Arnold Sommerfeld Center for Theoretical Physics (ASC), Ludwig-Maximilians-Universit\"at M\"unchen, Theresienstr. 37, M\"unchen D-80333, Germany}
\affiliation{Munich Center for Quantum Science and Technology (MCQST), Schellingstr. 4, D-80799 M\"unchen, Germany}

\author{Eugene Demler}
\affiliation{Institute for Theoretical Physics, ETH Zurich, 8093 Zurich, Switzerland}

\author{Jad C.~Halimeh}
\affiliation{Department of Physics and Arnold Sommerfeld Center for Theoretical Physics (ASC), Ludwig-Maximilians-Universit\"at M\"unchen, Theresienstr. 37, M\"unchen D-80333, Germany}
\affiliation{Munich Center for Quantum Science and Technology (MCQST), Schellingstr. 4, D-80799 M\"unchen, Germany}

\author{Fabian Grusdt}
\email{fabian.grusdt@physik.uni-muenchen.de}
\affiliation{Department of Physics and Arnold Sommerfeld Center for Theoretical Physics (ASC), Ludwig-Maximilians-Universit\"at M\"unchen, Theresienstr. 37, M\"unchen D-80333, Germany}
\affiliation{Munich Center for Quantum Science and Technology (MCQST), Schellingstr. 4, D-80799 M\"unchen, Germany}

\begin{abstract}
Gauge fields coupled to dynamical matter are ubiquitous in many disciplines of physics, ranging from particle to condensed matter physics, but their implementation in large-scale quantum simulators remains challenging.
Here we propose a realistic scheme for Rydberg atom array experiments in which a \Ztwo{}~gauge structure with dynamical charges emerges on experimentally relevant timescales from only local two-body interactions and one-body terms in two spatial dimensions.
The scheme enables the experimental study of a variety of models, including $(2+1)$D \Ztwo{}~lattice gauge theories coupled to different types of dynamical matter and quantum dimer models on the honeycomb lattice, for which we derive effective Hamiltonians.
We discuss ground-state phase diagrams of the experimentally most relevant effective \Ztwo{}~lattice gauge theories with dynamical matter featuring various confined and deconfined, quantum spin liquid phases.
Further, we present selected probes with immediate experimental relevance, including signatures of disorder-free localization and a thermal deconfinement transition of two charges.
\end{abstract}

\date{\today}

\maketitle
\section{Introduction}
It has been a long sought goal to faithfully study lattice gauge theories (LGTs) with dynamical matter in the realm of strong coupling.
Since their discovery, \Ztwo{}~LGTs have sparked the interest of physicists from various different fields including high-energy~\cite{Wilson1974}, condensed matter~\cite{Wegner1971, FradkinSusskind1978, Kogut1979} or biophysics~\cite{LammertRokhsarToner1993}.
The seminal work by Fradkin and Shenker~\cite{FradkinShenker1979} in 1979 predicted the existence of two phases in their model, in which \Ztwo{}~charged particles are either confined or deconfined in $(2+1)$D.
This insight made it a particularly promising candidate theory that could capture some of the essential physics of quark confinement in QCD~\cite{Wilson1974} while hosting a much simpler gauge group.
Likewise, it provides one of the most fundamental instances of the Higgs mechanism.
Since then the study of \Ztwo{}~LGTs has inspired physicists because of their intimate relation to topological order~\cite{Wen2007}, quantum spin liquids~\cite{ReadSachdev1991, Sachdev2019} and quantum information~\cite{Kitaev2003}, to name a few.
While the physics of these models could give insights into outstanding problems, e.g., how to define confinement in the presence of dynamical matter, the numerical (e.g. Refs.~\cite{Trebst2007, Vidal2009, Tupitsyn2010, Gazit2017, Borla2022}) and experimental exploration is at the same time very challenging beyond $(1+1)$D (e.g. Refs.~\cite{Schweizer2019, Barbiero2019, Homeier2021, Zohar2021}).

The experimental developments over the past years have driven the field of analog quantum simulation towards exploring many-body physics in system sizes out of reach for any numerical simulation and offering a new toolbox to approach complex, physical phenomena such as quantum spin liquids~\cite{Semeghini2021}.
The difficulty to implement gauge constraints and robustness against ever-present gauge-breaking errors in analog quantum simulators, however, has hindered the field to push forward into the aforementioned direction and a scalable, reliable implementation of LGTs with dynamical matter in (2+1)D remains a central goal.

\begin{figure*}[t]
\includegraphics[width=\textwidth]{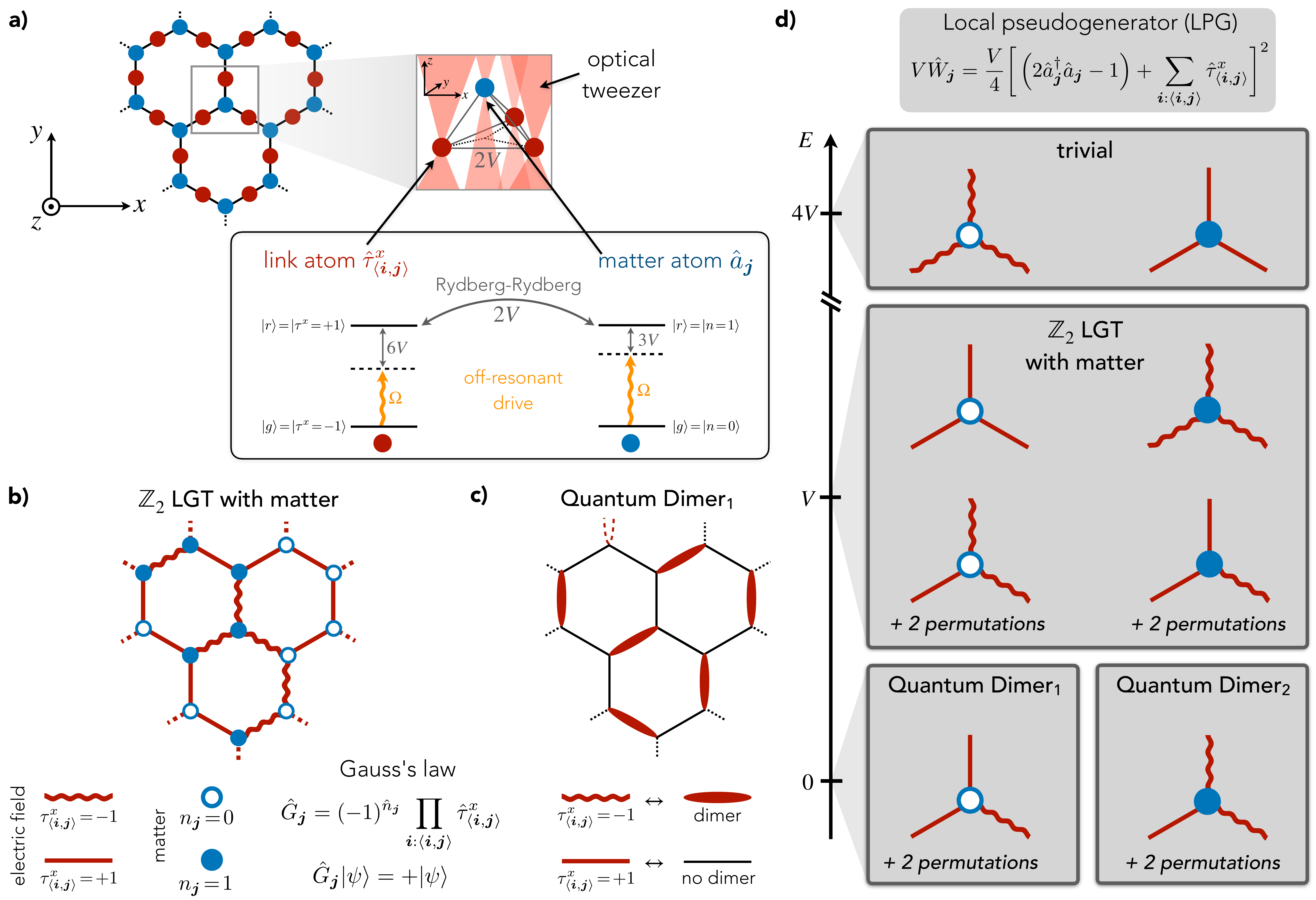}
\caption{\textbf{Constraint-based implementation of \Ztwo{}~mLGT with qubits.} The \Ztwo{}~gauge structure emerges from the dominant local-pseudogenerator (LPG) interaction on the honeycomb lattice introduced in panel~\textbf{a)}. A vertex contains matter~$\aj$ qubits (blue) and shares link~$\tauxij$ qubits (red) with neighboring vertices. All qubits connected to a vertex interact pairwise with strength~$2V$. In a Rydberg atom array experiment the qubits are implemented by individual atoms in optical tweezers, which are assigned the role of matter or link depending on the position in the lattice. Here, the ground- and Rydberg state of the atoms, $\ket{g}$ and $\ket{r}$, encode qubit states, which are coupled by an off-resonant drive~$\Omega$ to induce effective interactions. To realize equal strength nearest neighbor, two-body Rydberg-Rydberg interactions, the matter atoms can be elevated out of plane. In panel~\textbf{b)} we introduce the notation for the \Ztwo{}~mLGT, for which the Hilbert space constraint is given by Gauss's law~$\Gj=+1$. We illustrate the electric field~$\tau^x_{\ij}=+1$ ($\tau^x_{\ij}=-1$) with flat (wavy) red lines and the matter site occupation $n_{\j}=0$ ($n_{\j}=1$) with empty (full) blue dots. Panel~\textbf{c)} shows the notation for the QDM subspace with exactly one dimer per vertex. Panel~\textbf{d)} illustrates how the distinct subspaces are energetically separated by the LPG term~$V\Wj$. The two quantum dimer subspaces are disconnected when the matter is static, which can be exactly realized by the absence of matter atoms in panel~\textbf{a)} and setting $(2\adj\aj-1)=\pm 1$ in $V\Wj$. }
\label{fig1}
\end{figure*}

The rich structure of gauge theories emerges from locally constraining the Hilbert space.
This constraint can be formulated by Gauss's law, which requires all physical states~$\ket{\psi}$ to fulfill~$\Gj \ket{\psi} = g_{\bm{j}}\ket{\psi}$.
For the \Ztwo{}~LGT with dynamical matter (\Ztwo{}~mLGT) we consider in this work the symmetry generators~$\Gj$ are given by
\begin{align}
    \Gj = (-1)^{\nj} \prod_{\bm{i}:\ij}\tauxij, \label{eq:Goperator}
\end{align}
where~$\nj=\adj\aj$ is the number operator for (hard-core) matter on site~$\j$ and the Pauli matrix~$\tauxij$ defines the electric field on the link between site~$\bm{i}$ and~$\j$; hence \mbox{$g_{\j} = \pm 1$}.
Our starting point throughout this work are link and site qubits on a two-dimensional honeycomb lattice, see Fig.~\ref{fig1}a.

We propose to realize matter and link variables as qubits, implementable e.g. by the ground~$\ket{g}$ and Rydberg~$\ket{r}$ states of atoms in optical tweezers~\cite{Labuhn2016, Bernien2017, Keesling2019, Browaeys2020, Ebadi2021, Semeghini2021}, see Fig.~\ref{fig1}a-c.
Thus, the product in Eq.~(\ref{eq:Goperator}) measures the parity of qubit excitations of matter and links around vertex~$\j$.

By encoding the degrees-of-freedom in qubits the enlarged Hilbert space contains physical ($g_{\j} = +1$) and unphysical ($g_{\j} = -1$) states:
The latter do not fulfill Gauss's law.
Since any local perturbations present in a realistic quantum simulation experiment mix the two subspaces, quantum simulations can become unreliable, effectively breaking gauge-invariance.
Nevertheless, by energetically separating the physical from unphysical states transitions into the latter can be suppressed and the gauge structure emerges from the enlarged Hilbert space.

The simplest way, theoretically, to achieve such gauge protection, is by adding $-V\sum_{\j}\Gj$ to the Hamiltonian with large~\hbox{$V>0$}~\cite{Halimeh2020PRLRealiability, Halimeh2021PRXQ, HalimehHauke2022ReviewLinearProtection}.
But since this would require \textit{strong four-body interactions}, it is experimentally not feasible in current experimental platforms.

Here we demonstrate that simple two-body Ising-type interactions, which are readily available in e.g. Rydberg tweezer arrays~\cite{Labuhn2016, Bernien2017, Keesling2019, Browaeys2020, Ebadi2021, Semeghini2021}, combined with longitudinal and weak transverse fields provide a minimal set of ingredients which allow to robustly implement a variety of LGTs with dynamical matter~\cite{Sachdev2019}.
The scheme we propose not only offers inherent protection against arbitrary gauge-breaking errors; it also provides a surprising degree of flexibility, including cases with global conserved particle number, global number-parity conservation, and quantum dimer models on a bipartite lattice which map to $U(1)$~gauge theories.

In the following, we show that readily available Ising-type two-body interactions, in addition to local fields, are sufficient to protect Gauss's law on experimentally relevant timescales by employing the so-called local pseudogenerator (LPG) method~\cite{Halimeh2022LPG}.
Moreover, we show that the proposed protection scheme provides a generic means to engineer a variety of effective \Ztwo{}~mLGT Hamiltonians by weakly driving the qubits.
As an example, we demonstrate how this allows to realize the celebrated Fradkin-Shenker model~\cite{FradkinShenker1979}, and discuss the phase diagrams of several related effective Hamiltonians.
Finally, we elaborate on some realistic experimental probes that we view as most realistic in state-of-the-art quantum simulators.

\section{Results}
\textbf{Local pseudogenerator on the honeycomb lattice.--}
The main ingredient of the experimental scheme proposed in this Article is the local pseudogenerator (LPG) interaction term~$V\Wj$.
As shown in Fig.~\ref{fig1}a, $V\Wj$ consists of equal-strength~$2V$ interactions among all qubits (matter and gauge) around vertex~$\j$, taking the form
\begin{align} \label{eq:LPG}
    V\Wj =  \frac{V}{4} \left[ \left( 2\nj - 1 \right) +  \sum_{\bm{i}: \ij}\tauxij \right]^2.
\end{align}
We assume that $V$ defines the largest energy scale in the problem, which separates the Hilbert space into constrained subspaces.
This overcomes the most challenging step, imposing different gauge constraints in the emerging subspaces (Supplementary note~1).

We obtain three distinct eigenspaces of the LPG term:
1)~Two (distinct) quantum dimer model (QDM) subspaces with static matter at low-energy, 2)~physical states of a \Ztwo{}~mLGT at intermediate energies, and 3)~trivial, polarized states at high energy, see Fig.~\ref{fig1}b-d.

The LPG method requires that~$V\Wj$ acts identical to the full protection term on all physical states in the target gauge sector, i.e.\ $\Wj\ket{\psi} = \Gj\ket{\psi}$.
For unphysical states, instead, the LPG term splits into many manifolds that can be energetically above and below the target sector~\cite{Halimeh2022LPG}.
This construction allows to reduce experimental complexity from four- to two-body interactions.

Experimentally, we propose to implement strong LPG terms in the Hamiltonian such that quantum dynamics are constrained to remain in LPG eigenspaces by large energy barriers enabling the large-scale quantum simulation of \Ztwo{}~mLGTs in (2+1)D.
To introduce constraint-preserving dynamics within the LPG subspaces, the latter are coupled by weak on-site driving terms of strength~$\Omega \ll V$ as discussed below.
Through the constrained dynamics, a \Ztwo{}~mLGT emerges in an intermediate-energy eigenspace of~$V\Wj$, which is accessible in quantum simulation platforms and which distinguishes our work from previous studies on emergent gauge symmetries, e.g.~\cite{Hermele2004, Glaetzle2014, Samajdar2023}.

The LPG method is built upon stabilizing a high-energy sector of the spectrum, which comes with the caveat that a few unphysical states are resonantly coupled when considering the entire lattice.
In particular, there is a subset of unphysical states that violate Gauss's law on four vertices with energy lowered on three vertices and raised on one vertex; hence these states are on resonance with physical states.
However, numerical simulations in small systems suggest that these gauge-breaking terms only play a subdominant role and gauge-invariance remains intact (Supplementary note~2).

Ultimately, the problem of resonances with a few unphysical states can be remedied by promoting $V \rightarrow V_{\j}$ to be site-dependent such that high-energy sectors can be faithfully protected~\cite{Halimeh2021StabilizingDFL,Halimeh2021EnhancingDFL} against potential gauge non-invariant processes described above (see Methods section).
Site-dependent protection terms do not require any additional experimental capabilities in our protocol described below.
Even more, experimental imperfections inherently give disorder stabilizing the gauge sectors further.
It is also important to note that the presence of only weak disorder (compared to the energy scale~$V$) is enough, which does not alter the effective couplings in the emergent gauge-invariant effective Hamiltonian.

In the following, we introduce the microscopic model that we propose to implement in an experiment.
From the microscopic model, effective Hamiltonians for the \Ztwo{}~mLGT and QDM subspaces can be derived by a Schrieffer-Wolff transformation (Supplementary note~2 and~4). On realistic timescales of experiments, the effective models are gauge-invariant by construction and studied further below.

\textbf{Experimental realization in Rydberg atom arrays.--}
Here, we propose the microscopic model~$\H^{\mathrm{mic}}$ which can be directly implemented in state-of-the-art Rydberg atom arrays in optical tweezers, see Fig.~\ref{fig1}a.

The constituents are qubits, which can be modeled by the ground~$\ket{g}$ and Rydberg~$\ket{r}$ states of individual atoms.
As shown in Fig.~\ref{fig1}a, we label the atoms as \textit{matter atom} or \textit{link atom} depending on their position on the lattice.
The \Ztwo{}~gauge structure then emerges from nearest-neighbor Ising interactions~$V$ realized by Rydberg-Rydberg interactions and hence the real space geometric arrangement plays a key role.
The dynamics is induced by a weak transverse field~$\Omega_{m}$~($\Omega_{l}$), which corresponds to a homogeneous drive between the ground and Rydberg states of the matter (link) atoms.
Moreover, tunability of parameters defining the phase diagram is achieved by a longitudinal field or detuning~$\Delta_{m}$~($\Delta_{l}$) of the weak drive.

The interesting physics emerges in different energy subsectors of the LPG protection term~$\propto V\Wj$ in Eq.~(\ref{eq:LPG}); in particular the \Ztwo{}~mLGT is a sector in the middle of the spectrum of~$\H^{\mathrm{mic}}$.
The suitability for Rydberg atom arrays comes from the flexibility in geometric arrangement required for the LPG term as well as from the natural energy scales~$V \gg \Omega$ in the system, which we use to derive the effective models below, see Eqs.~(\ref{eq:modelH}) and~(\ref{eq:HQDM}).

Matter atoms~$\aj$ form the sites of a honeycomb lattice and we map the empty~$\ket{n_{\j}=0}$ (occupied~$\ket{n_{\j}=1}$) state on the ground state~$\ket{g}_{\j}$ (Rydberg state~$\ket{r})_{\j}$) of the atoms.
Link atoms~$\tauxij$ are located on the links of the honeycomb lattice, i.e.\ a Kagome lattice, and analogously we map the $\tau^x_\ij=+1$ ($\tau^x_\ij=-1$) state on the atomic state $\ket{g}_\ij$ ($\ket{r}_\ij=\ad_\ij\ket{g}_\ij$).
Moreover, we want the matter and link atoms to be in different layers and those layers should be vertically slightly apart in real space to ensure equal two-body interactions between matter and link atoms (Supplementary note~5).
Using the out-of-plane direction has the advantage that it only requires atoms of the same species and with the same internal states.
However, the equal strength interaction can also be achieved in-plane by using e.g. two atomic species or different (suitable) internal Rydberg states for the matter and link atoms.

We first propose a non gauge-invariant microscopic Hamiltonian from which we later derive an effective model with only gauge-invariant terms.
To lowest order in perturbation theory and on experimentally relevant  timescales, the system evolves under an emergent gauge-invariant Hamiltonian.
The microscopic Hamiltonian is given by
\begin{align}
\begin{split} \label{eq:micHam}
    \H^{\mathrm{mic}} &= V\sum_{\j}\Wj
    - \Delta_m\sum_{\j}\nj
    - \frac{\Delta_l}{2}\sum_{\ij} \tauxij\\
    &+ \Omega_m\sum_{\j}\left( \aj + \adj \right)
    + \Omega_l\sum_{\ij} \left( \a_\ij + \ad_\ij \right),
\end{split}
\end{align}
where bosonic operators $\adj$ and $\a_\ij^{(\dagger)}$ annihilate (create) excitations on the matter and link atoms, respectively; $\Wj$ is the LPG term introduced in the main text Eq.~\eqref{eq:LPG}.
The last two terms describe driving of matter ($\ket{g}_{\j} \leftrightarrow \ket{r}_{\j}$) and link atoms ($\ket{g}_\ij \leftrightarrow \ket{r}_\ij$) in the rotating frame.
Rewriting~(\ref{eq:micHam}) in the atomic basis yields Rydberg-Rydberg interactions of strength~$2V$ and renormalized, large detunings~$\tilde{\Delta}_m = -3V + \Delta_m$ and~$\tilde{\Delta}_l = -3V + \Delta_l$.
In a Rydberg setup the driving terms can be realized by an external laser, which couples ~$\ket{g} \leftrightarrow \ket{r}$, while the detunings~$\Delta_m$,~$\Delta_l$ of the laser relative to the resonance frequency controls the electric field~$\Delta_l$ and chemical potential~$\Delta_m$ in the rotating frame.

In the limit $\Omega_m,\,\Omega_l \ll V$, the energy subspaces defined by the LPG term~$V\Wj$, Eq.~(\ref{eq:LPG}), are weakly coupled by the drive to induce effective interactions and it is convenient but not required to choose~$\Omega_m=\Omega_l=\Omega$.
The \Ztwo{}~mLGT emerges as an intermediate-energy eigenspace of the LPG term~$V\Wj$.
The effective interactions in the constrained \Ztwo{}~mLGT and QDM subspaces of~$\Wj$ can be derived by a Schrieffer-Wolff transformation (Supplementary note~2 and~4) and yielding the models discussed in the next section.

In the experiment we propose, the Rydberg-Rydberg interactions are not only restricted to nearest neighbours but are long ranged.
We emphasize that beyond nearest neighbour interactions are inherently gauge invariant and hence do neither influence the LPG gauge protection scheme nor the Schrieffer-Wolff transformation.
However, the long-range interactions can have strong influence on the \Ztwo{}~invariant dynamics.
While the interaction strength decreases as $1/R^6$, where $R$ is the distance between atoms, the interaction is still comparable to the effective perturbative dynamics (Supplementary note~5).
We note that the dynamics might be slowed down but the qualitative features of the \Ztwo{}~mLGT remain intact.

\begin{figure}[t!]
\includegraphics[width=0.4\textwidth]{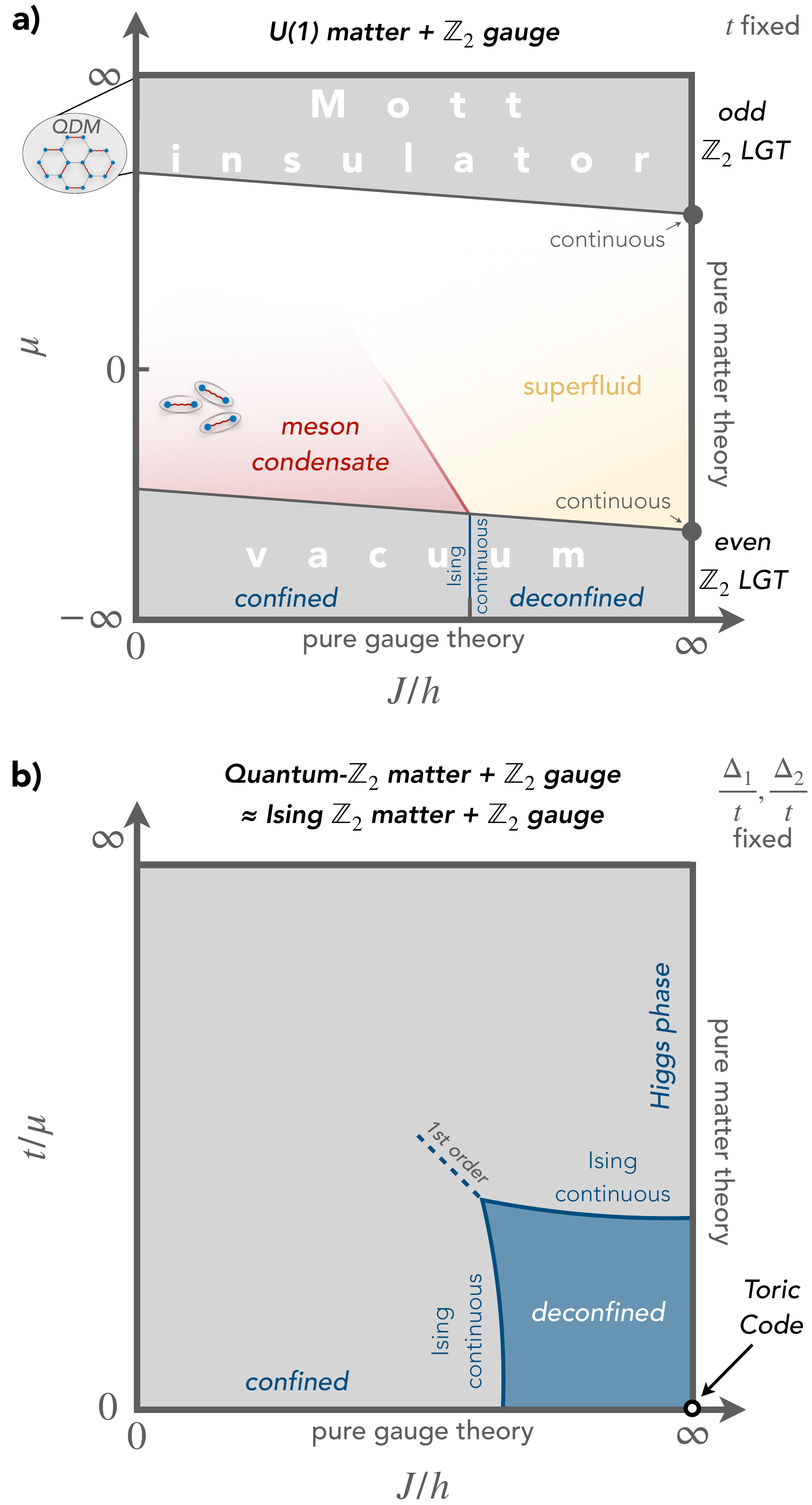}
\caption{\textbf{Conjectured ground-state phase diagrams.} We show two qualitative sketches of phase diagrams for the effective model~(\ref{eq:modelH}). In panel~\textbf{a)}, we consider $U(1)$ matter~($\Delta_1=\Delta_2=0$) coupled to a dynamical \Ztwo{}~gauge field as discussed in the main text. Along the vertical direction the filling is tuned, which yields an even (odd) \Ztwo{}~pure gauge theory in the vacuum (Mott insulator) illustrated by the grey regions. In between the matter and gauge degrees-of-freedom interplay, for which we examined the limiting cases. Above the deconfined region, we expect a superfluid regime (yellow), while above the confined region composite mesons of \Ztwo{}~charges may condense (red).
In panel~\textbf{b)}, we show the phase diagram for an Ising \Ztwo~LGT as proposed by Fradkin and Shenker~\cite{FradkinShenker1979}. The $2$D quantum Hamiltonian of the Ising \Ztwo{}~mLGT has equal hopping~$t$ and pairing~$\Delta_1$ strength and can thus be mapped on a classical $3$D Ising theory. Because our model with quantum \Ztwo{}~matter coupled to dynamical \Ztwo{}~gauge fields has slight anisotropy between hopping and pairing,~$t\neq\Delta_1$, as well as additional anomalous pairing terms~$\Delta_2$, the classical mapping can only work approximately.
We anticipate that the phase diagram should be qualitatively very similar to panel~\textbf{b)}.
}
\label{fig2}
\end{figure}

\textbf{Effective \Ztwo{}~mLGT model.--}
A model is locally \Ztwo{}~invariant if its Hamiltonian~$\H$ commutes with all symmetry generators~$\Gj$, i.e. $[\H,\Gj]=0$ for all~$\j$.
This ensures that all dynamics is constrained to the physical subspace without leaking into unphysical states.
In Eq.~(\ref{eq:LPG}), the target sector is~$g_{\j}=+1$ for all~$\j$ but our scheme can be easily adapted for any $\{g_{\j}\}_{\j}$ (Supplementary note~1).

In the presence of strong LPG protection, the system is energetically enforced to remain in a target gauge sector and unphysical states are only virtually occupied by the drive~$\Omega$.
To be precise, resonant couplings to unphysical sectors are suppressed by the (experimentally feasible) disorder protection scheme discussed above and in the Methods section.
Otherwise emergent gauge-breaking terms appear in third-order perturbation theory.
However, in small systems we have numerically confirmed that even without disorder in the LPG terms Gauss's law is well conserved (Supplementary note~2), which in larger systems we expect to crossover to an \textit{approximate} gauge invariance.
In the following we assume disorder protection or small systems, where leading order gauge-breaking terms are absent or can be neglect, respectively.

For the proposed on-site driving terms discussed above and shown in Fig.~\ref{fig1}a, we derive the following effective Hamiltonian from the microscopic model~(\ref{eq:micHam}) in the intermediate-energy LPG eigenspace (Supplementary note~2):
\begin{widetext}
\begin{align}
\begin{split} \label{eq:modelH}
    \H^{\mathrm{eff}}_{\mathbb{Z}_2} &= \sum_{\ij}\left( t\ad_{\bm{i}} \tauzij \aj
    + \Delta_1 \ad_{\bm{i}} \tauzij \adj
    + \Delta_2 \ad_{\bm{i}} \tauxij\tauzij \adj
    + \mathrm{H.c.} \right)
    - J\sum_{\includegraphics[width=6pt]{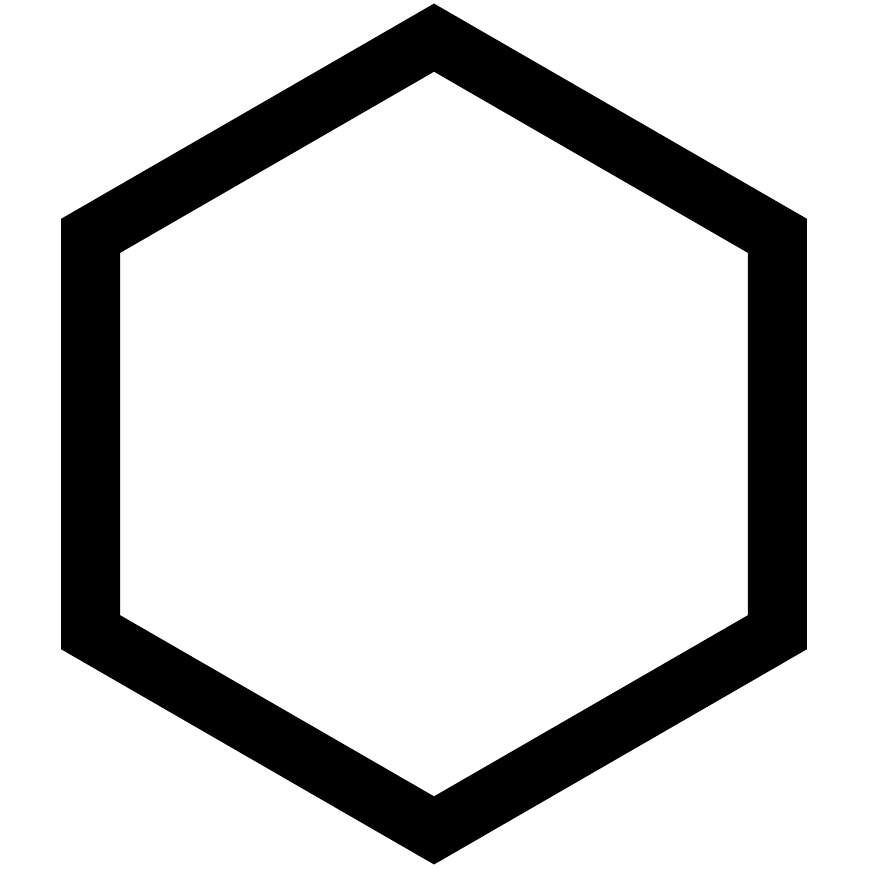}} \prod_{\ij \in\includegraphics[width=6pt, trim= 0 30pt 0 0]{plaquette.pdf}} \tauzij
    - h\sum_{\ij} \tauxij
    - \mu\sum_{\j}  \nj. 
\end{split}
\end{align}
\end{widetext}
The first terms in Eq.~(\ref{eq:modelH}) describe gauge-invariant hopping of matter excitations with amplitude~$t$ and (anomalous) pairing~$\propto \Delta_1$ ($\propto \Delta_2$).
The term~$\propto J$ is the magnetic plaquette interaction on the honeycomb lattice.
The last two terms are referred to as electric field term~$h$ and chemical potential~$\mu$, respectively.
Note that deriving Hamiltonian~(\ref{eq:modelH}) from the microscopic model in Eq.~(\ref{eq:micHam}) yields additional higher-order terms~$\propto \hat{\tau}^x\hat{\tau}^x,\,\hat{\tau}^x\hat{n}$, etc.
In the effective model $\H^{\mathrm{eff}}_{\mathbb{Z}_2}$ we treat these higher-order terms on a mean-field level of the electric field and matter density (Supplementary note~2).
Moreover, we emphasize that the effective model is solely derived from the microscopic Hamiltonian, which only requires a simple set of one- and two-body interactions between the constituents.

For any site~$\j$, one can take \mbox{$\a_{\j} \rightarrow -\a_{\j}$} and \mbox{$\tauzij \rightarrow -\tauzij$}; hence the effective Hamiltonian~(\ref{eq:modelH}) has a local \Ztwo{}~symmetry, $[\H^\mathrm{eff}_{\mathbb{Z}_2},\Gj] = 0~\forall \j$, qualifying it as \Ztwo{}~mLGT in $(2+1)$D. 
In particular, in our proposed scheme we do not have to apply involved steps to engineer \Ztwo{}-invariant interactions but rather we exploit the intrinsic gauge protection by dominant LPG terms, which enforces any weak perturbation to yield an effective \Ztwo{}~mLGT.
This approach also inherently implies robustness against gauge-symmetry breaking terms in experimental realizations.

In the following, we discuss the rich physics of the effective model~(\ref{eq:modelH}).
However, due to the complexity of the system, it is challenging to conduct faithful numerical studies in extended systems.
As a first step, we examine well-known limits of the model and conjecture $T=0$ phase diagrams of the effective Hamiltonian when the \Ztwo{}~gauge field is coupled to $U(1)$ or quantum-\Ztwo{} dynamical~matter, respectively.
We note that the strength of the plaquette interaction can only be estimated (Supplementary note~2) and competes with the long-range Rydberg interactions. 
Moreover, the disorder protection scheme underlying the derivation of the effective Hamiltonian ensures gauge-invariance of the leading order contributions but higher-order gauge breaking terms can in principle appear and affect the physics at very long timescales.

Our effective model describes the physics of experimental system sizes and timescales; the efficiency of the LPG gauge protection in the thermodynamic limit is a subtle open question.
Hence, in the following we discuss phases of the effective model~\eqref{eq:modelH} that may (or may not) emerge from the microscopic model~\eqref{eq:micHam}. 

\textit{$U(1)$~matter.--}
By fixing the number of matter excitations in the system, i.e.\ $\Delta_1=\Delta_2=0$ in Hamiltonian~(\ref{eq:modelH}), the model has a global $U(1)$~symmetry of the matter (hard-core) bosons, which can be achieved by choosing the detuning at the matter sites~$\Delta_m$ comparable to~$V$ in our proposed experimental scheme Eq.~(\ref{eq:micHam}).
Here, we consider the phase diagram when the filling of matter excitations is controlled by the chemical potential~$\mu$.
To map out different possible phases, we fix the hopping~$t$ and study limiting cases.

First, we consider the pure gauge theory with no matter excitations ($\mu \rightarrow -\infty$), see Fig.~\ref{fig2}a (bottom).
The Hamiltonian then reduces to the pure Ising LGT~\cite{Wegner1971} with matter vacuum - an even \Ztwo{} LGT.
The dual of this model exhibits a continuous (2+1)D Ising phase transition, corresponding to a confined (deconfined) phase below (above) a critical~$(J/h)_c$, respectively~\cite{Wegner1971, Kogut1979}.
At the toric code point ($J/h=\infty$) the system is exactly solvable~\cite{Chandran2013} and the gapped ground state has topological order.

Because for~$J/h=\infty$ the gauge field has no fluctuations, we can fix the gauge by setting~$\tau^z_\ij=+1$ and map out the pure matter theory in Fig.~\ref{fig2}a (right).
For finite~$\mu$ we find a model with free hopping of hard-core bosons, for which the filling can be tuned by changing the chemical potential~$\mu$.
Hence, for increasing~$\mu$ and results based on the square lattice~\cite{Bernardet2002, Melko2004} we expect two continuous phase transitions: vacuum-to-superfluid and superfluid-to-Mott insulator.
The Mott insulator phase is an odd \Ztwo{}~LGT because the matter is static and acts as background charge and thus can be treated as a pure gauge theory with $g_{\j} = -1$~\cite{Sachdev2019}.
In the opposite limit~$J/h=0$, the same Mott state gives rise to a hard-core quantum dimer constraint for the \Ztwo{}~electric field lines.
On the square lattice, the quantum dimer model and odd \Ztwo{}~LGT exhibit a phase transition from a confined to deconfined phase~\cite{Borla2022}.
The honeycomb lattice and next-nearest neighbor Rydberg-Rydberg interactions might feature additional symmetry-broken phases.
Hence it requires a sophisticated analysis to map out the substructure of the Mott insulating phase in Fig.~\ref{fig2}a.

In the limit of low fillings and small but finite $J/h \ll 1$, the matter excitations form two-body mesonic bound states~\cite{Borla2022}, which are \Ztwo{}-charge neutral and can be considered as point-like particles.
We can derive an effective meson model yielding hard-core bosons on the sites of a Kagome lattice (Supplementary note~3).

At $T=0$ and sufficiently low densities, the mesons can condense and spontaneously break the emergent global~$U(1)$ symmetry associated with meson number conservation.
To determine the phase boundary of the meson condensate, we consider a single pair of matter excitations doped into the vacuum.
This pair cannot alter the pure gauge phases and thus the two charges can be considered as probes for the (de)confined regime.
For the latter, the matter excitations are bound into mesons, in contrast to free excitations above the deconfined regime.
Hence, the effective description of bound mesonic pairs breaks down at the phase transition of the pure gauge theory indicating the phase boundary of the meson condensate phase at small filling.

At higher densities, dimer-dimer interactions and fluctuations of the gauge field play a role, requiring a more sophisticated analysis to predict the ground state.
We emphasize that the rich physics in this model emerges from the gauge constraint generated by the LPG terms.
Moreover, we note that by lifting the hard-core boson constraint, which is beyond our experimental scheme, the model maps onto a classical XY model coupled to a \Ztwo{}~gauge field~\cite{Sachdev2019}.
This model has been studied on the square lattice in the context of topological phases of matter~\cite{Sachdev2019} and high-Tc superconductivity~\cite{SenthilFisher2000, Sedgewick2002, Podolsky2005}, to name a few.

\textit{Classical mapping.--}
For~$t=\Delta_1$ and~$\Delta_2=0$ the model is well-studied and maps onto a classical Ising lattice gauge theory coupled to Ising \Ztwo{}~matter~\cite{FradkinShenker1979}.
In our experimental proposal~$\Delta_1$ and~$\Delta_2$ cannot be independently tuned, but due to the relevance of the model and its proximity to our effective model we briefly summarize the most important results for the square lattice here, see Fig.~\ref{fig2}b.

In the limit with frozen gauge fields (pure matter axis, $J/h = \infty$) the resulting pure matter theory corresponds to a transverse field Ising model with a global \Ztwo{}~symmetry, which maps to a classical $3$D Ising model and exhibits a continuous phase transition.
On the pure gauge axis ($t/\mu = 0$) the model exhibits a topological phase transition without local order parameters~\cite{Wegner1971}.
Instead, the scaling of non-local Wegner-Wilson loops with their area/perimeter distinguishes the confined from the deconfined phase.
Remarkably, the pure gauge model is also dual to a classical $3$D Ising model, rendering the pure gauge axis dual to the pure matter axis.
The same pure gauge phases are realized for~$\mu \rightarrow -\infty$ in the case with $U(1)$ matter.

For more general~$J/h$, the model's self-duality yields a symmetry in the phase diagram, which allows to study the pure gauge and matter theory in Fig.~\ref{fig2}b but does not reveal the interior away from the axis. 
Fradkin's and Shenker's accomplishment was to show the existence of two distinct, extended phases: the confined and deconfined ``free charge'' phase, which have been confirmed numerically \cite{Vidal2009, Tupitsyn2010}. 
From today's perspective, the latter would be characterized as topological phase of matter in the toric code universality class.

\textit{Quantum-\Ztwo{}~matter.--}
Now, we consider the full effective Hamiltonian~(\ref{eq:modelH}), where hopping and pairing are anisotropic~$t \neq \Delta_1$ and the pairing strength can depend on the electric field configuration~$\Delta_2 \neq 0$, and relate it to Fig.~\ref{fig2}b.
Here, the pure matter theory can no longer be mapped on the classical $3$D Ising model.
Hence, we introduce the term quantum-\Ztwo{}~matter, which emphasizes the matter's \Ztwo{}~symmetry group but points out that a mapping to a known classical model is lacking.

We note that close to the toric code point ($J/h=\infty$ and $t/\mu=0$) in Fig.~\ref{fig2}b, the expectation value of the electric field vanishes,~$\langle \tauxij \rangle=0$, and thus in mean-field approximation the anomalous terms should be negligible and renormalize the pairing~$\Delta_1 \rightarrow \tilde{\Delta}_1$.
For the pure gauge theory it has been shown~\cite{Trebst2007} that the expectation value~$\langle \tauxij \rangle$ continuously changes by tuning the electric field term~$h$.
Hence, by performing a mean-field approximation in the electric field, the quantum-\Ztwo{}~mLGT maps onto the classical Ising \Ztwo{}~mLGT (Supplementary note~2~C).

Due to its proximity to the Ising \Ztwo{}~mLGT and its common symmetries generated by the proposed LPG term, we anticipate that the phase diagram of the quantum-\Ztwo{}~mLGT shares all essential features of the Ising \Ztwo{}~mLGT as shown in Fig.~\ref{fig2}b.

\textbf{Quantum dimer model (QDM).--}
Rokhsar and Kivelson introduced the QDM in the context of \mbox{high-$T_c$} superconductivity, which has the constraint that exactly one dimer is attached to each vertex~\cite{Rokhsar1988, Moessner2010}.
The QDM is an odd \Ztwo{}~LGT, i.e. a pure gauge theory with~$g_{\j}=+1$ replaced by $g_{\j}=-1~\forall\j$, with~$h \rightarrow \infty$, and its fundamental monomer excitations are gapped and can only be created in pairs.

Our proposed scheme allows to directly implement the gauge constraint of the QDM experimentally by preparing the system in the ground-state manifold of the LPG term as shown in Fig.~\ref{fig1}b and d.
Note that the LPG term splits the ground-state manifold into two distinct subspaces, QDM\textsubscript{1} and QDM\textsubscript{2}, which can be seen by entirely removing the matter atoms and setting~$\nj=0,1$ in Eq.~(\ref{eq:LPG}), such that only the link atom Kagome lattice remains; hence it can be implemented in-plane.
A dimer then corresponds to either~$\tau^x_\ij=-1$ (QDM\textsubscript{1}) or $\tau^x_\ij=+1$ (QDM\textsubscript{2}).
Due to the LPG protection the QDM subspaces are energetically protected and monomer excitations cost a finite energy~$2V$.

By weakly driving the system, the motion of virtual, gapped monomer pairs perturbatively induces plaquette terms of strength~$J_{\mathrm{QDM}}$, and we can derive an effective model (Supplementary note~4) given by
\begin{align} \label{eq:HQDM}
    \H^{\mathrm{eff}}_{\mathrm{QDM}} = -J_{\mathrm{QDM}} \sum_{\includegraphics[width=6pt]{plaquette.pdf}}\prod_{\ij \in \raisebox{-.15\height}{\includegraphics[width=6pt]{plaquette.pdf}}} \tauzij + K\sum_{\text{NNN}}\hat{\tau}^x_{\langle \bm{i},\j\rangle}\hat{\tau}^x_{\langle \bm{m},\bm{n}\rangle}.
\end{align}
Here, the NNN link-link interaction~$K$ can be tuned by the blockade radius of the Rydberg-Rydberg interactions.

Experimental~\cite{Semeghini2021} and theoretical~\cite{Verresen2021, Samajdar2021, Giudici2022, Samajdar2023} studies of QDMs in Rydberg atom arrays for different geometries and parameters regimes have shown to be an promising playground to probe \Ztwo{}~spin liquids.
Our proposed setup is a promising candidate to further study QDMs due to its versatility and its inherent protection by the LPG term and the phase diagram of Hamiltonian~(\ref{eq:HQDM}) remains to be explored

Here, we examine two limiting cases of Hamiltonian~(\ref{eq:HQDM}).
For $J_{\mathrm{QDM}}/K \gg 1$, the system is in the so-called plaquette phase~\cite{Moessner2001}, which is characterized by a maximal number of flippable plaquettes and resonating dimers.
On the other hand, for $J_{\mathrm{QDM}}/K \ll 1$ we find a classical Ising antiferromagnet on the Kagome lattice with NN and NNN interactions from the hard-core dimer constraint and $K$-term, respectively.

\textbf{Experimental probes.--}
\begin{figure*}[t]
\includegraphics[width=0.75\textwidth]{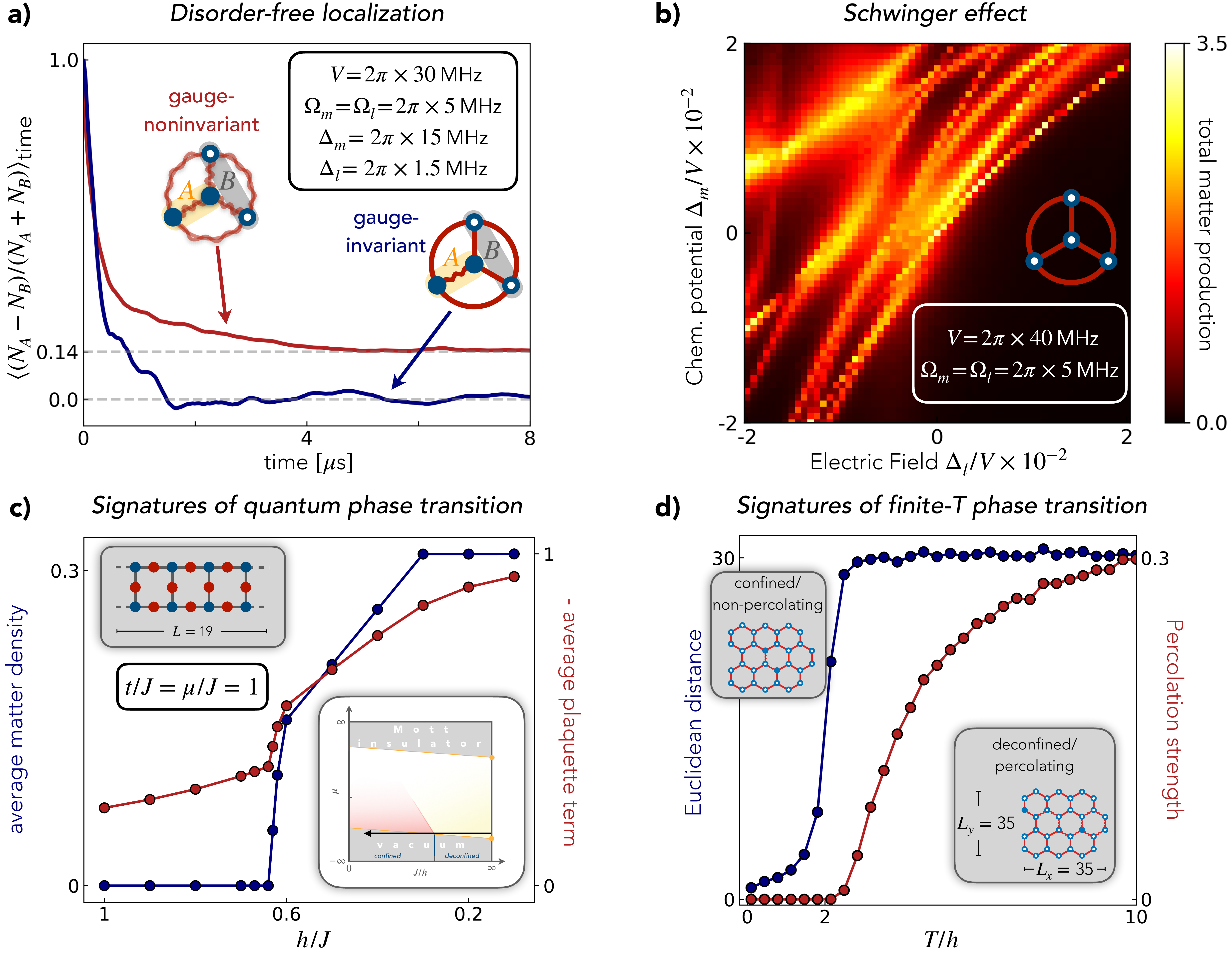}
\caption{\textbf{Experimental probes.} We analyze several observables that could be probed experimentally. Panel~\textbf{a)} and~\textbf{b)} show results from ED simulations of the time-evolution of the microscopic model~(\ref{eq:micHam}) with experimentally realistic parameters in a system with coordination number~$z=3$ (see inset). In panel~\textbf{a)} we observe disorder-free localization by initializing the system in a gauge-invariant (blue curve) and gauge-noninvariant (red curve) initial state with two matter excitations localized in subsystem A and calculating the time-averaged imbalance between subsystem A and B as shown. In panel~\textbf{b)}, we probe the Schwinger effect by quenching the vacuum state with the microscopic model for different experimentally relevant parameters: matter detuning~$\Delta_m$ (chemical potential) and link detuning~$\Delta_l$ (electric field). We find lines of resonance, where the production of matter excitations out of the vacuum is large. In panel~\textbf{c)} we plot the average $U(1)$~matter density (blue curve) obtained from DMRG calculations on a ladder with~$J<0$. We can qualitatively understand the sharp decay of matter as a transition into the vacuum phase as discussed in Fig.~\ref{fig2}a. Additionally, a kink in the plaquette expectation value (red curve) signals a phase transition. In panel~\textbf{d)}, we use two fluctuating test charges to probe a temperature-induced deconfinement transition in a classical limit of our effective model using Monte Carlo simulations. Both in the percolation strength (red curve) and the Euclidean distance of two matter excitations (blue curve), we find that above a certain temperature~$T/h$ the system undergoes a percolation transition. }
\label{fig3}
\end{figure*}
In the following, we discuss potential signatures of the rich physics that can be readily explored with the proposed experimental setup Eq.~(\ref{eq:micHam}).

\textit{Disorder-free localization.--}
Recently, the idea of disorder-free localization (DFL), where averaging over gauge sectors induces disorder, has sparked theoretical interest~\cite{Smith2017, Smith2018}.
DFL is an example where the entire \Ztwo{}~mLGT Hilbert space participates in the dynamics including sectors with~$g_{\j} \neq +1$.
It has been demonstrated that the $(2+1)$D $U(1)$ quantum link models can show DFL~\cite{Karpov2021, Chakraborty2022}; further it was proposed that in a $(1+1)$D \Ztwo{}~LGT, LPG protection leads to enhanced localization~\cite{Halimeh2021EnhancingDFL}.
However, experimental evidence is still lacking.
The scheme we propose is suitable to experimentally study ergodicity breaking without disorder in a strongly interacting $(2+1)$D system with $U(1)$~matter.

In Fig.~\ref{fig3}a we show results of a small-scale exact diagonalization (ED) study using realistic parameters for the experimentally relevant microscopic Hamiltonian (Supplementary note~6).
The system is prepared in two different initial states: 1) A gauge-invariant state~$\ket{\psi^\mathrm{inv}}$, and 2) a gauge-noninvariant state~$\ket{\psi^\mathrm{ninv}}$, both with (without) localized matter excitations in subsystem~$A$ ($B$).

We find distinctly different behaviours for the time-averaged matter occupation imbalance between subsystem~$A$ and~$B$ (Supplementary note~6):
While the gauge-invariant state~$\ket{\psi^\mathrm{inv}}$ thermalizes, the gauge-noninvariant state~$\ket{\psi^\mathrm{ninv}}$ breaks ergodicity on experimentally relevant timescales.
Experimentally much larger systems can be addressed.

\textit{Schwinger effect.--}
The Schwinger effect describes the creation of pairwise matter excitations from vacuum in strongly-coupled gauge theories~\cite{Martinez2016}.
Here, we use the Schwinger effect to test the validity of our LPG scheme.
Starting from the microscopic model~(\ref{eq:micHam}), we time-evolve the vacuum state with no matter excitations and extract the maximum number of created matter excitations in the initial gauge sector~$g_{\j}=+1~\forall \j$.
As shown in Fig.~\ref{fig3}b, by tuning the electric field and chemical potential we find resonance lines, where many matter excitations are produced in the system, and we verify that gauge-invariant processes dominate (Supplementary note~7).

\textit{Phase transitions in a ladder geometry.--} 
Our proposed scheme is suitable for any geometry with coordination number~$z=3$; hence one can experimentally study square ladders of coupled $1$D chains.
Here, we have examined the ground state of Hamiltonian~\eqref{eq:modelH} with $U(1)$~matter using the density matrix renormalization group (DMRG) technique~\cite{Schollwoeck2011} (Supplementary note~8) on a ladder and we find signatures of a quantum phase transition.
As shown in Fig.~\ref{fig3}c, both the average density of matter excitations and the plaquette terms, which are experimentally directly accessible by projective measurements, change abruptly by tuning the electric field~$h$ indicating a transition into the vacuum phase.
We emphasize that the ladder geometry is different from the $(2+1)$D model studied in Fig.~\ref{fig2}a, however numerical simulations suggest the presence of a phase transition and hence the ladder geometry offers a numerically and experimentally realistic playground for future studies of our model.

\textit{Thermal deconfinement from string percolation.--}
We examine a temperature-induced deconfinement transition in a classical limit of our effective model~\eqref{eq:modelH}, which neglects charge and gauge dynamics~$t=\Delta_{1,2}=J=0$.
We use Monte Carlo simulations on a $35\!\times\!35$ honeycomb lattice (Supplementary note~9).

To study thermal deconfinement, we consider exactly two matter excitations which, due to Gauss's law, have to be connected by a string~$\Sigma$ of electric field lines; i.e. $\Sigma$ is a path of links with electric fields $\tau^x_\ij = -1$ for~$\ij \in \Sigma$.
This setting can be used as a probe of a deconfined (confined) phase, in which the \Ztwo{}~matter is free (bound)~\cite{Hahn2022}.

To determine the classical equilibrium state, we note the following:
1) Due to the electric field term~$h$ in the Hamiltonian, a string of flipped electric fields~$\tau^x_\ij=-1$ costs an energy~$2h \cdot \ell $, where~$\ell$ is the length of the string.
2) Gauss's law enforces that at least one string is connected to each matter excitation.

Hence, in the classical ground state the two matter excitations form a mesonic bound state on nearest neighbor lattice sites.
Therefore, the matter excitations are confined by a linear string potential.
In the co-moving frame of one matter excitation, this model can approximately be described as a particle in a linear confining potential.

At non-zero temperature~$T>0$, the entropy contribution to the free energy~$F=E-TS$ must also be considered.
Even though the electric field term~$h$ yields an approximately linear string tension, the two charges can separate infinitely in thermal equilibrium provided that~$E(\ell)<T\log(N_{\ell})$ for~$\ell \rightarrow \infty$, where $\log(N_{\ell})=S$ denotes the entropy~$S$ of all the string states~$N_{\ell}$ with length~$\ell$ (setting $k_B=1$) and~$E(\ell)$ is their typical energy~\cite{Hahn2022}.
This happens beyond a critical temperature~$T>T_c$, when a percolating net of \Ztwo{}~electric strings forms.

At the critical temperature~$T_c$ we anticipate a thermal deconfinement transition, where matter excitations become free \Ztwo{}~charges (bound mesons) for~$T>T_c$ ($T<T_c$).
To study this transition we use the percolation strength -- a measure for the spatial extend of a global string net (see Methods) -- as an order parameter for the deconfined phase. 
For experimentally realistic parameters, we find a sharp transition for both the percolation strength and Euclidean distance between two matter excitations around $(T/h)_c \approx 2$ as shown in Fig.~\ref{fig3}d.
Although our classical simulation neglects quantum fluctuations, we expect that the revealed finite-temperature deconfinement transition is qualitatively captured.

For a finite density of matter excitations in the system, the Euclidean distance is not a reasonable measure anymore.
However, we speculate that a percolation transition might be related to (de)confinement at finite densities.
How this transition is related to the quantum deconfinement transition at~$T=0$~\cite{Mildenberger2022, Halimeh2022_PRXQ}, driven by quantum fluctuations, will be subject of our future research.
Hence, experimentally exploring this transition not only in the classical case, but also in the presence of quantum fluctuations could give insights in the mechanism of charge (de)confinement.

\section{Conclusion}
We introduced an experimentally feasible protection scheme for \Ztwo{}~mLGTs and QDMs in $(2+1)$D based on two-body interactions, where the \Ztwo{}~gauge structure emerges from well-defined subspaces at high and low energy, respectively.
The scheme not only allows reliable quantum simulation of gauge theories but provides an accessible approach to engineer gauge-invariant Hamiltonians.
We derived an effective \Ztwo{}~mLGT, Eq.~(\ref{eq:modelH}), and QDM, Eq.~(\ref{eq:HQDM}), and discussed some of their rich physics.
In particular, we suggested several experimental probes, for which we provide numerical analysis using ED of the experimentally relevant microscopic model~(\ref{eq:micHam}) as well as DMRG and Monte Carlo simulations of the effective models.
Experimentally, we anticipate that significantly larger systems are accessible.

Our proposed scheme is not only suitable and realistic to be implemented in Rydberg atom arrays, see Eq.~(\ref{eq:micHam}), but it is also of high interest for future theoretical and numerical studies.
Hard-core bosonic matter coupled to \Ztwo{}~gauge fields in $(2+1)$D plays a role in theoretical models, e.g. in the context of high-Tc superconductivity~\cite{SenthilFisher2000}.
While certain limits such as the fine-tuned, classical limit studied by Fradkin and Shenker~\cite{FradkinShenker1979} or coupling to fermionic matter~\cite{Gazit2017, Borla2022} are well-understood, surprisingly little is known about the physics of our proposed model.
What are the implications of anisotropic hopping and pairing~$t \neq \Delta_1$ or anomalous pairing terms~$\Delta_2$, i.e. when the classical mapping fails?
How can (de)confinement in the presence of dynamical matter be captured?
Is disorder-free localization a mechanism for ergodicity breaking in $(2+1)$D?
The possibility to study these questions experimentally will spark future theoretical interest.

\textbf{Acknowledgments.--}
We thank M. Aidelsburger, D. Bluvstein, D. Borgnia, N.C. Chiu, S. Ebadi, M. Greiner, J. Guo, P. Hauke, J. Knolle, M. Lukin, N. Maskara, R. Sahay, C. Schweizer, R. Verresen and T. Wang for fruitful discussions.
L.H. acknowledges support from the Studienstiftung des
deutschen Volkes.
This research was funded by the European Research Council (ERC) under the European Union’s Horizon 2020 research and innovation programm (Grant Agreement no 948141) — ERC Starting Grant SimUcQuam, by the Deutsche Forschungsgemeinschaft (DFG, German Research Foundation) under Germany's Excellence Strategy -- EXC-2111 -- 390814868 and via Research Unit FOR 2414 under project number 277974659, by the NSF through a grant for the Institute for Theoretical Atomic, Molecular, and Optical Physics at Harvard University and the Smithsonian Astrophysical Observatory, and by the ARO grant number W911NF-20-1-0163.

\section*{Methods}

\textbf{Local pseudogenerators for \Ztwo{}~mLGTs.--}
The implementation of LGTs in quantum simulation platforms have two inherent challenges to overcome:
\begin{enumerate}
    \item The physical Hilbert space of gauge theories is highly constrained and given by the gauge constraint~$\Gj\ket{\psi^\mathrm{physical}}=g_{\j}\ket{\psi^\mathrm{physical}}$. In contrast the Hilbert space of the experimental setup is larger and also contains unphysical states~$\ket{\psi^\mathrm{unphysical}}$, which do not satisfy Gauss's law. Therefore, the dynamics of the system is fragile in the presence of experimental errors which couple physical and unphysical states. However, it has been shown that this can be reliably overcome by energetically gapping the physical from unphysical states using stabilizer/protection terms in the Hamiltonian~\cite{Halimeh2021PRXQ, HalimehHauke2022ReviewLinearProtection}. These strong stabilizer terms can be understood as \textit{strong projectors} onto its energy eigenspaces, which are chosen to be the physical subsectors of a \Ztwo{}~gauge theory in our case; hence the effective dynamics is constraint to quantum Zeno subspaces~\cite{Facchi2002}. Note that here the quantum Zeno effect is fully determined by a unitary time-evolution and not driven by dissipation, in agreement with the original effect~\cite{Facchi2002}.
    
    The obvious choice of such a protection term is the symmetry generator, Eq.~(\ref{eq:Goperator}). However, this requires strong and hence unfeasible multi-body interactions. In contrast, the LPG term~$\Wj$, Eq.~(\ref{eq:LPG}), only contains two- and one-body terms and is engineered such that an energy gap between the physical and unphysical states is introduced under the reasonable condition that only one (target) gauge sector is protected. In particular, the LPG term in the~$2$D honeycomb lattice fulfills the condition
    \begin{align}
        V\Wj\ket{\psi^\mathrm{physical}} &= +V\ket{\psi^\mathrm{physical}}\\
        V\Wj\ket{\psi^\mathrm{unphysical}} &= \begin{cases} +4V\ket{\psi^\mathrm{unphysical}} \\ ~ \\ 0V\ket{\psi^\mathrm{unphysical}}\end{cases},
    \end{align}
    where~$V$ is the strength of the LPG term. The spectrum of~$\Wj$ for the gauge choice~$g_{\j}=+1$ is illustrated in Fig.~\ref{fig1}c.
    
    \item To study gauge theories, a $\mathbb{Z}_2$-invariant Hamiltonian has to be engineered first, e.g. the Hamiltonian~(\ref{eq:modelH}) discussed in the main text. In our scheme we exploit the LPG term with its large gap between energy sectors to construct an effective Hamiltonian perturbatively as explained in Supplementary note~2.
\end{enumerate}

To faithfully stabilize large systems for -- in principle -- infinitely long times, we want to discuss the stabilization of high-energy sectors by considering undesired instabilities/resonances in the spectrum~$V\sum_{\j} \Wj$. 
The eigenvalues of $V\Wj$ are $w_{\j}=(0,\,V,\,4V)$ and we want to protect a sector with intermediate energies.
If the interaction strength~$V$ is equally strong at each vertex gauge-symmetry breaking can occur.
For example, by exciting vertex $\j_0$ and simultaneously de-exciting three vertices $\j_1$, $\j_2$ and $\j_3$.
This process has a net energy difference of $\Delta E= +3V - 3\cdot V=0$ and the resonance between the two states can lead to an instability towards unphysical states, hence gauge-symmetry breaking (Supplementary note~2~G).

Therefore, the LPG method without disorder cannot energetically protect against \textit{some} states that break Gauss's law on four vertices.
An efficient way to stabilize the gauge theory even against such scenarios is to introduce disorder in the coupling strengths by $\W=\sum_{\j}V_{\j}\Wj$ with $V_{\j} = V + \delta V_{\j}$.
The couplings~$\delta V_{\j}$ are random and form a so-called compliant sequence~\cite{Halimeh2021PRXQ, Halimeh2022LPG}.
In $1$D systems, this has been shown to faithfully protect \Ztwo{}~LGTs also for extremely long times, see Ref.~\cite{Halimeh2022LPG} for a detailed discussion of (non)compliant sequences.
Moreover, we note that for small system sizes and experimentally relevant timescales even noncompliant sequences such as the simple choice~$V_{\j}=V~\forall\j$ lead to only small errors (Supplementary note~2~G).

\begin{figure}[t]
\includegraphics[width=0.45\textwidth]{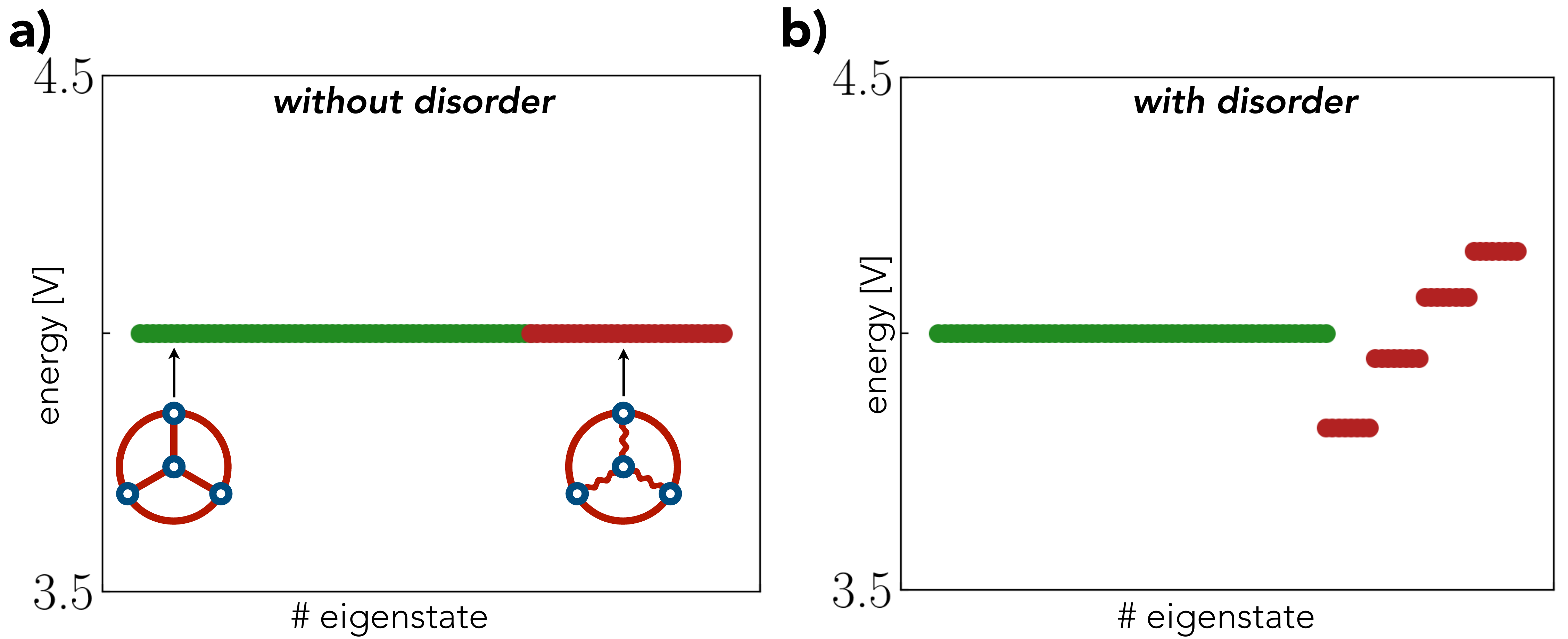}
\caption{\textbf{Disorder-based protection scheme.} We calculate the spectrum of the minimal model studied in Fig.~\ref{fig3}a)-b) with~$\Omega=0$ and plot all eigenstates around energy~$E=4V$. Green (red) dots are states that fulfil (break) Gauss's law as illustrated with two examples in the inset of panel~\textbf{a)}. Without disorder, i.e.\ $V_{\j}=V$ for all~$\j$, the physical and unphysical states are on resonance. In panel~\textbf{b)}, we show the effect of disordered protection terms $V_{\j} = V + \delta V_{\j}$, which only shifts the unphysical states out of resonance and hence fully stabilizes the gauge theory. We note that even without disorder, the emergent gauge structure is remarkably robust~(Supplementary note~2~G). }
\label{figDisorder}
\end{figure}

For our $(2+1)$D model, we illustrate the effect of disordered protection terms in Fig.~\ref{figDisorder}, which shows that only the gauge non-invariant states are shifted out of resonance.
Moreover, we propose to use weak disorder such that the overall perturbative couplings remain unchanged in leading order.
We emphasize that the disorder scheme does not require any additional experimental capabilities but only arbitrary control over the geometry as well as local detuning patterns.
Even more, an experimental realization will always encounter slight disorder, i.e.\ the gauge non-invariant processes might already be sufficiently suppressed in experiment.

We further note that the example above, where Gauss's law is violated on four vertices, yields gauge-breaking terms in third-order perturbation theory.
Ensuring that none of the protection terms~$V_{\j}$ have gauge-breaking resonances within such a nearest-neighbour cluster, these terms can be suppressed.
However, now it remains space for fifth-order breaking terms on next-nearest neighbour vertices.
Hence, the non-resonance condition is now desired on a larger cluster and so forth. 
Therefore, systematically choosing the disorder potentials can suppress gauge-breaking terms to arbitrary finite order and stabilize gauge invariance up to exponential times.
Its fate in the thermodynamic limit, however, is an open question beyond the scope of this study.

\textbf{Percolating strings from classical Monte Carlo.--}
The finite temperature percolation transition in Fig.~\ref{fig3}d is obtained from classical Monte Carlo simulations on the honeycomb lattice with matter and link variables.
In this section, we discuss the percolation strength order parameter~\cite{Essam1980} and details of the numerical simulations in more detail.

The classical model we consider is motivated by the microscopic Hamiltonian~(\ref{eq:micHam}) and its effective model~(\ref{eq:modelH}) - in particular we used the precise effective model as derived in Eq.~(S13) of Supplementary note~2 for~$\Omega/V=1/8$,~$\Delta_m=V/2$ and~$\Delta_l/V \approx 0.044$.
For elevated temperatures~$T \lesssim V$, we expect that classical fluctuations dominate in the system while the Gauss's law constraint is still satisfied due to the LPG protection.
Therefore, we neglect quantum fluctuations and set~$t=\Delta_1=\Delta_2=J=0$.
Hence, the resulting matter-excitation conserving Hamiltonian is purely classical and a configuration is fully determined by the distribution of matter and electric field lines under the Gauss's law constraint, i.e.~$\{ (n_{\j}, \tau^x_\ij)~~|~~(-1)^{n_{\j}}=g_{\j}\prod_{\bm{i}: \ij}\tau^x_\ij~\forall \j \}$ and we consider the sector with~\mbox{$g_{\j}=+1~\forall \j$}.

From the numerical Monte Carlo simulation, we want to quantify the features discussed in the main text: 1) string net formation and 2) bound versus free matter excitations.
To this end, we define the percolation strength as the number of strings in the largest percolating cluster of \Ztwo{}~electric strings, normalized to the system size.
Furthermore, we consider the Euclidean distance between two matter excitation and show that an abrupt change of behaviour in this quantity indicates the disappearance of the bound state.

The Monte Carlo simulations are performed on a $35\!\times\!35$ honeycomb lattice (in units of lattice spacing) using classical Metropolis-Hastings sampling (Supplementary note~9).
Further analysis of the obtained samples allows to extract the number of strings in the largest percolating cluster to calculate the percolation strength.
As shown in Fig.~\ref{fig3}d, we find that for low temperatures~$T$ the percolation strength vanishes.
At a critical temperature~$(T/h)_c\approx 2$, the percolation strength abruptly increases, i.e. the string net percolates.
Moreover, at the same critical temperature~$(T/h)_c\approx 2$ the Euclidean distance shows a drastic change of behavior and saturates at about~$30$ for high temperatures.
This saturation can be explained by the finite system size.

\section{Data availability}
The datasets generated and/or analysed during the current study are available from
the corresponding author on reasonable request.

\section{Code availability}
The data analysed in the current study has been obtained using the open-source tenpy package; this DMRG code is available via GitHub at \url{https://github.com/tenpy/tenpy} and the documentation can be found at \url{https://tenpy.github.io/#}.
The code used in the exact diagonalization and Monte Carlo studies are available from
the corresponding author on reasonable request.

\section{Author Contributions}
LH, JCH and FG devised the initial concept.
LH proposed the idea for the two-dimensional model, worked out the main analytical calculations and performed the exact diagonalization studies.
LH, AB and FG proposed the experimental scheme.
SL performed the Monte Carlo simulations.
AB conducted the DMRG calculations. 
All authors contributed substantially to the analysis of the theoretical results and writing of the manuscript.

\section{Competing interests}
Authors declare that they have no competing interests.

\renewcommand\refname{Reference}

\newpage~
\newpage~

\setcounter{equation}{0}
\setcounter{figure}{0}
\setcounter{table}{0}
\setcounter{section}{0}
\setcounter{page}{1}
\renewcommand{\theequation}{S\arabic{equation}}
\renewcommand{\thefigure}{S\arabic{figure}}
\renewcommand{\thetable}{S\Roman{table}}

\title{Supplementary Information: Realistic scheme for quantum simulation of $\mathbb{Z}_2$ lattice gauge theories with dynamical matter in $(2+1)$D}
\maketitle

\onecolumngrid

\definecolor{verylightgray}{rgb}{0.93, 0.93, 0.93}
\begin{table*}[h!!]
\begin{tabularx}{\textwidth}{|c|X|c|c|}\specialrule{.2em}{.1em}{.1em} 
    \rowcolor{verylightgray}\multicolumn{1}{|c|}{\multirow{1}{*}{SI}}
    & \multicolumn{1}{c|}{\multirow{1}{*}{Summary}}
    & \multicolumn{1}{c|}{\multirow{1}{*}{Model}}
    & \multicolumn{1}{c|}{\multirow{1}{*}{Main results}} \\
    \specialrule{.25em}{.1em}{.1em} 

    \multicolumn{1}{|c|}{\multirow{2}{*}{\ref{app:LPG}}}
    & \multicolumn{1}{X|}{\multirow{2}{*}{\parbox{\linewidth}{Local pseudogenerator method for odd \Ztwo{}~mLGTs and for QDMs}}}
    & \multicolumn{1}{c|}{\multirow{1}{*}{\Ztwo{}~mLGT}}
    & \multicolumn{1}{c|}{\multirow{2}{*}{Fig.~\ref{fig1}b-c}} \\
    & & \multicolumn{1}{c|}{\multirow{1}{*}{QDM}} &\\
    \specialrule{.25em}{.1em}{.1em} 
    
    \multicolumn{1}{|c|}{\multirow{2}{*}{\ref{app:EffHam}}}
    & \multicolumn{1}{X|}{\multirow{2}{*}{\parbox{\linewidth}{General procedure to derive the effective \Ztwo{}~mLGT Hamiltonians from the microscopic model}}}
    & \multicolumn{1}{c|}{\multirow{2}{*}{\Ztwo{}~mLGT}}
    & \multicolumn{1}{c|}{\multirow{2}{*}{Eqs.~\eqref{eq:modelH} and~\eqref{eq:micHam}}} \\
    & &  &\\
    \specialrule{.005em}{.1em}{.1em} 
    
    \multicolumn{1}{|c|}{\multirow{2}{*}{\ref{app:effHamU1}/\ref{app:PlaquetteU1}}}
    & \multicolumn{1}{X|}{\multirow{2}{*}{\parbox{\linewidth}{Effective Hamiltonian / Plaquette interactions}}}
    & \multicolumn{1}{c|}{\multirow{2}{*}{$U(1)$~matter}}
    & \multicolumn{1}{c|}{\multirow{2}{*}{Figs.~\ref{fig2}a and~\ref{fig3}c-d, Eq.~\eqref{eq:modelH} }} \\
     &  &  &\\
    \specialrule{.005em}{.1em}{.1em} 
    
    \multicolumn{1}{|c|}{\multirow{2}{*}{\ref{app:effHamNoU1}/\ref{app:PlaquetteZ2}}}
    & \multicolumn{1}{X|}{\multirow{2}{*}{\parbox{\linewidth}{Effective Hamiltonian / Plaquette interactions}}}
    & \multicolumn{1}{c|}{\multirow{2}{*}{quantum-\Ztwo{}~matter}}
    & \multicolumn{1}{c|}{\multirow{2}{*}{Fig.~\ref{fig2}b, Eq.~\eqref{eq:modelH}}} \\
    & &  &\\
    \specialrule{.005em}{.1em}{.1em} 
    
    \multicolumn{1}{|c|}{\multirow{2}{*}{\ref{app:EDEffHam}/\ref{app:MicVsEff}}}
    & \multicolumn{1}{X|}{\multirow{1}{*}{\parbox{\linewidth}{Exact diagonalization studies of the microscopic and effective models}}}
    & \multicolumn{1}{c|}{\multirow{1}{*}{quantum-\Ztwo{} and}}
    & \multicolumn{1}{c|}{\multirow{2}{*}{Eqs.~\eqref{app:eqEffHamZ2mLGTU1matter} and~\eqref{app:eqEffHamZ2mLGTZ2matter} }} \\
    & &  \multicolumn{1}{c|}{\multirow{1}{*}{$U(1)$~matter}} &\\
    \specialrule{.005em}{.1em}{.1em} 
    
    \multicolumn{1}{|c|}{\multirow{2}{*}{\ref{app:GaugeBreaking}}}
    & \multicolumn{1}{X|}{\multirow{2}{*}{\parbox{\linewidth}{Gauge non-invariant processes}}}
    & \multicolumn{1}{c|}{\multirow{1}{*}{quantum-\Ztwo{} and}}
    & \multicolumn{1}{c|}{\multirow{2}{*}{ Eq.~\eqref{eq:modelH} and Fig.~\ref{app:MicVsEff}a-b }} \\
    & &  \multicolumn{1}{c|}{\multirow{1}{*}{$U(1)$~matter}} &\\
    \specialrule{.25em}{.1em}{.1em} 

    \multicolumn{1}{|c|}{\multirow{2}{*}{\ref{app:effHamU1DimerPhase}}}
    & \multicolumn{1}{X|}{\multirow{2}{*}{\parbox{\linewidth}{Derivation of the effective meson model}}}
    & \multicolumn{1}{c|}{\multirow{1}{*}{\Ztwo{}~mLGT}}
    & \multicolumn{1}{c|}{\multirow{2}{*}{Fig.~\ref{fig2}a}} \\
    & & \multicolumn{1}{c|}{\multirow{1}{*}{$U(1)$~matter}} &\\
    \specialrule{.25em}{.1em}{.1em} 
    
    \multicolumn{1}{|c|}{\multirow{2}{*}{\ref{app:EffHamQDimer}}}
    & \multicolumn{1}{X|}{\multirow{2}{*}{\parbox{\linewidth}{Derivation of the effective quantum dimer model (QDM) incl. plaquette interactions}}}
    & \multicolumn{1}{c|}{\multirow{2}{*}{QDM}}
    & \multicolumn{1}{c|}{\multirow{2}{*}{Eq.~\eqref{eq:HQDM}}} \\
    & & &\\
    \specialrule{.25em}{.1em}{.1em} 
    
    \multicolumn{1}{|c|}{\multirow{2}{*}{\ref{app:expReal}}}
    & \multicolumn{1}{X|}{\multirow{2}{*}{\parbox{\linewidth}{Details about the experimental realization in Rydberg atom arrays}}}
    & \multicolumn{1}{c|}{\multirow{1}{*}{\Ztwo{}~mLGT}}
    & \multicolumn{1}{c|}{\multirow{2}{*}{Fig.~\ref{fig1}a, Eq.~\eqref{eq:micHam}}} \\
    & & \multicolumn{1}{c|}{\multirow{1}{*}{QDM}} &\\
    \specialrule{.25em}{.1em}{.1em} 
    
    \multicolumn{1}{|c|}{\multirow{2}{*}{\ref{app:DFL}}}
    & \multicolumn{1}{X|}{\multirow{2}{*}{\parbox{\linewidth}{Disorder-free localization in the Mercedes star model (main text) and $1$D Zig-Zag chain (SI only); exact diagonalization}}}
    & \multicolumn{1}{c|}{\multirow{1}{*}{\Ztwo{}~mLGT}}
    & \multicolumn{1}{c|}{\multirow{2}{*}{Fig.~\ref{fig3}a, Eq.~\eqref{eq:micHam}}} \\
    & & \multicolumn{1}{c|}{\multirow{1}{*}{$U(1)$~matter}} &\\
    \specialrule{.25em}{.1em}{.1em}
    
    \multicolumn{1}{|c|}{\multirow{2}{*}{\ref{app:Schwinger}}}
    & \multicolumn{1}{X|}{\multirow{2}{*}{\parbox{\linewidth}{Details about the Schwinger effect; exact diagonalization}}}
    & \multicolumn{1}{c|}{\multirow{1}{*}{\Ztwo{}~mLGT}}
    & \multicolumn{1}{c|}{\multirow{2}{*}{Fig.~\ref{fig3}b, Eq.~\eqref{eq:micHam}}} \\
    & & \multicolumn{1}{c|}{\multirow{1}{*}{quantum-\Ztwo{}~matter}} &\\
    \specialrule{.25em}{.1em}{.1em}
    
    \multicolumn{1}{|c|}{\multirow{2}{*}{\ref{app:DMRG}}}
    & \multicolumn{1}{X|}{\multirow{2}{*}{\parbox{\linewidth}{Density matrix renormalization group (DMRG) in the ladder}}}
    & \multicolumn{1}{c|}{\multirow{1}{*}{\Ztwo{}~mLGT}}
    & \multicolumn{1}{c|}{\multirow{2}{*}{Figs.~\ref{fig2}a and~\ref{fig3}c, Eq.~\eqref{eq:modelH}}} \\
    & & \multicolumn{1}{c|}{\multirow{1}{*}{$U(1)$~matter}} &\\
    \specialrule{.25em}{.1em}{.1em}
    
    \multicolumn{1}{|c|}{\multirow{2}{*}{\ref{app:Deconfinement}}}
    & \multicolumn{1}{X|}{\multirow{2}{*}{\parbox{\linewidth}{Classical Monte Carlo simulations on the $2$D honeycomb lattice}}}
    & \multicolumn{1}{c|}{\multirow{1}{*}{\Ztwo{}~mLGT}}
    & \multicolumn{1}{c|}{\multirow{2}{*}{Fig.~\ref{fig3}d, Eq.~\eqref{app:eqEffHamZ2mLGTU1matter}}} \\
    & & \multicolumn{1}{c|}{\multirow{1}{*}{$U(1)$~matter}} &\\
    \specialrule{.25em}{.1em}{.1em}

\end{tabularx}
\caption{Overview and Summary of the Supplementary Information.}
\end{table*}

\section{Local pseudogenerator on the $2$D honeycomb lattice}
\label{app:LPG}

In the following, we discuss local pseudogenerators (LPG) for arbitrary \Ztwo{}~mLGT gauge sectors as well as for QDMs.

\subsection{LPG for \Ztwo{}~mLGTs and $g_j=-1$}
\label{app:LPGZtwo}
\begin{figure*}[t]
\includegraphics[width=\textwidth]{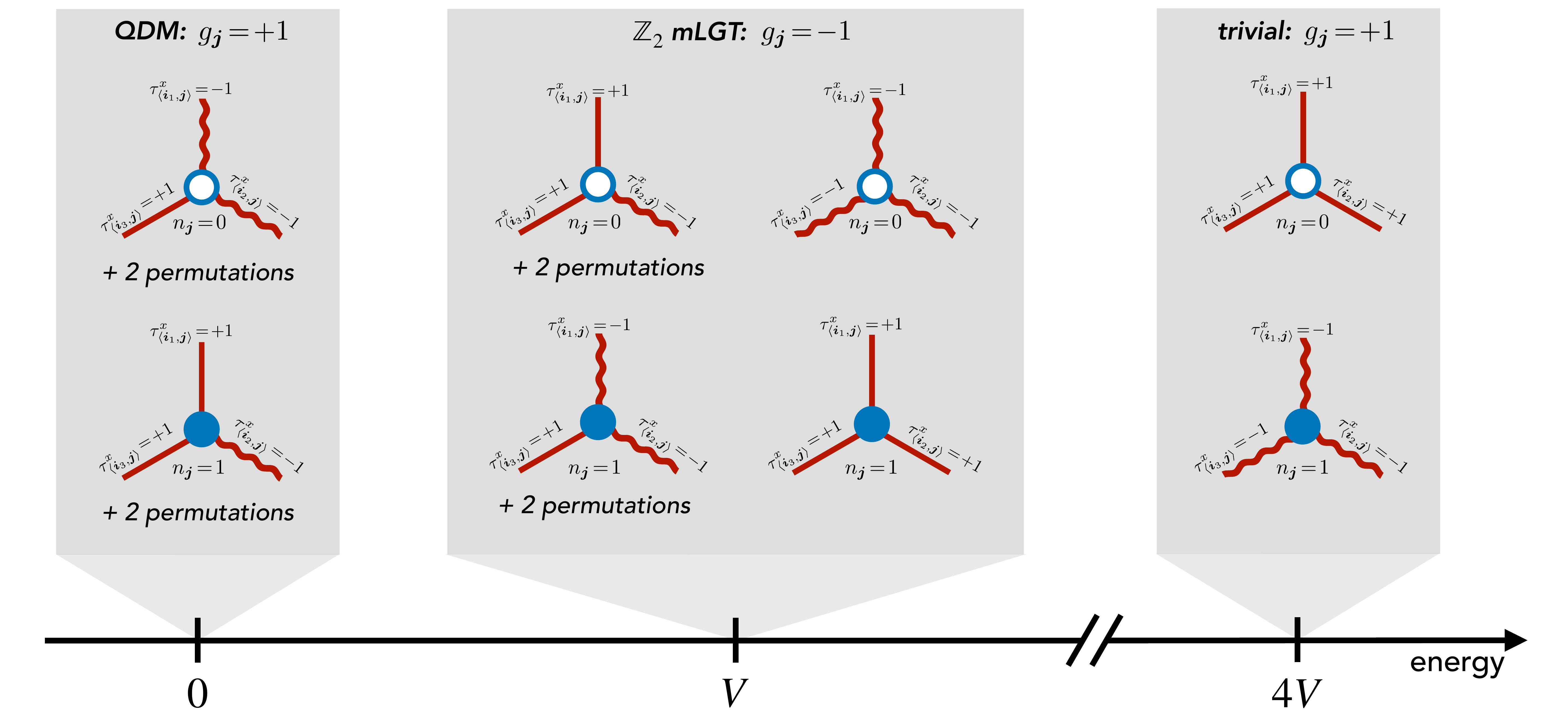}
\caption{\textbf{Spectrum of the local pseudogenerator for $g_{\j}=-1$.} We show the LPG term for \Ztwo{}~mLGTs with $g_{\j}=-1$.   }
\label{app:figLPG}
\end{figure*}

The LPG term in the main text, Eq.~(\ref{eq:LPG}), can be easily generalized to protect any of the two~$g_{\j}=\pm1$ sectors by choosing:
\begin{align} \label{app:eqLPGneg}
    V\W^{g}_{\j} =  \frac{V}{4} \left[ g_{\j}\left( 2\nj - 1 \right) +  \sum_{\bm{i}: \ij}\tauxij \right]^2.
\end{align}
The case~$g_{\j}=+1$ is shown and discussed in the main text, Fig.~\ref{fig1}c, while the case $g_{\j}=-1$ is illustrated in Fig.~\ref{app:figLPG}.

\subsection{Quantum Dimer Models}
\label{app:LPGQDM}
Rokshar and Kivelson~\cite{Rokhsar1988} introduced the QDM as a toy model to study short-range resonating valence bond (RVB) states on the square lattice.
Their model has two phases: a columnar and a staggered phase.
At the phase transition, the so-called Rokshar-Kivelson point, the model becomes exactly solvable and has deconfined monomer excitations.
The experimental challenge is to impose the hard-core dimer constraint and to gap out monomers -- the fundamental, fractionalized excitations of the system.
Here, the LPG term overcomes both challenges.

As shown in Fig.~\ref{fig1}c the ground-state manifold of the LPG term allows for six different configurations per vertex~$\j$.
The subsector with $n_{\j}=0$ ($n_{\j}=1$) should be called QDM\textsubscript{1} (QDM\textsubscript{2}) and we want the two subsectors to be decoupled.
This can be exactly fulfilled by entirely eliminating the local matter degrees-of-freedom, i.e.\ experimentally only the link atoms on the Kagome lattice are implemented, see Fig.~\ref{fig1}a.
Hence, the LPG term for the two subsectors read
\begin{align}
    V\Wj^{\mathrm{QDM}_1} &= \frac{V}{4} \left[ \sum_{\bm{i}: \ij}\tauxij + 1\right]^2\\
    V\Wj^{\mathrm{QDM}_2} &= \frac{V}{4} \left[   \sum_{\bm{i}: \ij}\tauxij - 1\right]^2
\end{align}

In contrast to the \Ztwo{}~mLGT, we note that the QDM\textsubscript{1} (QDM\textsubscript{2}) subspaces are now the lowest-energy eigenspaces of the LPG term.
Therefore, any state violating the hard-core dimer constraint has a larger energy, which qualifies the LPG term as a full-protection scheme~\cite{Halimeh2021PRXQ} for QDMs.

\section{Derivation of the effective \Ztwo~\lowercase{m}LGT Hamiltonian}
\label{app:EffHam}
\begin{figure*}[t]
\includegraphics[width=\textwidth]{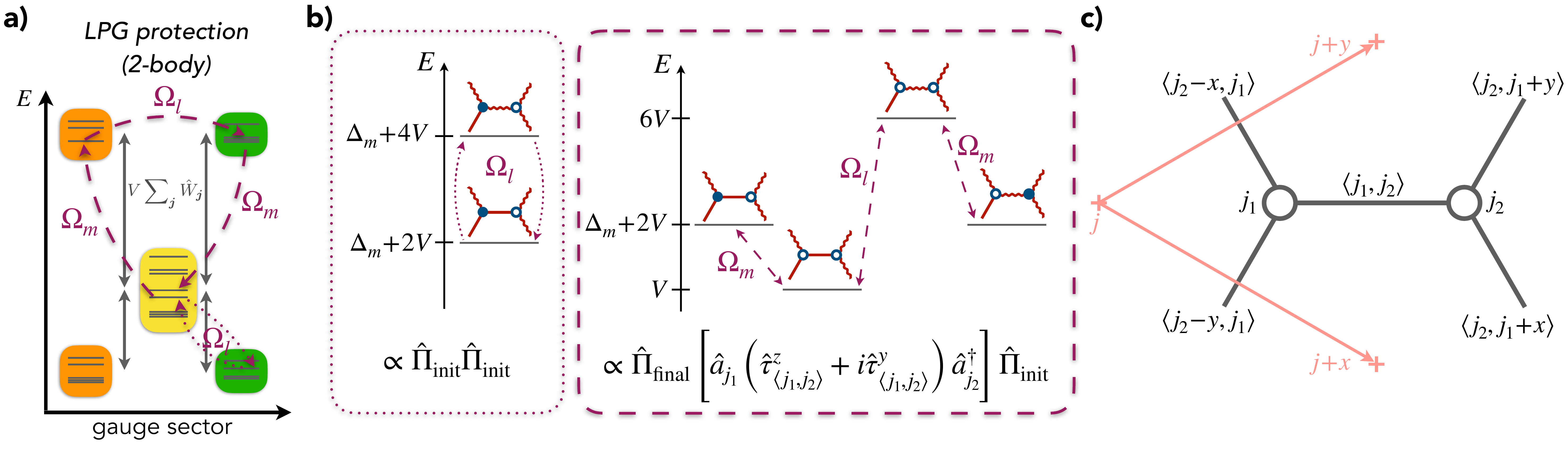}
\caption{\textbf{Perturbative derivation of the effective \Ztwo~mLGT Hamiltonian.} From the perspective of \Ztwo~mLGTs, the LPG protection term energetically splits a target gauge sector (yellow) from other sectors (orange, green) as shown in panel~\textbf{a)}. A gauge-noninvariant perturbation~$\H^\mathrm{drive}$ with strength~$\Omega_m, \Omega_l \ll V$ leads to virtual processes to unphysical sectors of the Hilbert space, which can be treated in perturbation theory and which ultimately yield the effective Hamiltonian~$\H^{\mathrm{eff}}_{\mathbb{Z}_2}$ in the main text. In panel~\textbf{b)}, we illustrate an example for a second-order (left) and third-order (right) process. By using projection operators on the initial (final) state, $\hat{\Pi}_\mathrm{init}$ ($\hat{\Pi}_\mathrm{final}$), the operator form of~$\H^\mathrm{eff}_{\mathbb{Z}_2}$ can be determined. Panel~\textbf{c)} introduces the notation for sites and links on the $2$D honeycomb lattice with lattice vectors shown in light red.}
\label{app:figEffHam}
\end{figure*}

In this section, we explain the derivation of the effective Hamiltonian~(\ref{eq:modelH}) in terms of a Schrieffer-Wolff transformation~\cite{Schrieffer1966}.
The derivation of the effective QDM is discussed in SI~\ref{app:EffHamQDimer}.
Starting point is the experimentally motivated microscopic Hamiltonian (\ref{eq:modelH}),
\begin{alignat}{2}
    &\H^\mathrm{mic} &&= \H^{\mathrm{LPG}} + \H^{\mathrm{detuning}} + \H^{\mathrm{drive}} \label{app:eqHmic1} \\
    &\H^\mathrm{LPG} &&= V\sum_{\j}\Wj \label{app:eqHmic2} \\
    &\H^{\mathrm{detuning}} &&= -\Delta_m\sum_{\j}\nj - \frac{\Delta_l}{2}\sum_{\ij} \tauxij \label{app:eqHmic3} \\
    &\H^{\mathrm{drive}} &&= \Omega_m\sum_{\j}\left( \aj + \adj \right) + \Omega_l\sum_{\ij} \left( \a_\ij + \ad_\ij \right), \label{app:eqHmic4}
\end{alignat}
where~$\H_0 = \H^\mathrm{LPG}+\H^{\mathrm{detuning}}$ is the unperturbed Hamiltonian and~$\H^{\mathrm{drive}}$ is a small perturbation~\cite{Yang2019}, i.e.~$V \gg \Omega_m, \Omega_l$, see Fig.~\ref{app:figEffHam}a.
Note that the perturbation is a gauge-symmetry breaking term, $[\H^\mathrm{drive},\Gj] \neq 0~\forall \j$.
However a state prepared in the physical subspace, $g_{\j}=+1~\forall\j$, will only virtually occupy unphysical states under~$\H^\mathrm{mic}$ because of the large energy gap~$V$ between the sectors in the limit of weak driving,~$\Omega_m, \Omega_l \ll V$.

Hamiltonian~$\H_0$ is diagonal in the matter density and electric field basis and hence the unperturbed eigenstates are product states~ $\ket{\alpha}=\bigotimes_{\j}\ket{n_{\j}}\bigotimes_{\ij}\ket{\tau^x_{\ij}}$.
Since~$\H^{\mathrm{drive}}$ only contains off-diagonal elements, there are no first-order contributions, $\bra{\alpha}\H^{\mathrm{drive}}\ket{\alpha}=0$.
The derivation of the second- and third-order terms are explained in the following together with an explicit example, see Fig.~\ref{app:figEffHam}b.
We note that the second- and third-order contributions require to calculate $16 + 32 + 3 \cdot2 \cdot 16 = 144$ amplitudes.

The second-order terms are given by
\begin{align} \label{appEq:SW2nd}
    \bra{\beta}\H^\mathrm{eff}_{\mathrm{2nd}}\ket{\alpha} = \frac{1}{2}\sum_{\delta}\bra{\beta}\H^{\mathrm{drive}}\ket{\delta}\bra{\delta}\H^{\mathrm{drive}}\ket{\alpha}\left( \frac{1}{E_\beta-E_\delta} + \frac{1}{E_\alpha-E_\delta} \right),
\end{align}
where $\ket{\alpha}$ ($\ket{\beta}$) are the initial (final) state and~$\ket{\delta}$ are virtual states.
Because~$\H^{\mathrm{drive}}$ has only off-diagonal elements, it always couples to states outside the physical energy sector and hence in second-order the initial and final state coincide,~$\ket{\alpha}=\ket{\beta}$, in order to remain within the same energy subspace.

In Fig.~\ref{app:figEffHam}b (left) we show one example process for the parameters~$V=|\Delta_m| \gg \Delta_l,\,\Omega_m,\,\Omega_l$ (see below).
While the amplitude of the process can be calculated using Eq.~(\ref{appEq:SW2nd}), the operator form can be expressed in terms of projectors, which for the example in Fig.~\ref{app:figEffHam}b left is given by
\begin{align}
\hat{\Pi}_{\mathrm{init}} &= \ket{\alpha}\bra{\alpha}\\
    &=2^{-5}
    \hat{n}_{j_1}
    (1+\hat{\tau}^x_{\langle {j_2}-x,{j_1} \rangle})
    (1-\hat{\tau}^x_{\langle {j_2}-y,{j_1} \rangle})
    (1+\hat{\tau}^x_{\langle {j_1},{j_2} \rangle})
    (1-\hat{\tau}^x_{\langle {j_2}+y,{j_1} \rangle})
    (1-\hat{\tau}^x_{\langle {j_2}+x,{j_1} \rangle})
    (1-\hat{n}_{j_2})
\end{align}
and hence only diagonal terms appear in second-order perturbation theory.
Here, we have used the notation introduced in Fig.~\ref{app:figEffHam}c: $j_\nu=(j_x,j_y,\nu)$ corresponds to an explicit site on the honeycomb lattice with two-basis unit cell and $\langle j_2,j_1+x \rangle$ or $\langle j_2-y,j_1 \rangle$ describe links, where $x$ and $y$ are the unit vectors and $\nu=1,2$ is an intracell index.
The notation~$\ij$ should still be used when all links are addressed.

In third-order perturbation theory, coupling between different states,~$\ket{\alpha} \neq \ket{\beta}$, occurs, which yields dynamical hopping and pairing terms.
The coupling elements in the effective Hamiltonian can be calculated by evaluating
\begin{align}
\begin{split} \label{app:eq3rdOrder}
    \bra{\beta}\H^\mathrm{eff}_{\mathrm{3rd}}\ket{\alpha} = \frac{1}{3}\sum_{\delta,\delta'}&\bra{\beta}\H^{\mathrm{drive}}\ket{\delta}\bra{\delta}\H^{\mathrm{drive}}\ket{\delta'}\bra{\delta'}\H^{\mathrm{drive}}\ket{\alpha} \\
    &\times\left[ \frac{1}{(E_{\delta}-E_{\delta'})(E_{\delta'}-E_{\alpha})}
    + \frac{1}{(E_{\beta}-E_\delta)(E_{\delta}-E_{\delta'})} 
    - \frac{2}{(E_{\beta}-E_\delta)(E_{\delta'}-E_{\alpha})} \right],
\end{split}
\end{align}
where the sum runs over two virtual states~$\ket{\delta},\,\ket{\delta'}$.
As shown in the example in Fig.~\ref{app:figEffHam}b (right), we can write down the operator corresponding to the coupling~(\ref{app:eq3rdOrder}) by projecting onto the initial state $\hat{\Pi}_\mathrm{init}$, then acting with an operator coupling to the final state followed by a projection on the latter by~$\hat{\Pi}_\mathrm{final}$.
In our example, the projector reads
\begin{align}
\begin{split}
    \hat{\Pi}_{\mathrm{final}} &= \ket{\beta}\bra{\beta}\\
    &=2^{-5}
    (1-\hat{n}_{j_1})
    (1+\hat{\tau}^x_{\langle {j_2}-x,{j_1} \rangle})
    (1-\hat{\tau}^x_{\langle {j_2}-y,{j_1} \rangle})
    (1-\hat{\tau}^x_{\langle {j_1},{j_2} \rangle})
    (1-\hat{\tau}^x_{\langle {j_2}+y,{j_1} \rangle})
    (1-\hat{\tau}^x_{\langle {j_2}+x,{j_1} \rangle})
    \hat{n}_{j_2}.
\end{split}
\end{align}

Executing the above steps to all states in the target energy subspace yield the effective Hamiltonian~(\ref{eq:modelH}).
Note that the plaquette terms are not appearing directly in third-order perturbation but would require to go to sixth-order perturbation theory.
Hence, we discuss them separately in SI~\ref{app:PlaquetteU1} and~\ref{app:PlaquetteZ2}.
First, we want to give an explicit expression of the effective Hamiltonian up to third-order perturbation theory and distinguish the cases with and without global~$U(1)$ symmetry in SI~\ref{app:effHamU1} and \ref{app:effHamNoU1}, respectively.

\subsection{$U(1)$ matter: $V=2|\Delta_m| \gg \Delta_l,\,\Omega_m,\,\Omega_l$}
\label{app:effHamU1}
\begin{table*}[!htbp]
    \centering
    \renewcommand{\arraystretch}{1.3}
    \begin{tabularx}{\linewidth}{|c|X|X|X|X|}\specialrule{.35em}{.1em}{.1em} 
    \multicolumn{1}{|c|}{\multirow{2}{*}{ampl.}}
    & \multicolumn{1}{c|}{\multirow{2}{*}{$\H^\mathrm{eff}$}}
    & \multicolumn{1}{c|}{\multirow{2}{*}{}}
    & \multicolumn{1}{c|}{\multirow{2}{*}{$U(1)$ matter}}
    & \multicolumn{1}{c|}{\multirow{2}{*}{Quantum-\Ztwo{}~matter}} \\
    & & & &\\
    \specialrule{.35em}{.1em}{.1em} 
    
    & & & &\\
    \multicolumn{1}{|c|}{\multirow{3}{*}{$t$}}
    & \multicolumn{1}{c|}{\multirow{3}{*}{$\begin{aligned}\sum\limits_{\ij}\left( \ad_{\bm{i}}\hat{\tau}^z_{\ij} \aj +\mathrm{H.c.} \right)\end{aligned}$}}
    & \multicolumn{1}{c|}{\multirow{3}{*}{\includegraphics[width=24pt, trim= 0 30pt 0 0]{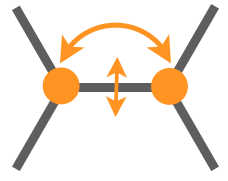}}}
    & \multicolumn{1}{c|}{\multirow{3}{*}{$\dfrac{4 \Omega_l \Omega_m^2 \left( 3+\Delta_m^2/V^2 \right)}{(9V^2-\Delta_m^2)(1-\Delta_m^2/V^2)^2}$}}
    & \multicolumn{1}{c|}{\multirow{3}{*}{$\dfrac{4\Omega_l \Omega_m^2}{3V^2}$}} \\
    & & & &\\[20pt]\hline
    
    & & & &\\
    \multicolumn{1}{|c|}{\multirow{3}{*}{$\Delta_1$}}
    & \multicolumn{1}{c|}{\multirow{3}{*}{$\begin{aligned}\sum\limits_{\ij} \left( \ad_{\bm{i}}\hat{\tau}^z_{\ij} \adj +\mathrm{H.c.} \right)\end{aligned}$}}
    & \multicolumn{1}{c|}{\multirow{3}{*}{\includegraphics[width=24pt, trim= 0 30pt 0 0]{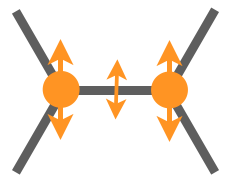}}}
    & \multicolumn{1}{c|}{\multirow{3}{*}{--}}
    & \multicolumn{1}{c|}{\multirow{3}{*}{$\dfrac{20\Omega_l \Omega_m^2}{9V^2}$}} \\
    & & & &\\[20pt]\hline
    
    & & & &\\
    \multicolumn{1}{|c|}{\multirow{3}{*}{$\Delta_2$}}
    & \multicolumn{1}{c|}{\multirow{3}{*}{$\begin{aligned}\sum\limits_{\ij} \left( \ad_{\bm{i}}\hat{\tau}^x_{\ij}\hat{\tau}^z_{\ij} \adj +\mathrm{H.c.} \right)\end{aligned}$}}
    & \multicolumn{1}{c|}{\multirow{3}{*}{\includegraphics[width=24pt, trim= 0 30pt 0 0]{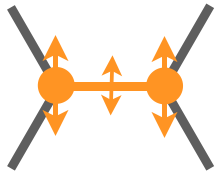}}}
    & \multicolumn{1}{c|}{\multirow{3}{*}{--}}
    & \multicolumn{1}{c|}{\multirow{3}{*}{$\dfrac{16\Omega_l \Omega_m^2}{9V^2}$}} \\
    & & & &\\[20pt]
    \specialrule{.2em}{.1em}{.1em} 
    
    & & & &\\
    \multicolumn{1}{|c|}{\multirow{3}{*}{$h$}}
    & \multicolumn{1}{c|}{\multirow{3}{*}{$\begin{aligned}\sum\limits_{\ij} \tauxij \end{aligned}$}}
    & \multicolumn{1}{c|}{\multirow{3}{*}{\includegraphics[width=24pt, trim= 0 30pt 0 0]{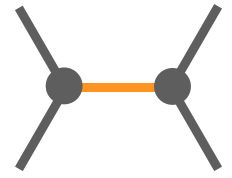}}}
    & \multicolumn{1}{c|}{\multirow{3}{*}{$\begin{gathered}\dfrac{2\Omega_m^2}{3V-2\Delta_m-\Delta_m^2/V}+\dfrac{\Delta_l}{2}\end{gathered}$}}
    & \multicolumn{1}{c|}{\multirow{3}{*}{$\dfrac{2\Omega_m^2}{3V}+\dfrac{\Delta_l}{2}$}} \\
    & & & &\\[20pt]\hline
    
    & & & &\\
    \multicolumn{1}{|c|}{\multirow{3}{*}{$\mu$}}
    & \multicolumn{1}{c|}{\multirow{3}{*}{$\begin{aligned}\sum\limits_{\j} \nj \end{aligned}$}}
    & \multicolumn{1}{c|}{\multirow{3}{*}{\includegraphics[width=20pt, trim= 0 30pt 0 0]{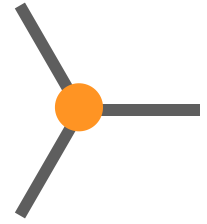}}}
    & \multicolumn{1}{c|}{\multirow{3}{*}{$\Delta_m + \dfrac{2\Omega_m^2\left( -7\Delta_m-\Delta_m^3/V^2 \right)}{9V^2-10\Delta_m^3/V +\Delta_m^3/V^2 }$}}
    & \multicolumn{1}{c|}{\multirow{3}{*}{$\dfrac{\Omega_l^2}{2V}+\Delta_m$}} \\
    & & & &\\[20pt]\hline
    
    & & & &\\
    \multicolumn{1}{|c|}{\multirow{3}{*}{$M$}}
    & \multicolumn{1}{c|}{\multirow{3}{*}{$\begin{aligned}\sum\limits_{\ij} \n_{\bm{i}}\nj \end{aligned}$}}
    & \multicolumn{1}{c|}{\multirow{3}{*}{\includegraphics[width=24pt, trim= 0 30pt 0 0]{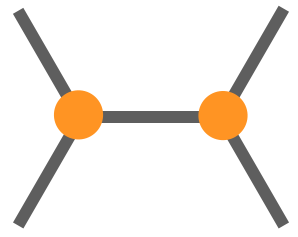}}}
    & \multicolumn{1}{c|}{\multirow{3}{*}{$\dfrac{\Omega_l^2}{3V}$}}
    & \multicolumn{1}{c|}{\multirow{3}{*}{$\dfrac{\Omega_l^2}{3V}$}} \\
    & & & &\\[20pt]\hline
    
    & & & &\\
    \multicolumn{1}{|c|}{\multirow{3}{*}{$\chi_1$}}
    & \multicolumn{1}{c|}{\multirow{3}{*}{$\begin{aligned}&\sum\limits_{j_1} \n_{j_1}\left( \taux{j_2}{j_1+x}+\taux{j_2}{j_1+y} \right)\\ + &\sum\limits_{j_2} \n_{j_2}\left( \taux{j_2-x}{j_1}+\taux{j_2-y}{j_1} \right)\end{aligned}$}}
    & \multicolumn{1}{c|}{\multirow{3}{*}{\includegraphics[width=24pt, trim= 0 30pt 0 0]{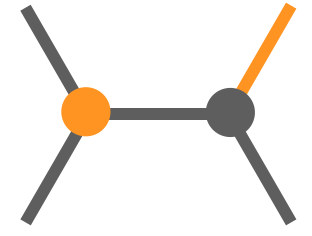}}}
    & \multicolumn{1}{c|}{\multirow{3}{*}{$\dfrac{\Omega_l^2}{6V}$}}
    & \multicolumn{1}{c|}{\multirow{3}{*}{$\dfrac{\Omega_l^2}{6V}$}} \\
    & & & &\\[20pt]\hline
    
    & & & &\\
    \multicolumn{1}{|c|}{\multirow{3}{*}{$\chi_2$}}
    & \multicolumn{1}{c|}{\multirow{3}{*}{$\begin{aligned}&\sum\limits_{j_1} \left( \taux{j_2-x}{j_1}+\taux{j_2-y}{j_1} \right)\left( \taux{j_2}{j_1+x}+\taux{j_2}{j_1-y} \right)\end{aligned}$}}
    & \multicolumn{1}{c|}{\multirow{3}{*}{\includegraphics[width=24pt, trim= 0 30pt 0 0]{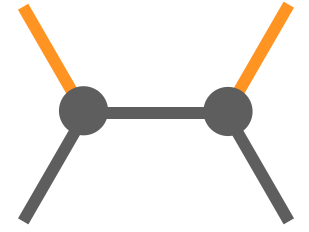}}}
    & \multicolumn{1}{c|}{\multirow{3}{*}{$\dfrac{\Omega_l^2}{12V}$}}
    & \multicolumn{1}{c|}{\multirow{3}{*}{$\dfrac{\Omega_l^2}{12V}$}} \\
    & & & &\\[20pt]\hline
    
    & & & &\\
    \multicolumn{1}{|c|}{\multirow{3}{*}{$\chi_3$}}
    & \multicolumn{1}{c|}{\multirow{3}{*}{$\begin{aligned}&\sum\limits_{j_1} \n_{j_1}\left( \taux{j_1}{j_2}+\taux{j_2-y}{j_1}+\taux{j_2-x}{j_1}\right)\\ + &\sum\limits_{j_2} \n_{j_2}\left( \taux{j_1}{j_2}+\taux{j_2}{j_1+y}+\taux{j_2}{j_1+x} \right)\end{aligned}$}}
    & \multicolumn{1}{c|}{\multirow{3}{*}{\includegraphics[width=20pt, trim= 0 30pt 0 0]{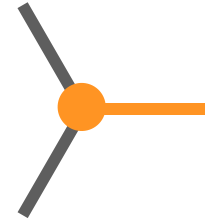}}}
    & \multicolumn{1}{c|}{\multirow{3}{*}{$-\dfrac{\Omega_l^2}{3V}+\dfrac{2\Omega_m^2(3-\Delta_m^2/V^2)}{9V-10\Delta_m^2/V+\Delta_m^3/V^2}$}}
    & \multicolumn{1}{c|}{\multirow{3}{*}{$-\dfrac{\Omega_l^2}{3V}+\dfrac{2\Omega_m^2}{3V}$}} \\
    & & & &\\[20pt]
    \specialrule{.2em}{.1em}{.1em} 
    
    & & & &\\
    \multicolumn{1}{|c|}{\multirow{3}{*}{const.}}
    & \multicolumn{1}{c|}{\multirow{3}{*}{$\mathbb{1}$}}
    & \multicolumn{1}{c|}{\multirow{3}{*}{\includegraphics[width=24pt, trim= 0 30pt 0 0]{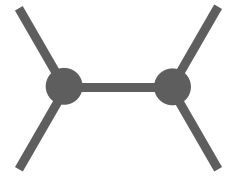}}}
    & \multicolumn{1}{c|}{\multirow{3}{*}{$\begin{gathered}\dfrac{\Omega_l^2}{6V}\left( \text{\# links}\right)  \\
    +\dfrac{\Omega_m^2\left( 2V-\Delta_m \right)}{\left( 3V-\Delta_m \right) \left( V+\Delta_m \right)}\left( \text{\# sites}\right)\end{gathered}$}}
    & \multicolumn{1}{c|}{\multirow{3}{*}{$\begin{aligned}&\dfrac{\Omega_l^2}{6V}\left( \text{\# links}\right)\\+&\dfrac{2\Omega_m^2}{3V}\left( \text{\# sites}\right)\end{aligned}$}} \\
    & & & &\\[20pt]\hline
    \end{tabularx}
    \caption{\textbf{Effective couplings derived from perturbation theory.} The LPG term defines energy subspaces, which can be weakly coupled by a drive, see Eqs.~(\ref{app:eqHmic1})-(\ref{app:eqHmic4}). The effective couplings can then be derived in terms of a Schrieffer-Wolff transformation, see SI~\ref{app:EffHam}, yielding Hamiltonians Eqs.~(\ref{app:eqEffHamZ2mLGTU1matter}) and~(\ref{app:eqEffHamZ2mLGTZ2matter}). The coupling amplitudes are plotted in Fig.~\ref{app:figEffCouplings}. }
    \label{app:TableEffHam}
\end{table*}

\begin{figure*}[t]
\includegraphics[width=\textwidth]{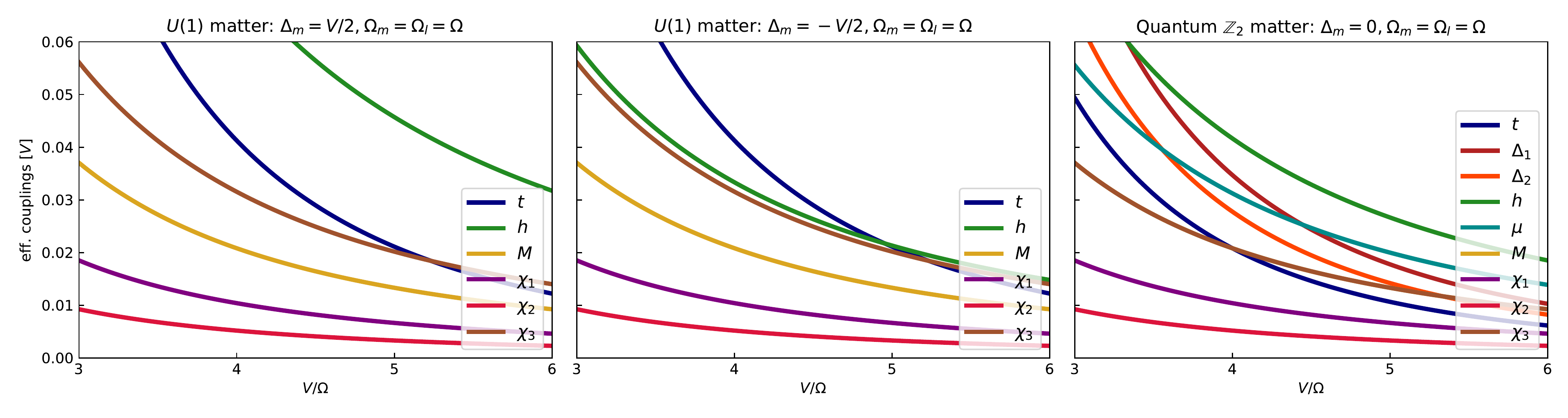}
\caption{\textbf{Effective couplings.} We plot the effective couplings for the \Ztwo{}~mLGT as derived in perturbation theory up to third order, see also Tab.~\ref{app:TableEffHam}. In panel~\textbf{a)} we show the $U(1)$~matter case for two different choices of matter detuning $\Delta_m=\pm V/2$. We do not plot the effective chemical potential term because it only contributes as a constant term in Hamiltonian~(\ref{app:eqEffHamZ2mLGTU1matter}). In panel~\textbf{b)}, we show the couplings for the effective quantum-\Ztwo{}~matter Hamiltonian, Eq.~(\ref{app:eqEffHamZ2mLGTZ2matter}). Note that \textit{small} detunings $\Delta_m$ and $\Delta_l$ can fully tune the chemical potential~$\mu$ and electric field~$h$ without affecting the other couplings in perturbation theory.}
\label{app:figEffCouplings}
\end{figure*}

To enforce conservation of matter excitations, we introduce an additional energy gap between different particle number sectors by choosing~$|\Delta_m|=V/2 \gg \Delta_l,\,\Omega_m,\,\Omega_l$.
This strong chemical potential term suppresses creation and annihilation of matter excitations induced by~$\H^\mathrm{drive}$.

The effective model for $U(1)$~matter coupled to a \Ztwo{}~gauge field in the sector~$g_{\j} = +1~\forall \j$ is given by
\begin{align}
\begin{split} \label{app:eqEffHamZ2mLGTU1matter}
    \hat{\tilde{H}}^{\mathrm{eff}}_{U(1)~\mathrm{matter}} &= 
    t\sum_{\ij} \raisebox{-1ex}{\includegraphics[width=20pt]{TabEffHam/hopping1.png}}
    - J\sum_{\includegraphics[width=6pt]{plaquette.pdf}} \raisebox{-1.2ex}{\includegraphics[width=20pt]{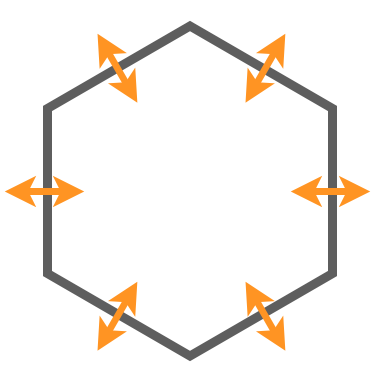}}
    - h\sum_{\ij} \raisebox{-1ex}{\includegraphics[width=20pt]{TabEffHam/elfield.png}}
    - \mu\sum_{\j} \raisebox{-1.25ex}{\includegraphics[width=15pt]{TabEffHam/chempot.png}}
    + M\sum_\ij \raisebox{-1ex}{\includegraphics[width=20pt]{TabEffHam/M.png}}\\
    &+\chi_1 \sum_{\ij} \bigg( \raisebox{-1ex}{\includegraphics[width=20pt]{TabEffHam/chi1.png}} +\raisebox{-1ex}{\includegraphics[width=20pt]{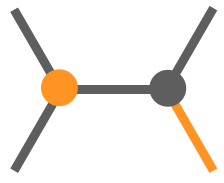}} +\raisebox{-1ex}{\includegraphics[width=20pt]{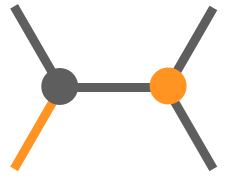}}
    +\raisebox{-1ex}{\includegraphics[width=20pt]{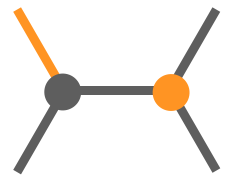}} \bigg)
   +\chi_2 \sum_{\ij} \bigg( \raisebox{-1ex}{\includegraphics[width=20pt]{TabEffHam/chi2.png}} +\raisebox{-1ex}{\includegraphics[width=20pt]{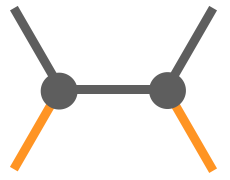}} +\raisebox{-1ex}{\includegraphics[width=20pt]{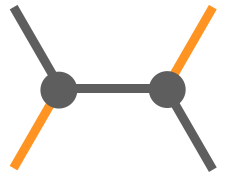}}
    +\raisebox{-1ex}{\includegraphics[width=20pt]{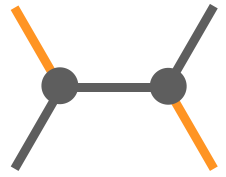}} \bigg)\\
    &+\chi_3 \sum_{\j} \bigg( \raisebox{-1.25ex}{\includegraphics[width=15pt]{TabEffHam/chi3.png}} +\raisebox{-1.25ex}{\includegraphics[width=15pt]{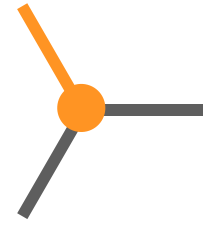}} +\raisebox{-1.25ex}{\includegraphics[width=15pt]{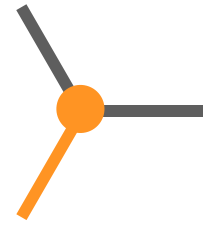}} \bigg)
    +\mathrm{const.}
\end{split}
\end{align}
The operator form and its corresponding coupling amplitudes for the second- and third order processes can be found in the fourth column of Tab.~\ref{app:TableEffHam} and are plotted in Fig.~\ref{app:figEffCouplings}a.
The plaquette interaction~$\propto J$ is a sixth-order perturbative term, which is discussed separately in SI~\ref{app:PlaquetteU1}.
Note that Gauss's law, $\Gj = +1$ has been used to simplify, collect and eliminate higher-order terms.

The terms~$\propto M$, $\propto \chi_1$, $\propto \chi_2$ and~$\propto \chi_3$ are (nearest neighbor density-density), (next-nearest neighbor density-electric field), (next-nearest neighbor electric field-electric field) and (nearest neighbor density-electric field) interactions, respectively.
In the main text, Eq.~(\ref{eq:modelH}), we treat these terms on mean-field level in the electric field $\tauxij$ and matter density~$\nj$, which is well-defined since both quantities are gauge invariant.
To be explicit, we perform for example a mean-field decoupling of~$M\sum_\ij\n_{\bm{i}} \nj \rightarrow M\langle \n_{\bm{i}} \rangle \sum_{\j} \nj$, which simplifies the effective Hamiltonian.

\subsection{Plaquette terms for $U(1)$~matter: $V=2|\Delta_m| \gg \Delta_l,\,\Omega_m,\,\Omega_l$}
\label{app:PlaquetteU1}

We want to perform an order of magnitude estimation of the plaquette interactions~$J$ in Eq.~(\ref{eq:modelH}).
The goal is to find the matrix elements corresponding to plaquette interactions e.g.~$J^{\mathrm{eff}}$(\raisebox{-.25\height}{\includegraphics[height=\baselineskip]{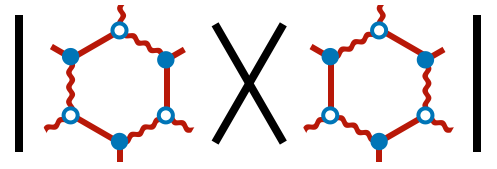}}+h.c.), which we derive by a Schrieffer-Wolff transformation from Eq.~(\ref{eq:micHam}) with~$\Omega_m=\Omega_l=\Omega$, see below.
In general, the effective coupling strengths $J_{n}^{\mathrm{eff}}=J^{\mathrm{eff}}(\{n_{\j} \},\{\tauxij \})$ depend on the configuration of matter and electric fields, yielding $n_\mathrm{max}=416$ independent $J_{n}^{\mathrm{eff}}$ after taking the 6-fold symmetry of the plaquette and Gauss's law into account.

\begin{figure*}[t]
\includegraphics[width=0.8\textwidth]{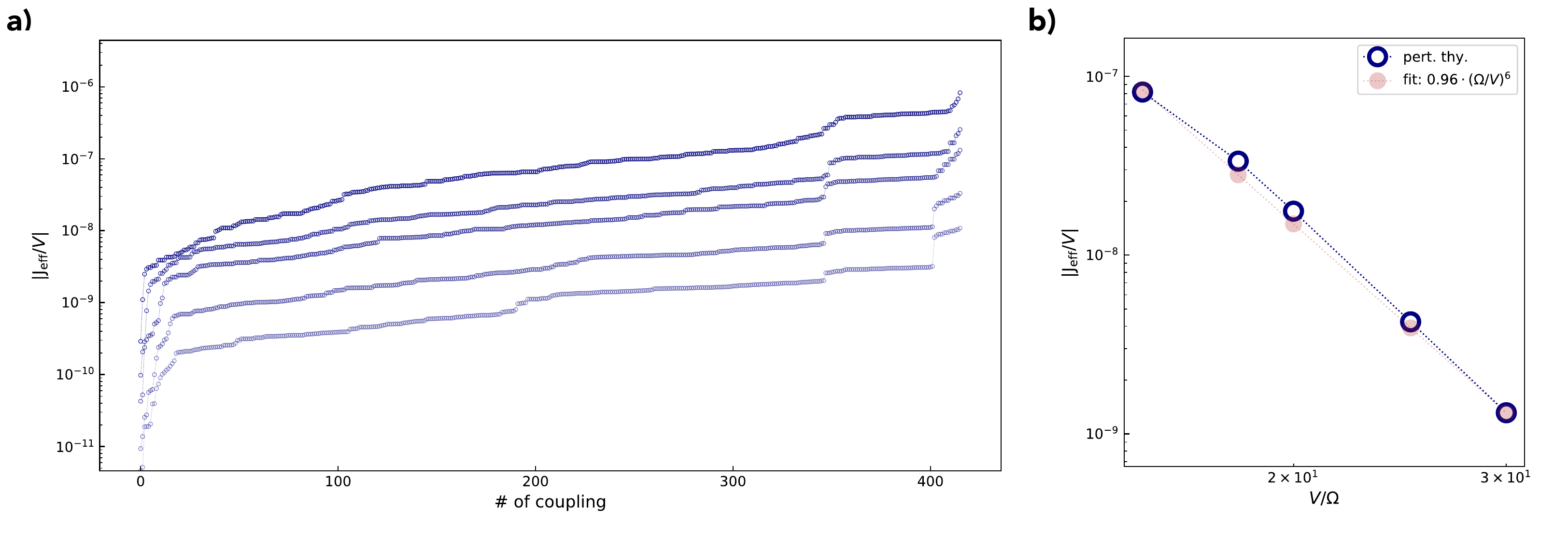}
\caption{\textbf{Estimation of plaquette terms ($U(1)$~matter).} The effective plaquette interaction derived by a Schrieffer-Wolff transformation depends on the matter and electric field configuration within each plaquette. In panel~\textbf{a)} we plot the absolute value of the coupling strength for all 416 different configurations for various driving strength~$\Omega/V$ and~$\Delta_m=V/2$ (from dark to bright shade: $V/\Omega = 15, 18, 20, 25, 30$). In panel~\textbf{b)}, we have averages (with the correct sign taken into account) over the 416 couplings elements for each driving strength~$\Omega/V$, which we plot on a log-log scale. The linear behaviour indicates a power-law behavior and we fit the expected sixth-order perturbation scaling. From the fit we can extract the prefactor, which yields the effective coupling $J_{U(1)}^{\mathrm{eff}}(\Delta_m=V/2)$, Eq.~(\ref{app:eqJeffPlaqU1}). }
\label{app:figPlaqTermsU1}
\end{figure*}

Hence, the effective plaquette interaction Hamiltonian takes the form
\begin{align} \label{app:eqHPlaqEffU1}
\begin{split}
    \H^{\mathrm{eff}}_{\raisebox{-.2\height}{\includegraphics[width=6pt]{plaquette.pdf}},\,U(1)} =
    -\sum_{\includegraphics[width=6pt]{plaquette.pdf}}
    &\Big[
    J_{1}^{\mathrm{eff}}\raisebox{-.3\height}{\includegraphics[height=\baselineskip]{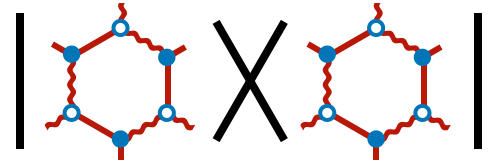}}+ 
    J_{2}^{\mathrm{eff}}\raisebox{-.3\height}{\includegraphics[height=\baselineskip]{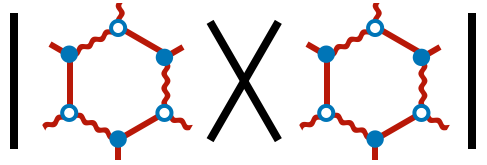}} +
    J_{3}^{\mathrm{eff}}\raisebox{-.3\height}{\includegraphics[height=\baselineskip]{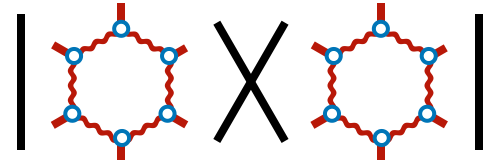}} + J_{4}^{\mathrm{eff}}\raisebox{-.3\height}{\includegraphics[height=\baselineskip]{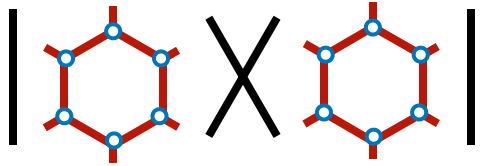}} + ...
    \Big]\times\\
    &\times\Big[ \prod_{\ij \in \raisebox{-.2\height}{\includegraphics[width=6pt]{plaquette.pdf}}}\tauzij  \Big]
    \Big[
    \raisebox{-.3\height}{\includegraphics[height=\baselineskip]{plaquette_stag1.png}} + \raisebox{-.3\height}{\includegraphics[height=\baselineskip]{plaquette_stag2.png}} +
    \raisebox{-.3\height}{\includegraphics[height=\baselineskip]{plaquette_pol1.png}} + \raisebox{-.3\height}{\includegraphics[height=\baselineskip]{plaquette_pol2.png}} + ...
    \Big]
\end{split}
\end{align}
The different coupling elements $J_{n}^{\mathrm{eff}}$ are calculated in degenerate perturbation theory (see below) and plotted in Fig.~\ref{app:figPlaqTermsU1}a for several driving strength~$\Omega/V$.
Because the plaquette interaction involves six links, we expect that the effective couplings scale as~$(\Omega/V)^6$.

Here, we want to estimate and simplify the plaquette interaction Eq.~(\ref{app:eqHPlaqEffU1}) by averaging over all configurations, i.e. we consider~$J_{U(1)}^{\mathrm{eff}} = 1/{n_\mathrm{max}}\sum_{n} J_{n}^{\mathrm{eff}}$.
To this end, we extract $J_{U(1)}^{\mathrm{eff}}=J_{U(1)}^{\mathrm{eff}}(\Omega/V,\Delta_m,V)$ and perform a fit with the expected scaling function $f(\Omega/V,\Delta_m,V)=\alpha(\Delta_m,V)\,(\Omega/V)^6$ as shown in Fig.~\ref{app:figPlaqTermsU1}b.
By extracting the fit parameter~$\alpha(\Delta_m,V)$ for $\Delta_m=V/2$, we can estimate the strength of the plaquette terms as
\begin{align}
    \H^{\mathrm{eff}}_{\raisebox{-.2\height}{\includegraphics[width=6pt]{plaquette.pdf}},\,U(1)} &\approx -J_{U(1)}^{\mathrm{eff}}(\Omega/V,\Delta_m=V/2)\sum_{\includegraphics[width=6pt]{plaquette.pdf}} \prod_{\ij \in \raisebox{-.2\height}{\includegraphics[width=6pt]{plaquette.pdf}}}\tauzij \\
    J_{U(1)}^{\mathrm{eff}}(\Omega/V,\Delta_m=V/2)&\approx -0.96\cdot \left(\Omega/V\right)^6 \label{app:eqJeffPlaqU1}
\end{align}

Let us now discuss the detailed derivation of Eq.~(\ref{app:eqHPlaqEffU1}) in terms of a Schrieffer-Wolff transformation.
The microscopic model is given by Hamiltonian~(\ref{app:eqHmic1}), where $V,|\Delta_m| \gg \Delta_l,\,\Omega_m,\,\Omega_l$ and~$\Omega_m=\Omega_l=\Omega$.
Further we set~$\Delta_l=0$.
The drive $\Omega$ couples the \Ztwo{}~mLGT sector to the gapped, virtual energy sectors defined by the LPG term.
Since we expect the effective plaquette terms to arise in sixth-order perturbation theory, we also need to consider couplings of~$\Omega$ \textit{within} the highly-degenerate virtual energy sectors.
Hence, it is required to apply degenerate perturbation theory and to diagonalize all energy sectors with respect to the perturbation~$\Omega$ first to lift the degeneracies and afterward perform standard perturbation theory.

To gain an intuitive understanding, we want to consider the following path in the perturbative calculation: for instance we start from a state with no matter excitations and all links in the $\tau^x_\ij=+1$ configuration.
Then, the drive flips one link, which costs an energy of~$-2V$ because Gauss's law is violated on two vertices.
Now, we can consecutively flip all links in clockwise direction.
Since these processes at the same time break and restore Gauss's law at different vertices, they are all degenerate and the denominators of the perturbative expansion vanish.
To circumvent this non-physical divergence, we first need to diagonalize the degenerate subspaces, which renormalizes all couplings and energy gaps.

The system is perturbed by a weak drive $\H^{\mathrm{drive}}$, Eq.~(\ref{app:eqHmic4}), and diagonalization of the degenerate subspaces yields the transformed Hamiltonian~$\hat{\tilde{H}}^\mathrm{mic} = \hat{U}^\dagger\hat{H}^\mathrm{mic}\hat{U}$ that is diagonal within the energy blocks but couples states from different blocks in a non-trivial way.
The off-diagonal terms in~$\hat{\tilde{H}}^\mathrm{mic}$ now become the perturbation~$\hat{\tilde{H}}^{\mathrm{drive}}$ in the new basis.
Note that the states have also transformed and should be denoted by $\ket{\tilde{\alpha}} = \hat{U}\ket{\alpha}$ in the new basis.

Since we have access to the full one-plaquette spectrum, we can now explicitly construct the unitary operator~$\hat{S}$ of the Schrieffer-Wolff transformation by calculating the matrix elements
\begin{align} \label{app:eqSWMatrixElements}
    \bra{\tilde{\beta}}\hat{S}\ket{\tilde{\alpha}} = \dfrac{\bra{\tilde{\beta}}\hat{\tilde{H}}^{\mathrm{drive}}\ket{\tilde{\alpha}}}{E_{\tilde{\beta}}-E_{\tilde{\alpha}}},
\end{align}
where~$\hat{\tilde{H}}^{\mathrm{drive}} = \hat{U}^\dagger\hat{H}^{\mathrm{drive}}\hat{U}$ and $E_{\tilde{\alpha}},\,E_{\tilde{\beta}}$ are the unperturbed energies in the transformed basis.
Because we completely diagonalized the degenerate subspace, divergences of the denominator only appear for uncoupled states, i.e.\ the nominator vanishes, for which we define the matrix element of~$\hat{S}$ to be zero.
In the Schrieffer-Wolff formalism we can now write down a well-defined expansion in~$\Omega/V$:
\begin{align}
    \H^{\mathrm{eff}}_{\includegraphics[width=6pt]{plaquette.pdf},U(1)} &= \sum_{n}\H^{(n)} \label{app:eqSWSum} \\
    \H^{(0)} &= \hat{\tilde{H}}^\mathrm{mic} - \hat{\tilde{H}}^{\mathrm{drive}}\\
    \begin{split}
    \H^{(1)} &=0\\
    &\vdots
    \end{split}\\
    \H^{(n)} &= \dfrac{n-1}{n!}\underbrace{\bigg[ \hat{S}, \big[ \hat{S},...,[ \hat{S},  \hat{\tilde{H}}^{\mathrm{drive}} ]...\big] \bigg]}_{(n-1)\text{-commutators}} \label{app:eqSWNth}
\end{align}
Note that in the transformed basis the energy denominator in Eq.~(\ref{app:eqSWMatrixElements}) can depend on~$V$ and~$\Omega$.
Since we require~$\Omega \ll V$, we can expand the expressions and find in leading order sixth-order contributions for any~$2 \leq n  \leq 6$.

In Fig.~\ref{app:figPlaqTermsU1}a, we plot the strength of all non-zero plaquette flip matrix elements in the gauge sector~$g_{\j}=+1$ for different driving strength~$\Omega/V$.
Note that the couplings can be positive and negative while we only plot their absolute value; In the average~$J_{U(1)}^{\mathrm{eff}}$ their signs are properly included, however.

\subsection{Quantum-\Ztwo{}~matter: $V \gg \Delta_m,\,\Delta_l,\,\Omega_m,\,\Omega_l$}
\label{app:effHamNoU1}
In this section, we discuss the derivation of the effective Hamiltonian~(\ref{eq:modelH}) with quantum-\Ztwo{}~matter coupled to a \Ztwo{}~gauge field.
In contrast to SI~\ref{app:effHamU1}, we do not enforce a global~$U(1)$ symmetry for the matter but otherwise the derivation is completely analogous.
This leads to the additional pairing terms~$\Delta_1,\,\Delta_2$ in Eq.~(\ref{eq:modelH}).
The effective model we find is invariant under the local transformation
\begin{align}
    \aj \longrightarrow -\aj ~~~~~~~~ \tauzij \longrightarrow - \tauzij~~\forall \bm{i}: \ij .
\end{align}
However, the $2$D quantum Hamiltonian cannot be mapped exactly on a classical $3$D Ising LGT~\cite{FradkinShenker1979}, which is origin of the term ``quantum-\Ztwo{}~mLGT''.

In the gauge sector $g_{\j} = +1~\forall \j$, the effective model reads
\begin{align}
\begin{split} \label{app:eqEffHamZ2mLGTZ2matter}
    \hat{\tilde{H}}^{\mathrm{eff},\,(3)}_{\mathbb{Z}_2} &= 
    t\sum_{\ij} \raisebox{-1ex}{\includegraphics[width=20pt]{TabEffHam/hopping1.png}}
    +\Delta_1 \sum_{\ij} \raisebox{-1ex}{\includegraphics[width=20pt]{TabEffHam/pairing1.png}}
    +\Delta_2 \sum_{\ij} \raisebox{-1ex}{\includegraphics[width=20pt]{TabEffHam/pairing2.png}}
    - J\sum_{\includegraphics[width=6pt]{plaquette.pdf}} \raisebox{-1.2ex}{\includegraphics[width=20pt]{TabEffHam/plaqInt.png}}
    - h\sum_{\ij} \raisebox{-1ex}{\includegraphics[width=20pt]{TabEffHam/elfield.png}}
    - \mu\sum_{\j} \raisebox{-1.25ex}{\includegraphics[width=15pt]{TabEffHam/chempot.png}}
    + M\sum_\ij \raisebox{-1ex}{\includegraphics[width=20pt]{TabEffHam/M.png}}\\
    &+\chi_1 \sum_{\ij} \bigg( \raisebox{-1ex}{\includegraphics[width=20pt]{TabEffHam/chi1.png}} +\raisebox{-1ex}{\includegraphics[width=20pt]{TabEffHam/chi1_2.png}} +\raisebox{-1ex}{\includegraphics[width=20pt]{TabEffHam/chi1_3.png}}
    +\raisebox{-1ex}{\includegraphics[width=20pt]{TabEffHam/chi1_4.png}} \bigg)
   +\chi_2 \sum_{\ij} \bigg( \raisebox{-1ex}{\includegraphics[width=20pt]{TabEffHam/chi2.png}} +\raisebox{-1ex}{\includegraphics[width=20pt]{TabEffHam/chi2_2.png}} +\raisebox{-1ex}{\includegraphics[width=20pt]{TabEffHam/chi2_3.png}}
    +\raisebox{-1ex}{\includegraphics[width=20pt]{TabEffHam/chi2_4.png}} \bigg)\\
    &+\chi_3 \sum_{\j} \bigg( \raisebox{-1.25ex}{\includegraphics[width=15pt]{TabEffHam/chi3.png}} +\raisebox{-1.25ex}{\includegraphics[width=15pt]{TabEffHam/chi3_2.png}} +\raisebox{-1.25ex}{\includegraphics[width=15pt]{TabEffHam/chi3_3.png}} \bigg)
    +\mathrm{const.}
\end{split}
\end{align}
The operator form and its corresponding second- and third-order coupling amplitudes can be found in the fifth column of Tab.~\ref{app:TableEffHam} and are plotted in Fig.~\ref{app:figEffCouplings}b, while the discussion of the sixth-order plaquette terms is dedicated to SI~\ref{app:PlaquetteZ2}.
Compared to~(\ref{app:eqEffHamZ2mLGTU1matter}), we now find pairing terms~$\Delta_1$ and $\Delta_2$, which also appear in Fradkin \& Shenker's Ising \Ztwo{}~mLGT in a similar fashion.
As explained in SI~\ref{app:effHamNoU1}, the terms~$\propto M$, $\propto \chi_1$, $\propto \chi_2$ and~$\propto \chi_3$ can be treated on mean-field level yielding the effective model~(\ref{eq:modelH}) discussed in the main text. 

In particular, the electric field term~$-h\sum_{\ij}\tauxij$ can be fine-tuned by changing the detuning~$\Delta_l$, which in the limit~$\Delta_l \ll V$ does not alter the results obtained from perturbation theory.
On mean-field level, this allows to tune the expectation value to $\langle \tauxij \rangle=-1/2$.
Then, the effective coupling renormalizes to $\tilde{\Delta}_1 = \Delta_1 - \langle \tauxij \rangle \Delta_2 = \Delta_1 - \Delta_2/2 = t$.
At this particular point, we retrieve the $(2+1)$D model studied by Fradkin \& Shenker~\cite{FradkinShenker1979}, where it is known to map on a classical $3$D \Ztwo{}~mLGT with continuous phase transitions in the Ising universality class.
Note that our model is defined on the honeycomb and not square lattice.
For a detailed discussion of the duality between a \Ztwo{}~mLGT on a honeycomb and triangular lattice, we refer to the Supplementary Information of Ref.~\cite{Samajdar2023}.
Because of this duality, the results obtained in Ref.~\cite{FradkinShenker1979} should be still valid, however the phase diagram might not be symmetric across the diagonal as illustrated for simplicity in Fig.~\ref{fig2}b.

\subsection{Plaquette terms for quantum-\Ztwo{}~matter: $V \gg \Delta_m,\,\Delta_l,\,\Omega_m,\,\Omega_l$}
\label{app:PlaquetteZ2}
\begin{figure*}[t]
\includegraphics[width=0.8\textwidth]{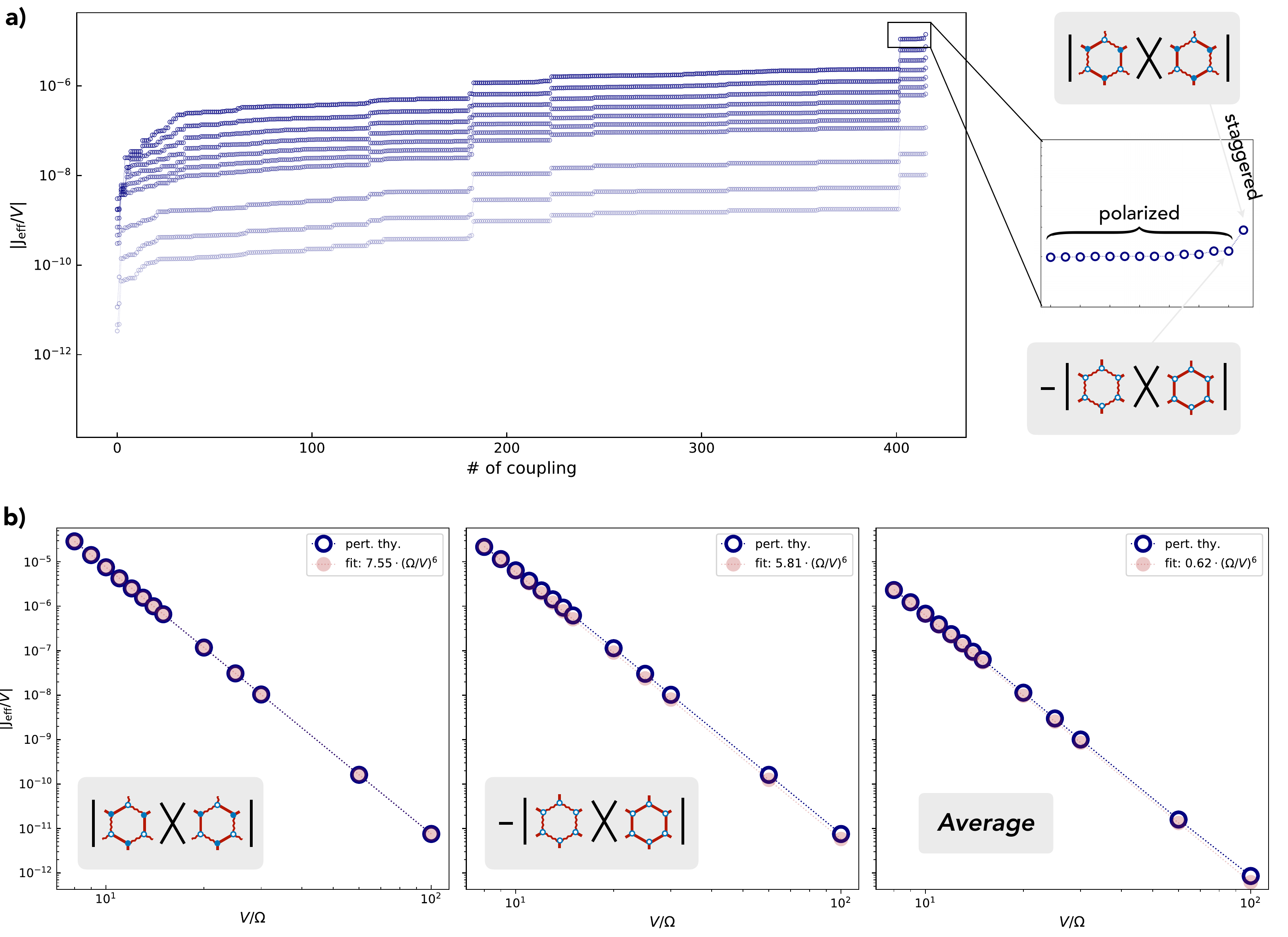}
\caption{\textbf{Estimation of plaquette terms (quantum-\Ztwo{}~matter).} The effective plaquette interaction derived by a Schrieffer-Wolff transformation depends on the matter and electric field configuration within each plaquette. In panel~\textbf{a)} we plot the absolute value of the coupling strength for all 416 different configurations for various driving strength~$\Omega/V$ (from dark to bright shade: $V/\Omega = 8, 10, 12, 14, 20, 30$). We find a plateau with strongest coupling for certain staggered and polarized electric field configurations as shown in the inset. Note the sign of $J^\mathrm{stag}_{\mathrm{eff}}$ differs from $J^\mathrm{pol}_{\mathrm{eff}}$. In panel~\textbf{b)} the absolute value of the effective coupling of the staggered, polarized and averaged configurations versus the driving strength~$V/\Omega$ on a log-log scale is shown. The linear behaviour indicates a power-law behaviour and we fit the expected sixth-order perturbation scaling. From the fit we can extract the prefactor, which yields the effective couplings Eqs.~(\ref{app:eqJeffPlaqStag}),\,(\ref{app:eqJeffPlaqPol}) and (\ref{app:eqJeffPlaqAvg}). }
\label{app:figPlaqTermsZ2}
\end{figure*}

Similar to the case with $U(1)$ matter discussed in SI~\ref{app:PlaquetteU1}, we want to estimate the strength of the plaquette terms~$J_{\mathbb{Z}_2}^{\mathrm{eff}}$ in the quantum-\Ztwo{}~matter model.
We perform a Schrieffer-Wolff transformation with $V \gg \Delta_m,\,\Delta_l,\,\Omega_m,\,\Omega_l$ and $\Omega_m=\Omega_l=\Omega$ and $\Delta_l=0$.
In Fig.~\ref{app:figPlaqTermsZ2}a, we plot the extracted coupling matrix elements between flippable plaquettes.
We find that there is a plateau with 14 distinct coupling elements, which are an order of magnitude larger than the remaining couplings.
As indicated in Fig.~\ref{app:figPlaqTermsZ2}b, these couplings correspond to 1) a staggered matter and electric field configuration with $J_\mathrm{stag}^{\mathrm{eff}}$ and to 2) configurations with a polarized electric field $J_\mathrm{pol}^{\mathrm{eff}}$, where all links are either~$\tau^x_{\ij}=+1$ or $\tau^x_{\ij}=-1$.
Note that these coupling elements might give rise to additional phases and we want to include them in the discussion of the plaquette terms here.
However, averaging over all coupling elements as in SI~\ref{app:PlaquetteU1} should give a useful estimation of the overall strength~$J_{\mathbb{Z}_2}^{\mathrm{eff}}$ of the plaquette terms.

As discussed in SI~\ref{app:PlaquetteU1}, we can extract the strength of the plaquette interaction by fitting the coupling elements for different driving strengths~$\Omega/V$.
We want to examine the three cases 1) staggered, 2) polarized and 3) averaged as shown in Fig.~\ref{app:figPlaqTermsZ2}c:
\begin{itemize}
    \item[(1)] For plaquettes with a staggered matter and electric field, we find
    \begin{align}
        \H^{\mathrm{eff}}_{\mathrm{stag},\mathbb{Z}_2} &= 
    -J_\mathrm{stag}^{\mathrm{eff}}\sum_{\includegraphics[width=6pt]{plaquette.pdf}} \P^\mathrm{stag}_{\includegraphics[width=6pt]{plaquette.pdf}} \Big( \prod_{\ij \in \raisebox{-.3\height}{\includegraphics[width=6pt]{plaquette.pdf}}}\tauzij  \Big) \P^\mathrm{stag}_{\includegraphics[width=6pt]{plaquette.pdf}}\\
    \P^\mathrm{stag}_{\includegraphics[width=6pt]{plaquette.pdf}} &= \raisebox{-.3\height}{\includegraphics[height=\baselineskip]{plaquette_stag1.png}} + \raisebox{-.3\height}{\includegraphics[height=\baselineskip]{plaquette_stag2.png}} + \raisebox{-.3\height}{\includegraphics[height=\baselineskip]{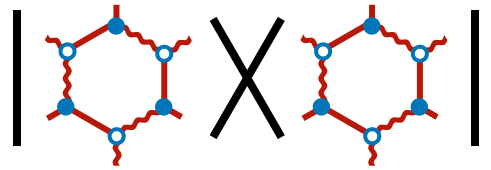}} + \raisebox{-.3\height}{\includegraphics[height=\baselineskip]{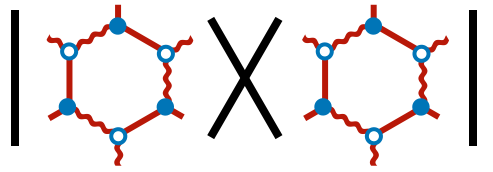}} \label{app:eqProjStag}\\
    J_\mathrm{stag}^{\mathrm{eff}} &\approx 7.55 \cdot (\Omega/V)^6 \label{app:eqJeffPlaqStag}
    \end{align}
    
     \item[(2)] For plaquettes with polarized electric field, we find
    \begin{align}
        \H^{\mathrm{eff}}_{\mathrm{pol},\mathbb{Z}_2} &= 
    -J_\mathrm{pol}^{\mathrm{eff}}\sum_{\includegraphics[width=6pt]{plaquette.pdf}} \P^\mathrm{pol}_{\includegraphics[width=6pt]{plaquette.pdf}} \Big( \prod_{\ij \in \raisebox{-.3\height}{\includegraphics[width=6pt]{plaquette.pdf}}}\tauzij  \Big) \P^\mathrm{pol}_{\includegraphics[width=6pt]{plaquette.pdf}}\\
    \P^\mathrm{pol}_{\includegraphics[width=6pt]{plaquette.pdf}} &= \prod_{\ij \in P} 2^{-1}\left( 1 - \tauxij  \right) + \prod_{\ij \in P} 2^{-1}\left( 1 + \tauxij  \right) \label{app:eqProjPol}\\
    J_\mathrm{pol}^{\mathrm{eff}} &\approx -5.81 \cdot (\Omega/V)^6 \label{app:eqJeffPlaqPol}
    \end{align}
    
    \item[(3)] By averaging over all couplings (as in SI~\ref{app:PlaquetteU1}), we find
    \begin{align}
    \H^{\mathrm{eff}}_{\raisebox{-.2\height}{\includegraphics[width=6pt]{plaquette.pdf}},\,\mathbb{Z}_2} &\approx -J_{\mathbb{Z}_2}^{\mathrm{eff}}\sum_{\includegraphics[width=6pt]{plaquette.pdf}} \prod_{\ij \in \raisebox{-.2\height}{\includegraphics[width=6pt]{plaquette.pdf}}}\tauzij \\
    J_{\mathbb{Z}_2}^{\mathrm{eff}}&\approx-0.62\cdot \left(\Omega/V\right)^6 \label{app:eqJeffPlaqAvg}
\end{align}
\end{itemize}

\subsection{Small-scale exact diagonalization study of the microscopic Hamiltonian}
\label{app:EDEffHam}
\begin{figure*}[t]
\includegraphics[width=\textwidth]{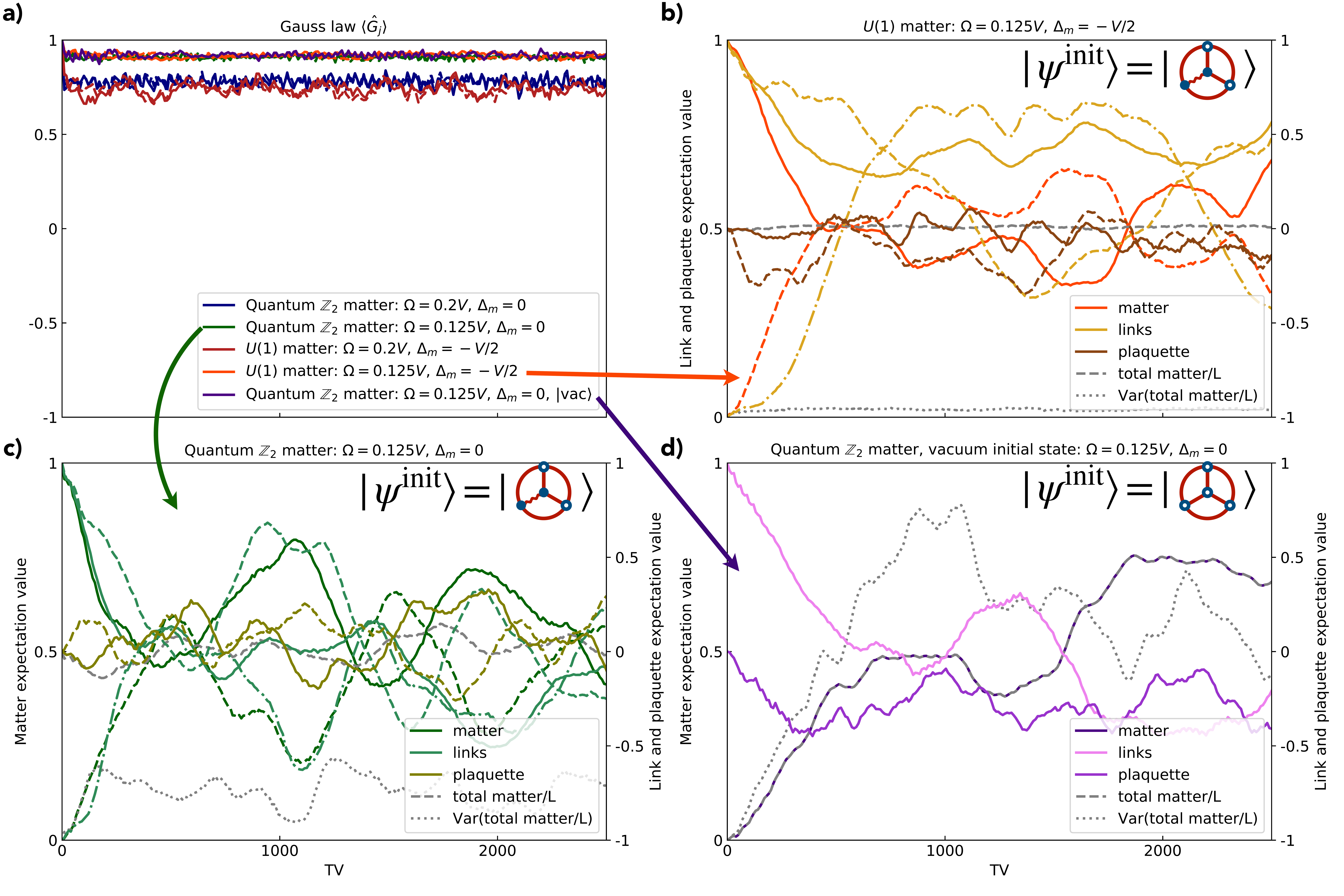}
\caption{\textbf{Dynamics of microscopic model for different parameters.} We initialize the system in a gauge-invariant state~$\Gj\ket{\psi^\mathrm{init}}=+\ket{\psi^\mathrm{init}}~\forall \j$ and time-evolve for time~$T$ under the microscopic Hamiltonian~(\ref{eq:micHam}) using exact diagonalization in a small-scale system, see inset in panel~\textbf{b)}-\textbf{d)}. We find strong dynamics of matter, link and plaquette degrees-of-freedom while the error in Gauss's law is small and remains constant for long, experimentally relevant timescales. The results are discussed in detail in SI~\ref{app:EDEffHam}. Note that the labels in the plots refer to the color scheme but not necessarily to the linestyle and we plot e.g.\ the expectation values of all four matter site with the same color but all four curves with a different linestyle (some curves overlap).}
\label{app:figMicHamDyamics}
\end{figure*}

In this section, we present results from time-evolution studies obtained by exact diagonalization of the full microscopic Hamiltonian~(\ref{eq:micHam}) in a minimal model with coordination number~$z=3$, i.e.\ four matter sites and six links as shown in the inset of Fig.~\ref{app:figMicHamDyamics}b-d.
While this model has a tetrahedron structure and triangular plaquettes, it is different from the honeycomb lattice.
However, because the model has coordination number~$z=3$ -- similar to the honeycomb lattice -- the physics of the LPG protection should be correctly modeled in this numerically feasible $2$D system.

We demonstrate that Gauss's law is indeed very well conserved,~$\langle \Gj \rangle \approx +1$, even for relatively strong drive~$\Omega/V$ (we set $\Omega_m=\Omega_l=\Omega$ throughout this section).
Moreover, the matter and link degrees-of-freedom show dynamics on the expected timescales.
The results are summarized in Fig.~\ref{app:figMicHamDyamics} and we want to elaborate on the different cases here:
\begin{itemize}
    \item Fig.~\ref{app:figMicHamDyamics}a: We plot the expectation value of Gauss's law after time-evolving different initial states and different parameters. If not specified otherwise, the initial state contains two localized matter excitations and fulfills Gauss's law, $\Gj\ket{\psi^\mathrm{init}}=+\ket{\psi^\mathrm{init}}$. While $\langle \Gj \rangle$ has an initial fast drop, the gauge-symmetry violation equilibrates around a constant value determined by the driving strength~$\Omega/V$. For  $\Omega/V=0.125$ ($\Omega/V=0.2$), the violation is about~$5\%$ ($15\%$).
    
    \item Fig.~\ref{app:figMicHamDyamics}b: We consider $U(1)$ matter, i.e.\ we have strong detuning/chemical potential~$\Delta_m=\pm V/2$ and plot the expectation values of the matter density~$\langle \nj \rangle$, the electric field~$\langle \tauxij \rangle$ and plaquette terms~$\langle \prod_{\ij \in P}\tauzij \rangle$ as well as the total number of matter excitations and its variance. Note that the total number of matter excitations only fluctuates marginally as anticipated for $U(1)$~matter. Calculating the effective hopping from Tab.~\ref{app:TableEffHam}, we expect oscillations with a timescale $TV = 2\pi \times 2520/13 \approx 1220$, which matches the timescales in Fig.~\ref{app:figMicHamDyamics}b approximately.
    
    \item Fig.~\ref{app:figMicHamDyamics}c: Next, we consider quantum-\Ztwo{}~matter, where pairs of matter excitations can be created and annihilated. Since the initial state has already two matter excitations (and two holes), the pair creation dynamics is not as heavy as in Fig.~\ref{app:figMicHamDyamics}d, where we start from vacuum. Because of the interplay between hopping and pairing, it is not straightforward to read off timescales from Rabi oscillation-like behaviour. From hopping and pairing, we would expect timescales of approximately~$TV\approx 2400$, respectively. However, we find an emergent timescale in this finite size model of about $TV=1000$. Note that in Hamiltonian~(\ref{app:eqEffHamZ2mLGTZ2matter}), we have (anomalous) pairing terms, which also influence the propagation of matter excitations.
    
    \item Fig.~\ref{app:figMicHamDyamics}d: Here, we initialize the system in the vacuum state and otherwise time-evolve with the same parameters as in Fig.~\ref{app:figMicHamDyamics}c. We find strong particle number fluctuations due to the creation of matter excitations. The expected timescale $TV\approx 800$ (on mean-field level) is in agreement with the overall timescale of oscillations we observe in the system.
\end{itemize}

\subsection{Microscopic versus effective model}
\label{app:MicVsEff}
\begin{figure*}[t]
\includegraphics[width=0.95\textwidth]{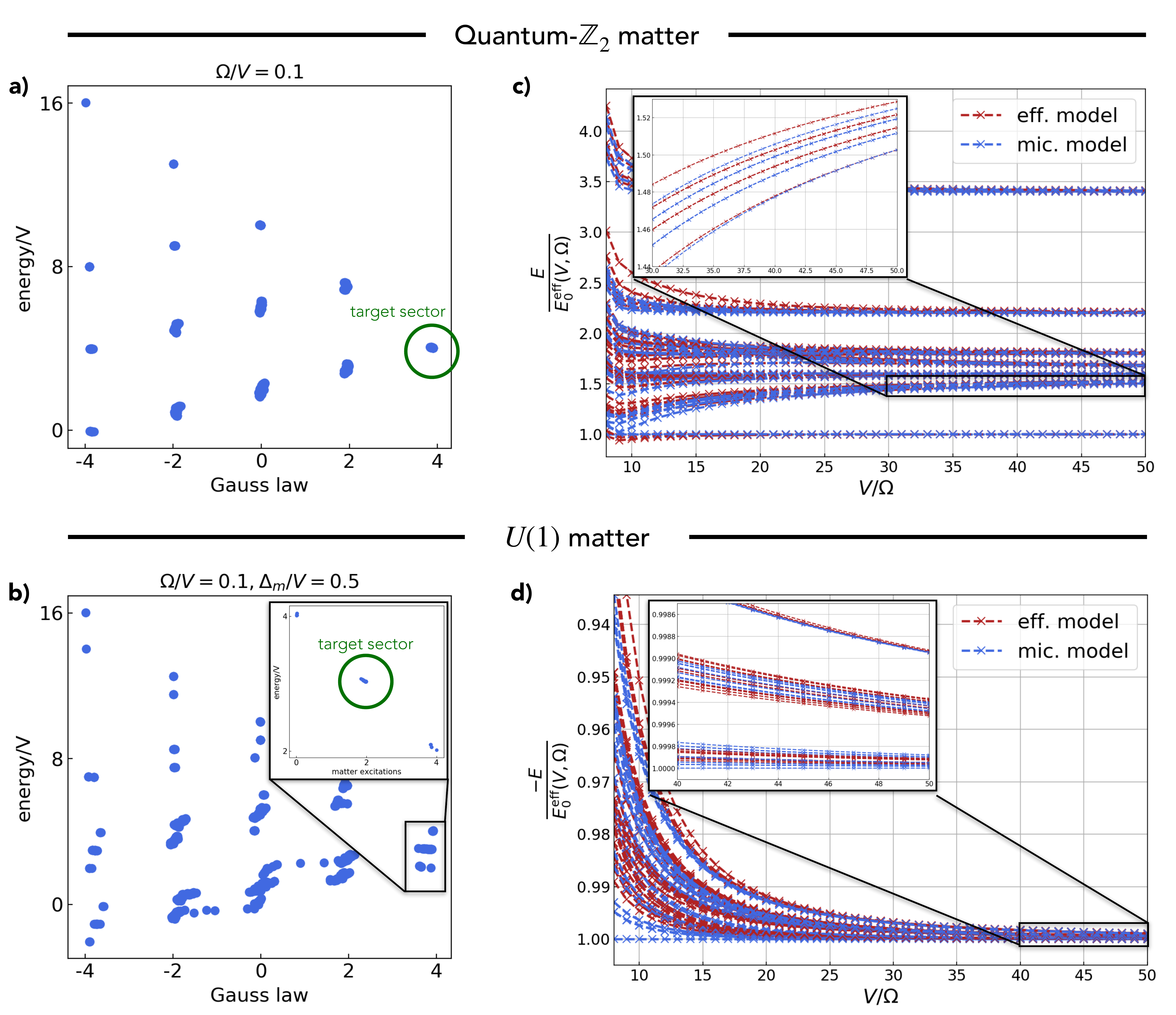}
\caption{\textbf{Microscopic versus effective model.} We show results of exact diagonalization studies in a minimal model and compare the microscopic model Eq.~(\ref{eq:micHam}) to the effective Hamiltonian with quantum-\Ztwo{}~matter, Eq.~(\ref{app:eqEffHamZ2mLGTZ2matter}), and $U(1)$~matter, Eq.~(\ref{app:eqEffHamZ2mLGTU1matter}). For the first (latter) case we choose~$\Delta_l=\Delta_m=0$ ($\Delta_l=0,\Delta_m=0.5V$). In panel~\textbf{a)} and~\textbf{b)}, we calculate the expectation value $g=\langle \sum_{\j} \Gj \rangle$ for each eigenvector of the microscopic model and our target sector is~$g=4$. In panel~\textbf{b)}, we additionally require the number of matter excitations to be conserved. For~$\Omega=0.1V$, we find that both the local and global symmetry emerge in the microscopic model. In panel~\textbf{c)} and \textbf{d)}, we now consider the eigenenergies in the target sector for different driving strength~$\Omega/V$ for the microscopic (blue) and effective (red) model. We find agreement of both spectra which supports the validity of our perturbative approach discussed in the main text and SI~\ref{app:EffHam}. }
\label{app:figMicVsEff}
\end{figure*}

In this section, we want to confirm the effective model by comparing the energy spectrum of the microscopic and effective model as a function of~$V$ and~$\Omega$ ($\Omega_m=\Omega_l=\Omega$) and show that for both quantum-\Ztwo{}~matter, Eq.~(\ref{app:eqEffHamZ2mLGTZ2matter}), and $U(1)$~matter, Eq.~(\ref{app:eqEffHamZ2mLGTU1matter}), the spectra converge in the limit~$V/\Omega\rightarrow \infty$ as expected from perturbation theory.
To this end, we perform exact diagonalization calculations of a minimal system (four matter sites and six links) as in SI~\ref{app:EDEffHam}.
We set the LPG protection strength to~$V=1$ and vary the drive~$\Omega/V$ in the microscopic model~(\ref{eq:micHam}) or correspondingly use the derived effective couplings, see Tab.~\ref{app:TableEffHam}.
Moreover, we set the link detuning~$\Delta_l=0$ and choose~$\Delta_m=0$ ($\Delta_m/V=0.5$) in the quantum-\Ztwo{}~matter ($U(1)$~matter) case.

As a first step, we need to identify the correct target sector of the microscopic model since this has no exact \Ztwo{} gauge symmetry or global $U(1)$~ symmetry.
Therefore, we diagonalize the microscopic Hamiltonian~(\ref{eq:micHam}) and calculate the expectation value of the symmetry generators~$g=\langle \sum_{\j} \Gj \rangle$ for each eigenvector as shown in Fig.~\ref{app:figMicVsEff}a and b.
Because we choose the LPG term to protect the target sector~$g_{\j}=+1~\forall \j$, we want~$g=4$ in our numerical study.
Additionally, for the case of $U(1)$~matter, we need to select a matter excitation sector by evaluating~$n=\langle \sum_{\j} \nj \rangle$ and we choose~$n=2$ in the following discussion.

Fig.~\ref{app:figMicVsEff} illustrates that the target gauge sectors for both cases, quantum-\Ztwo{} and $U(1)$~matter, form well-separated subspaces.
We want to emphasize the efficiency of our proposed LPG protection scheme:
As discussed in SI~\ref{app:LPG}, there are instabilities because we work in a high-energy sector of the LPG term.
These instabilities are resonant processes, where Gauss's law is violated in a way that on three vertices the LPG term lowers the energy while on one vertex the energy is increased.
If the instabilities would play a dominant role in the effective dynamics, we would expect no well-defined gauge sectors but a hybridization of all sectors which would broaden the clusters we find in Fig.~\ref{app:figMicVsEff}a and b.

As a next step, we show that the spectra in the target sectors of the effective and microscopic model converge as~$V/\Omega \rightarrow \infty$.
To this end, we diagonalize the microscopic model~(\ref{eq:micHam}) and the effective model for different~$V/\Omega$, which yields eigenenergies~$E^n_{\mathrm{mic}}(V,\Omega)$ and $E^n_{\mathrm{eff}}(V,\Omega)$.
To compare the spectrum at different driving strengths, we normalize the eigenenergies by the ground-state energy~$E^0_{\mathrm{eff}}(V,\Omega)$ of the corresponding effective model at each point~$V/\Omega$.
In Fig.~\ref{app:figMicVsEff}c and d, we plot the spectrum for the quantum-\Ztwo{} and $U(1)$ matter case as described above.
We find that by using the derived effective couplings in Tab.~\ref{app:TableEffHam} the effective models, Eqs.~\eqref{app:eqEffHamZ2mLGTZ2matter} and~\eqref{app:eqEffHamZ2mLGTU1matter}, very well describe the microscopic models justifying our perturbative analysis.
Note that we did not take the above derived plaquette interactions into account here, because the small-scale system we use in the exact diagonalization study has plaquettes with three edges instead of six edges on a honeycomb lattice.

\subsection{Gauge non-invariant processes}
\label{app:GaugeBreaking}
\begin{figure*}[t]
\includegraphics[width=0.55\textwidth]{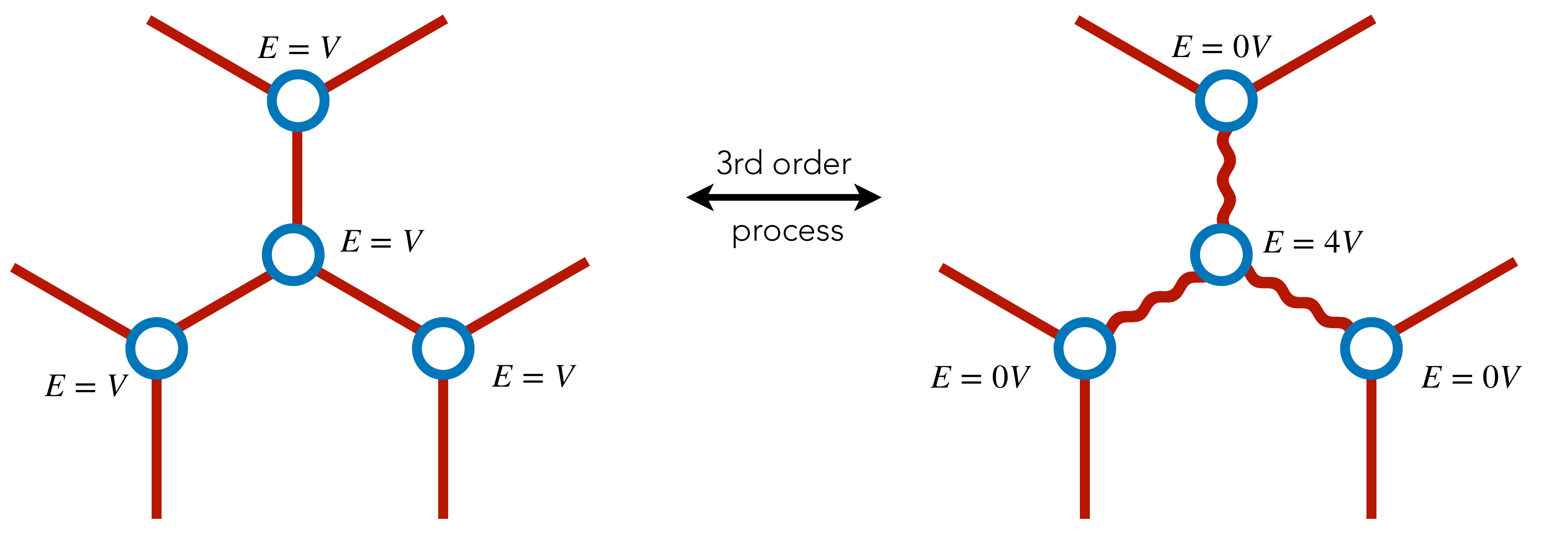}
\caption{\textbf{Third-order gauge-breaking process.} The state illustrated on the left (right) fulfills (breaks) Gauss's law at every vertex. The two states are coupled resonantly via a third-order process.}
\label{app:figGaugeBreaking}
\end{figure*}

So far, we have only considered resonant processes that conserve Gauss's law.
However, as discussed in the Methods section and Fig.~\ref{figDisorder}, the LPG method without disorder suffers from unwanted resonances with a few unphysical states.
Here, we want to discuss the effect of such resonances with respect to the numerical results from section SI~\ref{app:MicVsEff}.

As shown in Fig.~\ref{app:figGaugeBreaking}, it is possible to raise the energy by $+3V$ on one vertex and at the same time lowering the energy by~$-V$ on three neighbouring vertices. 
This process is resonantly coupled to the physical states in a third-order process if and only if the four vertices are arranged in a ``star'' geometry, see Fig.~\ref{app:figGaugeBreaking}.
Otherwise, this type of resonance only appears in fifth-order perturbation theory.
In the following, we argue that these processes do not alter the emergent gauge structure such that the effective model~\eqref{eq:modelH} is valid.

First, we note that the above processes can be entirely suppressed by applying weakly disordered protection terms,~$V \rightarrow V_{\j}+\delta V_{\j}$ and $\delta V_{\j} \ll V$.
The disorder only shifts the gauge non-invariant states out of resonance, see Fig.~\ref{figDisorder}, and its efficiency in $(1+1)$D has been demonstrated numerically~\cite{Halimeh2022LPG}.

We point out that our minimal model simulation in the Mercedes star is able to capture the third-order terms described in Fig.~\ref{app:figGaugeBreaking}.
Hence, the system is susceptible to resonant non-gauge invariant terms that potentially could lead to a complete breakdown of gauge invariance.
However, reconsideration of the numerical results presented in Fig.~\ref{app:figMicVsEff}a) and b) show a well-defined target sector~$g=+4$, which is energetically in resonance with a well-defined~$g=-4$ sector.
I.e.\ the two sectors only very weakly hybridize and the eigenstates of the microscopic Hamiltonian are almost exact Gauss's law eigenstates.
In contrast to the time-evolution of Gauss's law as shown in Fig.~\ref{app:figMicHamDyamics}a), which depends on the choice of the initial state, the plotted spectrum in Fig.~\ref{app:figMicVsEff}a) and b) is a very sensitive probe to validate the emergent gauge structure.
Moreover, this holds true for even stronger drivings~$\Omega/V=1/5$, which we use in further numerical simulations below.

Both thermalization dynamics as well as hybridization between the physical and unphysical sectors is highly suppressed despite comparable Hilbert space dimensions.
This robust gauge structure further suggest an additional mechanism that stabilizes the gauge sectors such as Hilbert space fragmentation, which should be investigated in future studies.

To summarize, we identify potential third-order processes and we present an easily implementable disorder-based protocol such that gauge invariance remains fully intact.
Furthermore, we observe from our numerical results that resonant physical and unphysical sectors show only very weak mutual coupling giving almost perfect gauge invariance even without disorder.

\section{Effective meson model}
\label{app:effHamU1DimerPhase}

For $U(1)$ matter coupled to a \Ztwo{}~gauge field, we predict the existence of a meson condensate phase, see Fig.~\ref{fig2}a.
Here, we want to derive an effective meson model, which captures the condensate phase.

\begin{figure*}[t]
\includegraphics[width=0.95\textwidth]{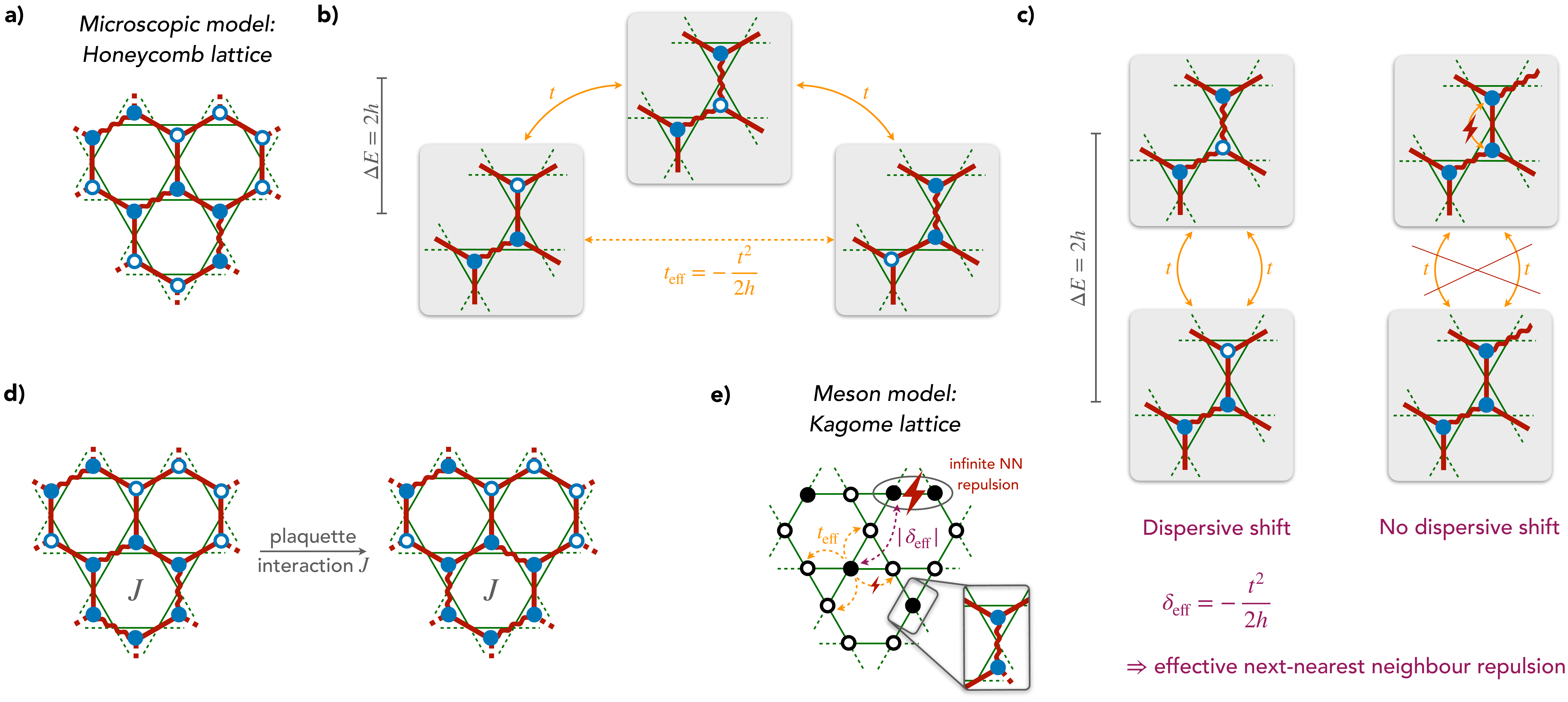}
\caption{\textbf{Effective meson model.} Panel~\textbf{a)}: For $J/h,t/h \rightarrow 0$, the matter excitations are tightly bound into mesonic pairs~$\hat{b}$ on the honeycomb lattice. These mesons are again hard-core bosons and \Ztwo{}~charge neutral. In panel~\textbf{b)} and~\textbf{c)}, we describe the leading order second-order processes derived from the \Ztwo{}~mLGT, which gives rise to hopping~$t_{\mathrm{eff}}$ of mesons as well as repulsive interactions~$|\delta_{\mathrm{eff}}|$ due to the absence of dispersive shifts for next-nearest neighbor mesons. Moreover, the plaquette interaction of the \Ztwo{}~mLGT yields to fluctuating mesons for plaquettes with exactly three mesons as depicted in panel~\textbf{d)}. Panel~\textbf{e)} summarizes the effective meson model. The model is described by hopping~$\propto t_{\mathrm{eff}}$ of hard-core bosons~$\hat{b}$ (black circles) on a Kagome lattice with infinitely strong nearest-neighbor repulsion (from the hard-core constraint on the honeycomb lattice), finite NNN repulsive interactions~$\propto |\delta_{\mathrm{eff}}|$ and plaquette interactions~$\propto J$.}
\label{app:figMesonModel}
\end{figure*}

In the limit~$J/h,t/h \rightarrow 0$ and dilute $U(1)$~matter in the ground state, electric field strings are minimized under the constraints imposed by Gauss's law, i.e. number of links with~$\tau^x_\ij=-1$ is minimized.
To fulfill Gauss's law~$g_{\j}=+1~\forall \j$ matter excitations are bound into pairs connected by an electric field string, see Fig.~\ref{app:figMesonModel}a.
Gauge-invariant hopping of matter excitations prolongs the string and thus kinetic energy~$t$ competes with the string tension~$h$.
Since~$h \gg t$ it is unfavourable for single matter excitations to be mobile, which justifies to describe the constituents as tightly bound mesons.

Nevertheless, the mesons can gain kinetic energy in two distinct processes: 1) a second-order hopping process~$t_\mathrm{eff}= -t^2/2h$, in which the entire pair moves from one link to a neighboring link as shown in Fig.~\ref{app:figMesonModel}b and 2) plaquette interactions~$\propto J$ induce fluctuations between the two different meson configurations on a plaquette as illustrated in Fig.~\ref{app:figMesonModel}d.
Additionally, the mesons gain dispersive energy shifts~$\delta_{\mathrm{eff}}=-t^2/2h$ if the matter excitation hops back and forth on neighboring sites, see Fig.~\ref{app:figMesonModel}c.
However, this process is only allowed for an empty neighboring site and therefore the dispersive energy shift leads to repulsive interactions between mesons.
To summarize, we find an effective model of \Ztwo{}~neutral, hard-core bosonic mesons~$\hat{b}_{\bm{n}}$ hopping on the sites of a Kagome lattice, with infinitely strong nearest neighbor (NN) repulsion, finite next-nearest neighbor (NNN) repulsion and plaquette interactions, see Fig.~\ref{app:figMesonModel}e.
The infinite repulsive term comes from the hard-core boson constraint of single matter excitations.
Therefore, the effective meson model is given by
\begin{align} \label{app:eqEffMeson}
    \H_{\mathrm{meson}} = t_\mathrm{eff}\sum_{\langle \bm{n},\bm{m} \rangle} \hat{P}_{\mathrm{NN}}\left( \hat{b}^\dagger_{\bm{n}}\hat{b}_{\bm{m}} + \mathrm{H.c.} \right)\hat{P}_{\mathrm{NN}} - J \sum_{P}\hat{P}_{\mathrm{NN}}\left( \raisebox{-1.25ex}{\includegraphics[width=50pt]{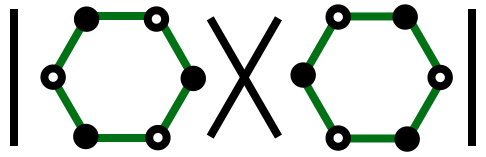}} + \mathrm{H.c.} \right)\hat{P}_{\mathrm{NN}}  + |\delta_{\mathrm{eff}}| \sum_{\langle\!\langle \bm{n},\bm{m} \rangle\!\rangle} \hat{b}^\dagger_{\bm{n}}\hat{b}_{\bm{n}} \hat{b}^\dagger_{\bm{m}}\hat{b}_{\bm{m}}
\end{align}
where~$\bm{n},\bm{m}$ denote sites of the Kagome lattice as shown in Fig.~\ref{app:figMesonModel}e and the projector~$\hat{P}_{\mathrm{NN}}$ ensures the constraint that no nearest neighbor mesons can exist.
Here, the notation~$\langle \cdot, \cdot \rangle$ ($\langle\!\langle \cdot, \cdot \rangle\!\rangle$) describes (next-)nearest neighbors on the Kagome lattice.

For experimentally relevant parameters, see SI~\ref{app:EffHam}, we can choose~$J/t \ll t/h$ and~$J/h \ll 1$ and thus neglect the plaquette interaction term.
In the limit of dilute mesons~$\langle \hat{b}^\dagger \hat{b}\rangle \approx 0$, we can treat~$\hat{P}_{\mathrm{NN}}$ on a mean-field level yielding free hard-core bosons on the Kagome lattice.
In the ground state the mesons~$\hat{b}$ condense as indicated in Fig.~\ref{fig2}a.

Taking the plaquette interactions into account, i.e. $J/t \approx t/h$, phase separation by clustering of mesons has been discussed~\cite{Borla2022} for spinless fermions on the square lattice.
However, the NNN repulsive interaction should suppress clustering and hence a more sophisticated analysis is required.

Away from the above discussed limit~$J/h \ll 1$, the meson pairs have some finite extend~$\ell^2$, which alters the effective model~(\ref{app:eqEffMeson}).
However, for a sufficiently dilute gas of matter excitations, i.e. $\ell^2 \ll 1/\langle \hat{b}^\dagger \hat{b} \rangle$, we expect the description of free hard-core bosons to be still valid which we indicate by the finite extend of the meson condensate phase in Fig.~\ref{fig2}a.
We note that the phase boundary is expected to end at the deconfinement-confinement transition of the vacuum since in the deconfined phase the picture of bound mesonic pairs breaks down.

Furthermore, at sufficiently large filling the interplay between kinetic energy and repulsion on the Kagome lattice might lead to additional, exotic phases of matter.
A more detailed phase diagram is beyond the scope of this Article and is a topic for future studies.

\section{Derivation of the effective Quantum Dimer Hamiltonian}
\label{app:EffHamQDimer}
\begin{figure*}[t]
\includegraphics[width=0.6\textwidth]{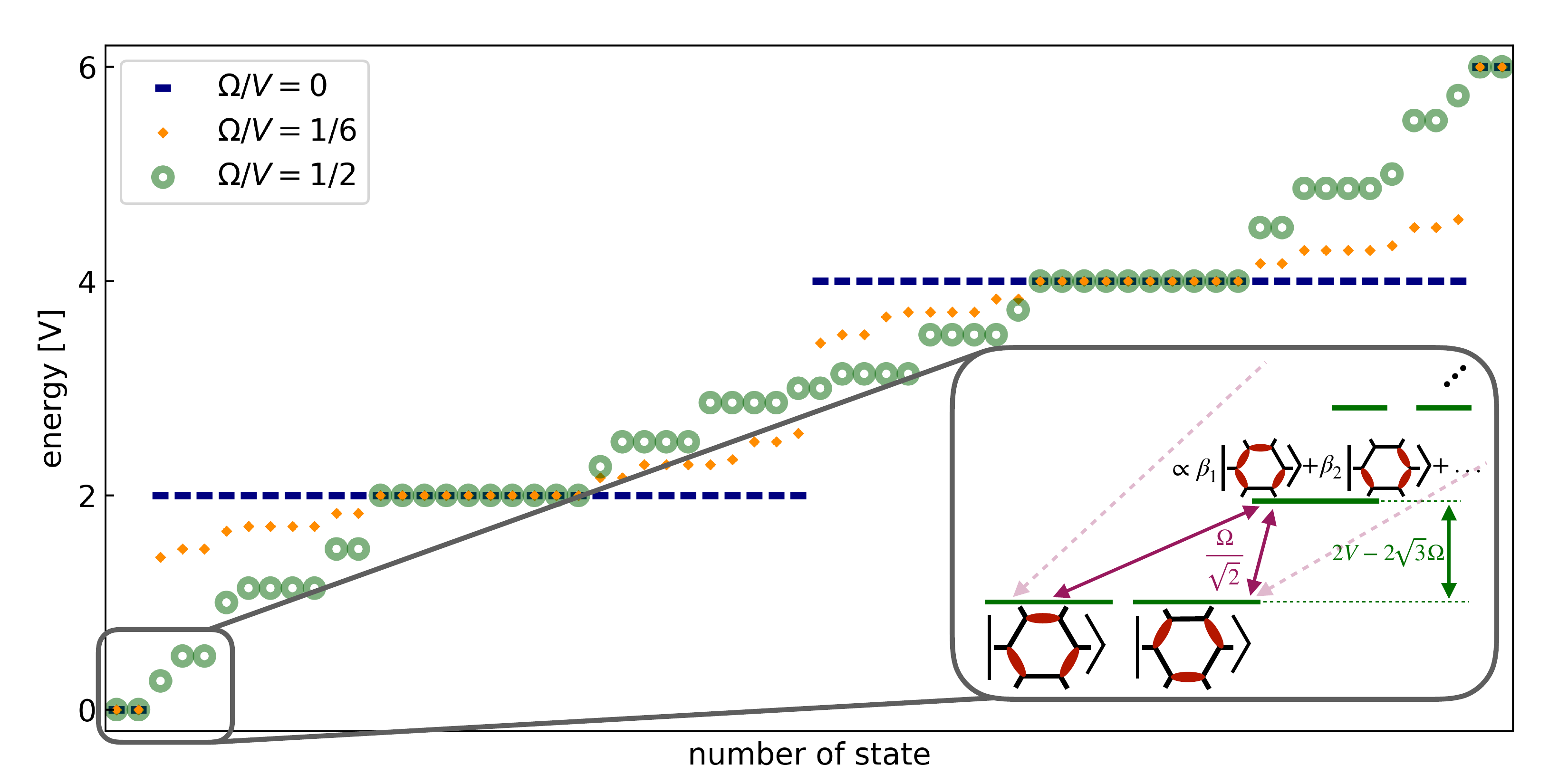}
\caption{\textbf{Derivation of plaquette terms for the QDM.} We show the spectrum of the microscopic model~\eqref{eq:micHam} on one plaquette for different driving strengths in the flippable plaquette subspace. The blue dashes show the unperturbed system. The two low-energy states are the two flippable plaquette configurations in the QDM sector. The first (second) excited manifold has two (four) monomers and the two high-energy states contain a maximum number of six monomers. To perform perturbation theory, we first need to diagonalize the degenerate subspaces because the drive couples within the two and four monomer excitation blocks, i.e.\ the drive can move around the monomer excitations without energy cost, thus making them mobile. The orange and green dots show the spectrum for different driving strength, where the blocks are diagonalized. However, there are still off-diagonal couplings between the blocks. These couplings are the starting point for the second step, which is the actual Schrieffer-Wolff perturbation theory. The inset illustrates the smallest energy gap and its renormalized coupling after the first step. This allows to compare the effective driving strength to the energy gap in order to determine a regime of validity for the perturbation theory. }
\label{app:figQDMPlaquette}
\end{figure*}

In this section, we want to derive the effective Hamiltonian~(\ref{eq:HQDM}) of the quantum dimer model,
\begin{align} \label{app:eqEffQDMHam}
    \H^{\mathrm{eff}}_{\mathrm{QDM}} = -J_{\mathrm{QDM}} \sum_{\includegraphics[width=6pt]{plaquette.pdf}}\prod_{\ij \in \raisebox{-.15\height}{\includegraphics[width=6pt]{plaquette.pdf}}} \tauzij + K\sum_{\text{NNN}}\hat{\tau}^x_{\langle \bm{i},\j\rangle}\hat{\tau}^x_{\langle \bm{m},\bm{n}\rangle}
\end{align}
from the microscopic model~(\ref{eq:micHam}).
Here, the physical subspace is given by the QDM\textsubscript{1} low-energy subspace of the LPG term, see Fig.~\ref{fig1}c.
Hence, for strong protection~$V \gg \Omega_m,\,\Omega_l$ states in the non-physical sectors are only virtually occupied and we can derive the effective model perturbatively, which yields the plaquette terms~$\propto J_{\mathrm{QDM}}$.

Let us first consider the simpler terms~$\propto K$.
These terms are introduced to drive potential quantum phase transitions in~$J/K$.
In our proposed scheme, the strong LPG terms arise from nearest-neighbor interactions.
However, the Rydberg-Rydberg interactions decay as~$R^{-6}$, where~$R$ is the distance between two atoms in optical tweezers.
Hence, there are small but finite next-nearest neighbor interactions between links of the honeycomb lattice (next-nearest neighbors on the Kagome lattice), which give rise to the term~$\propto K$ in Eq.~(\ref{app:eqEffQDMHam}).

Now, we want to elaborate on the degenerate perturbation theory to derive the plaquette terms in sixth order.
In the QDM subspace there are only two ``flippable'' configurations that can be coupled by plaquette terms and hence we can rewrite the interaction by
\begin{align} \label{app:eqFlippable}
    -J_{\mathrm{QDM}} \sum_{\includegraphics[width=6pt]{plaquette.pdf}}\prod_{\ij \in \raisebox{-.15\height}{\includegraphics[width=6pt]{plaquette.pdf}}}\tauzij = -J_{\mathrm{QDM}} \sum_{\includegraphics[width=6pt]{plaquette.pdf}}\left( \raisebox{-1.7ex}{\includegraphics[width=50pt]{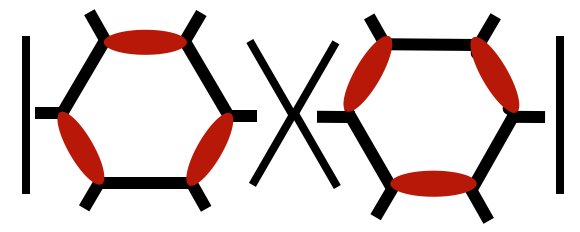}} +\mathrm{H.c.} \right),
\end{align}
where we have used the electric field string $\leftrightarrow$ dimer mapping shown in Fig.~\ref{fig1}b.
The two configurations shown in~(\ref{app:eqFlippable}) now span the low-energy manifold and are the starting point for our perturbation theory.
Above the low-energy manifold, we have three high-energy subspaces with energies $2V$, $4V$ and $6V$ since excitations (=monomers) can only be created in pairs.
The unperturbed subspaces are shown as blue dashes in Fig.~\ref{app:figQDMPlaquette}.

We perturb the system by a weak drive $\H^{\mathrm{drive}}$, Eq.~(\ref{app:eqHmic4}), coupling not only states between subsectors but also within the highly-degenerate manifolds with energy~$2V$ and $4V$.
Hence, we want to apply the same Schrieffer-Wolff formalism with degenerate subsectors as explained in SI~\ref{app:PlaquetteU1}.

In Fig.~\ref{app:figQDMPlaquette}, we show the full spectrum for different driving strengths~$\Omega/V$ ($\Omega_m=\Omega_l=\Omega$).
In general, the validity of a perturbation theory is determined by the coupling strength divided by the energy gap in the unperturbed system.
In degenerate perturbation theory, this quantity has to be evaluated after the transformation~$\hat{U}$.
As shown in the inset of Fig.~\ref{app:figQDMPlaquette}, the gap between the low-energy manifold and the first excited states becomes~$\tilde{V} = 2V-2\sqrt{3}\Omega$ and the matrix element between the two states is~$\tilde{\Omega} = \Omega/\sqrt{2}$.
Hence, we find
\begin{align}
    \dfrac{\tilde{\Omega}}{\tilde{V}} = \dfrac{\Omega}{2\sqrt{2}V-2\sqrt{6}\Omega}
\end{align}
and e.g. $\tilde{\Omega}/\tilde{V}=1/4$ for $\Omega/V \approx 1/3$, which allows to have relatively strong driving strength in the lab frame.

From $\hat{\tilde{H}}^{\mathrm{mic}}$, we can now calculate the Schrieffer-Wolff transformation as explained in SI~\ref{app:PlaquetteU1} [see Eqs.~(\ref{app:eqSWSum})-(\ref{app:eqSWNth})].
To summarize, by evaluating~(\ref{app:eqSWSum}) we can derive the leading order contribution of the plaquette interaction in the quantum dimer subspace and find
\begin{align}
    J_{\mathrm{QDM}} = \dfrac{917}{120}\dfrac{\Omega^6}{V^5},
\end{align}
which e.g.\ yields~$J_{\mathrm{QDM}}/V \approx 0.01$ for~$\Omega/V = 1/3$.
The effective coupling is surprisingly strong despite the small prefactor~$1/144$ in the perturbative expansion, Eq.~(\ref{app:eqSWNth}).

Intuitively, we can understand these strong many-body interactions to be induced by the highly mobile and gapped monomer excitations.

\section{Experimental realization}
\label{app:expReal}
\begin{figure*}[t]
\includegraphics[width=\textwidth]{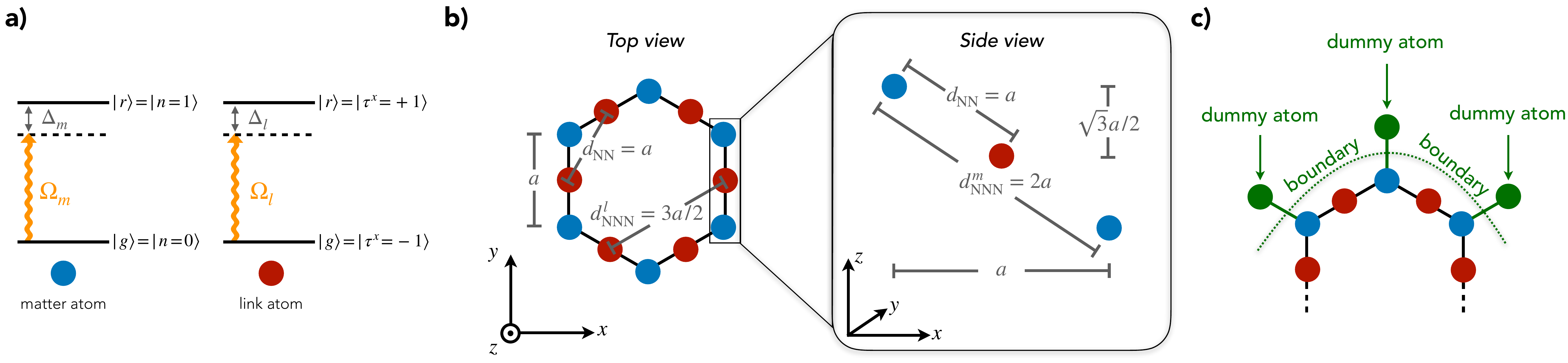}
\caption{\textbf{Rydberg atoms in tweezer array.} Panel~\textbf{a)}: The ground state~$\ket{g}$ and Rydberg state~$\ket{r}$ of the matter (link) atom is mapped on the matter field~$\hat{n}$ (electric field~$\hat{\tau}^x$) of the \Ztwo{}~mLGT. Panel~\textbf{b)}: To stabilize a gauge sector of the \Ztwo{}~mLGT or to enforce the hard-core dimer constraint in the QDM, we have introduced the LPG protection~(\ref{eq:LPG}) which requires two- and one-body interactions. The interaction strength can be adjusted by the geometry in a Rydberg atom array because the strength of the dipolar Rydberg-Rydberg interaction depends on the interatomic distance. Here, we show the suggested geometry to 1) realize the LPG term and 2) minimize the effects of long-range interactions. Panel~\textbf{c)}: A boundary in the finite size experimental setup cuts through links of the lattice. We introduce dummy atoms, which are not real and only used to illustrate the modification of the LPG term on the boundary. The LPG term on the boundary has less link atoms than in the bulk.
Hence, the detuning on the matter and link atoms along the boundary has to be adjusted. }
\label{app:figExpReal}
\end{figure*}

The scheme we propose in this Article is particularly suitable for Rydberg atom arrays because we require 1) control over the real-space configuration of atoms, 2) strong nearest-neighbor density-density interactions and 3) driving a two-level system.
The microscopic Hamiltonian has been introduced in the main text in Eq.~(\ref{eq:micHam}) and contains only nearest-neighbor interactions as well as an on-site drive, which is detuned from the ground state to Rydberg transition by~$\Delta_m$ and~$\Delta_l$, see Fig.~\ref{app:figExpReal}a.
The detuning has two contributions~$\Delta_{m,l}=-3V+\tilde{\Delta}_{m,l}$: on the one hand side the term~$-3V$ is essential for the LPG protection and appears when rewriting the Pauli spins as~$\tauxij = 2\n_\ij-1$.
On the other hand, the term~$\tilde{\Delta}_{m,l} \ll V$ enables to arbitrarily tune the electric field~$h$ and chemical potential~$\mu$ in Hamiltonian~(\ref{eq:modelH}).

The LPG term requires the Rydberg-Rydberg interaction between matter and link atoms at each vertex~$\j$ to have the same strength~$2V$, see Fig.~\ref{fig1}a.
The interaction strength between two atoms both excited to a Rydberg state scales as~$C_6/R^6$ for large distances, where $C_6$ is a constant containing details of the internal atomic structure and~$R$ is the distance between the two atoms.
Therefore, to have equal interaction strength between link and matter atoms, we require a tetrahedron geometry at each vertex as shown in Fig.~\ref{fig1}a and~\ref{app:figExpReal}b.
We define~$a$ as the length of the edge of the honeycomb lattice.
Link atoms are located on the center of edges and neighboring link atoms have a distance of~$a$ and thus we require the plane of matter and link atoms to have a distance of~$\sqrt{3}a/2$.
Since the honeycomb lattice is bipartite, we suggest to lift matter atoms in sublattice A (B) up (down) to decrease next-nearest neighbor interactions between matter atoms.
These undesired next-nearest neighbor interactions can be estimated, see Fig.~\ref{app:figExpReal}b, and we find that the nearest neighbor matter-matter interaction has strength~$V_{m-m}=V/64=0.02V$ and the next-nearest neighbor link-link interaction has strength~$V_{l-l}=64V/729 \approx 0.09V$.

Note that in an experimental setup the system has boundaries and here we want to discuss the LPG term at the boundary, see Fig.~\ref{app:figExpReal}c.
In particular, we want to consider the case where the boundary cuts through links of the honeycomb lattice, i.e.\ the boundary vertices have missing link atoms.
To this end, we introduce dummy atoms in Fig.~\ref{app:figExpReal}c which are not real atoms but instead substitute the missing link atom in the LPG term~Eq.~(\ref{eq:LPG}) by setting the corresponding dummy link to a constant value $\tau^x_{\mathrm{dummy}}=+1$.
This yields additional detuning terms for all atoms on the boundary vertices.
For technical purposes, instead of individually addressing the detuning on only boundary sites one could introduce real auxiliary atoms on the boundary, which are excited to a Rydberg state different from the matter and link atoms.
This allows to shift the Rydberg state of the matter and link atoms on the boundary, i.e. adding the required detuning, while the auxiliary atom is not affected by the driving field~$\Omega_{m,l}$.

\section{Disorder-free localization}
\label{app:DFL}
Disorder-free localization (DFL) is a phenomenon that has been studied in theories with local symmetries~\cite{Smith2017, Smith2018, Karpov2021,Halimeh2021StabilizingDFL, Halimeh2021EnhancingDFL,Chakraborty2022}.
Here, we use the microscopic model~(\ref{eq:micHam}) -- as it would be implemented in an experimental setup -- and show DFL behaviour in a small-scale exact diagonalization (ED) simulation for the case of $U(1)$~matter.
In particular, the observation of DFL would be an accessible experimental probe since it only requires to prepare the system in two different initial product states and time-evolve them under the microscopic Hamiltonian.

The key idea is the following:
Consider a system with two subsytems A and B and an initial state, where all matter sites in subsystem A (B) are occupied (empty).
We let the state time-evolve under a \Ztwo{}~invariant Hamiltonian and ask whether the matter excitations stay localized in subsystem A or delocalize equally across subsystem A and B.
Hence, the quantity of interest is the time-averaged imbalance~$\mathcal{I}(t)$ of matter excitations at time~$t$ between the subsytems given by
\begin{align} \label{app:eqImb}
    \mathcal{I}(t) = \frac{1}{Lt}\int_{0}^{t} \Big[ \big\langle \n_A(\tau) \big\rangle - \big\langle \n_B(\tau) \big\rangle \Big] d\tau,
\end{align}
where $\langle \n_{A (B)}(\tau) \rangle$ is the expectation value of total matter excitations in subsystem A (B) at time~$\tau$ and $L$ denotes the system size.

The eigenstate thermalization hypothesis (ETH) claims that~$\mathcal{I}(t)$ eventually approaches its thermal equilibrium value.
In DFL, this hypothesis is believed to be broken for gauge-noninvariant initial states $\ket{\psi^\mathrm{ninv}}$.
In the following sections we want to examine this behaviour on the example of 1) the minimal ``Mercedes star'' lattice with coordination number $z=3$ [$(2+1)D$] as presented in the main text and 2) on a Zig-Zag chain [$(1+1)D$], which would be experimentally easier to reach large system sizes.
In both examples, we performed ED studies of the microscopic Hamiltonian~(\ref{eq:micHam}) in small, numerically accessible systems.

\subsection{Mercedes star}
\label{app:DFLMercedes}
\begin{figure*}[t]
\includegraphics[width=0.8\textwidth]{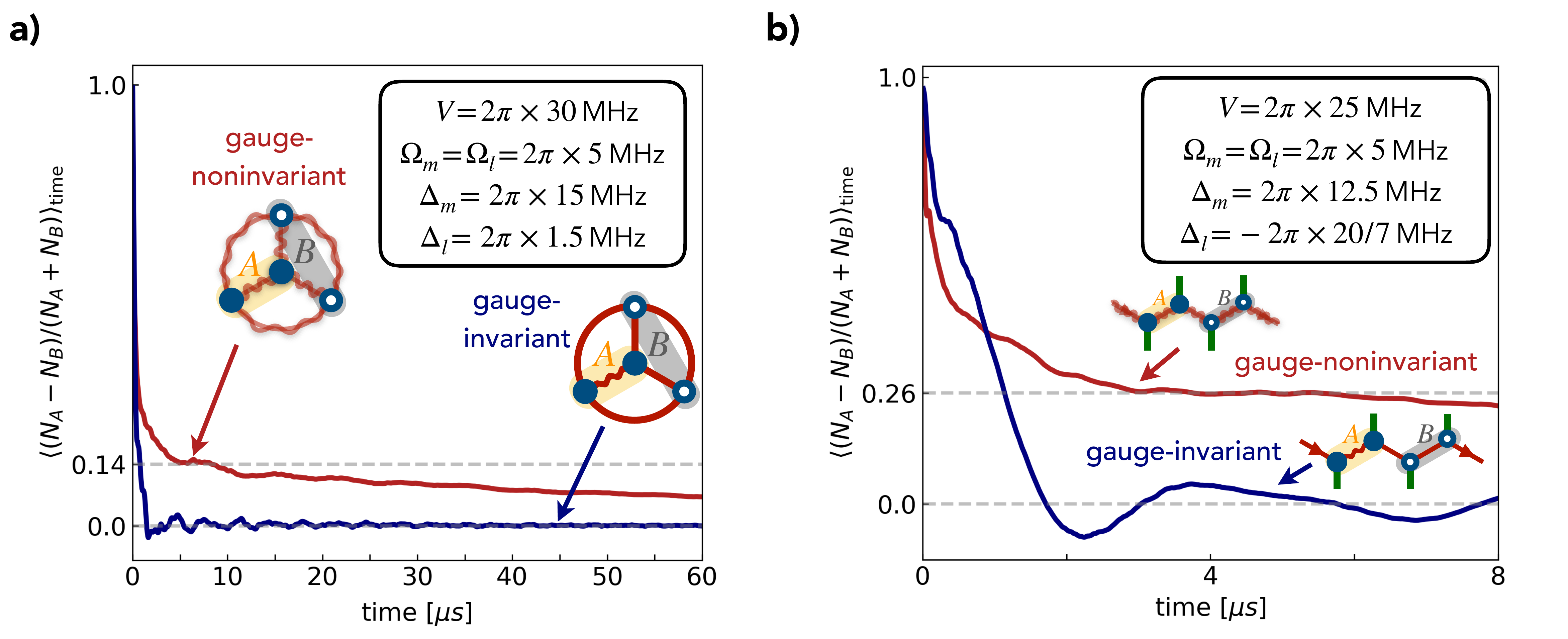}
\caption{\textbf{Disorder-free localization.} We show results from small-scale ED studies, where we time-evolve gauge-invariant~$\ket{\psi^\mathrm{inv}}$ and gauge-noninvariant initial states~$\ket{\psi^\mathrm{ninv}}$ under the microscopic Hamiltonian~(\ref{eq:micHam}). In panel~\textbf{a)} we show the same calculation as in Fig.~\ref{fig3}a but for much longer, experimentally inaccessible times. We see that the plateau at short times is a pre-thermal behaviour followed by several smaller plateaus. Eventually the imbalance of the gauge-noninvariant states slowly decays. This can be explained by the approximate but not exact local symmetry of the system. In panel~\textbf{b)}, we show analogous calculations to those in panel~\textbf{a)} or Fig.~\ref{fig3}a for a Zig-Zag chain. We find again distinctly different thermalization behaviour for the two different initial states. }
\label{app:figDFL}
\end{figure*}

Let us first consider the gauge-invariant initial state~$\ket{\psi^\mathrm{inv}}$ with $\Gj\ket{\psi^\mathrm{inv}}=+\ket{\psi^\mathrm{inv}}$ and matter excitations being distributed as described above. 
The Mercedes star model with four matter sites and six links is illustrated in the inset of Fig.~\ref{fig3}a.
For this initial state, we can compute the imbalance from the thermal ensemble as predicted by ETH (see also Ref.~\cite{Halimeh2021StabilizingDFL,Halimeh2021EnhancingDFL}, SI B) and find that indeed the system fully delocalizes~$\langle \mathcal{I} \rangle_\mathrm{thermal}=0$.
Comparing this to the ED results in Fig.~\ref{fig3}a, we find that the time-average imbalance quickly vanishes as expected.

The situation changes for the gauge-noninvariant initial state~$\ket{\psi^\mathrm{ninv}}$.
For this state the matter excitations should be located again only in subsystem A but the links should be in~$\tau^z_\ij=+1$ eigenstates as indicated in Fig.~\ref{fig3}a.
Therefore, $\ket{\psi^\mathrm{ninv}}$ is an equal superposition of all possible gauge sectors~$g_{\j}=\pm 1$.
While still $\langle \mathcal{I} \rangle_\mathrm{thermal}=0$, as we have verified numerically, we find that the state does not thermalize under time-evolution with~$\H_\mathrm{mic}$ as shown in Fig.~\ref{fig3}a.

For dynamics under a perfectly gauge-invariant Hamiltonian, $[\H,\Gj]=0~\forall \j$, the intuitive picture is the following:
When the system is initialized in a superposition of all gauge sectors, the system independently time evolves in each of these (uncoupled) gauge sectors.
In each gauge sector the system has different background charges, which affects the spreading of the matter excitations, and ultimately the average is taken over all these gauge sectors.
This effectively induces disorder in the system, which can lead to localization~\cite{Smith2017}.

While this interpretation holds for \Ztwo{}~gauge-invariant Hamiltonians, we want to note the differences to our scheme here.
Firstly, we generate the local symmetries in our system through local pseudogenerators~$\Wj$, which yield an even enriched symmetry structure because the system has three emerging local symmetry sectors as shown in Fig.~\ref{fig1}c.
This has been discussed previously~\cite{Halimeh2021EnhancingDFL} for $(1+1)$D systems. 
Secondly, the microscopic Hamiltonian explicitly breaks the local symmetry generated by the LPG terms,~$[\H_{\mathrm{mic}},\Wj] \neq 0~\forall \j$, to induce dynamics in the system (note that the gauge symmetry is only approximate and not exact).
Hence, the weak drive has to be considered as an error term, which eventually leads to thermalization of the system for long times.
However, at experimentally relevant timescales we find a clear pre-thermal plateau indicating DFL as shown in Fig.~\ref{app:figDFL}a.
The parameters used in the ED calculation are shown in the inset of Fig.~\ref{app:figDFL}a.

\subsection{Zig-Zag chain}
\label{app:DFLZigZag}
The Mercedes star model is a numerical toy model with~coordination number $z=3$.
A truly $(2+1)$D model with $z=3$ can be realized on the honeycomb lattice but requires a large number of qubits/atoms.
Therefore, as a first step to probe the proposed model, we suggest to implement a Zig-Zag chain with periodic boundary conditions, where each site of the chain is connected to a dummy atom, see Fig.~\ref{app:figDFL}b and SI~\ref{app:expReal}.
This additional dummy atom ensures coordination number $z=3$ such that the LPG protection scheme becomes fully applicable.

In Fig.~\ref{app:figDFL}b, we show results of an ED study in a Zig-Zag chain with four matter sites and four links with periodic boundary conditions.
Again, we can define a subsystem A (B), where matter excitations are located at time~$t=0$.
Furthermore, the gauge-invariant state~$\ket{\psi^\mathrm{inv}}$ and gauge-noninvariant state~$\ket{\psi^\mathrm{ninv}}$ differ by the configuration of the link atom.
For both initial states, we expect the thermal expectation value of the imbalance, Eq.~(\ref{app:eqImb}), to vanish~$\langle \mathcal{I} \rangle_\mathrm{thermal}=0$ 

Again, we find a clearly different behavior after time-evolving under the microscopic Hamiltonian~$\H_\mathrm{mic}$ and evaluating the time-averaged imbalance.
However, the observed dynamics is slower than in the Mercedes star model.
The parameters used in the ED calculation are shown in the inset of Fig.~\ref{app:figDFL}b.

\section{Schwinger effect}
\label{app:Schwinger}
\begin{figure*}[t]
\includegraphics[width=0.95\textwidth]{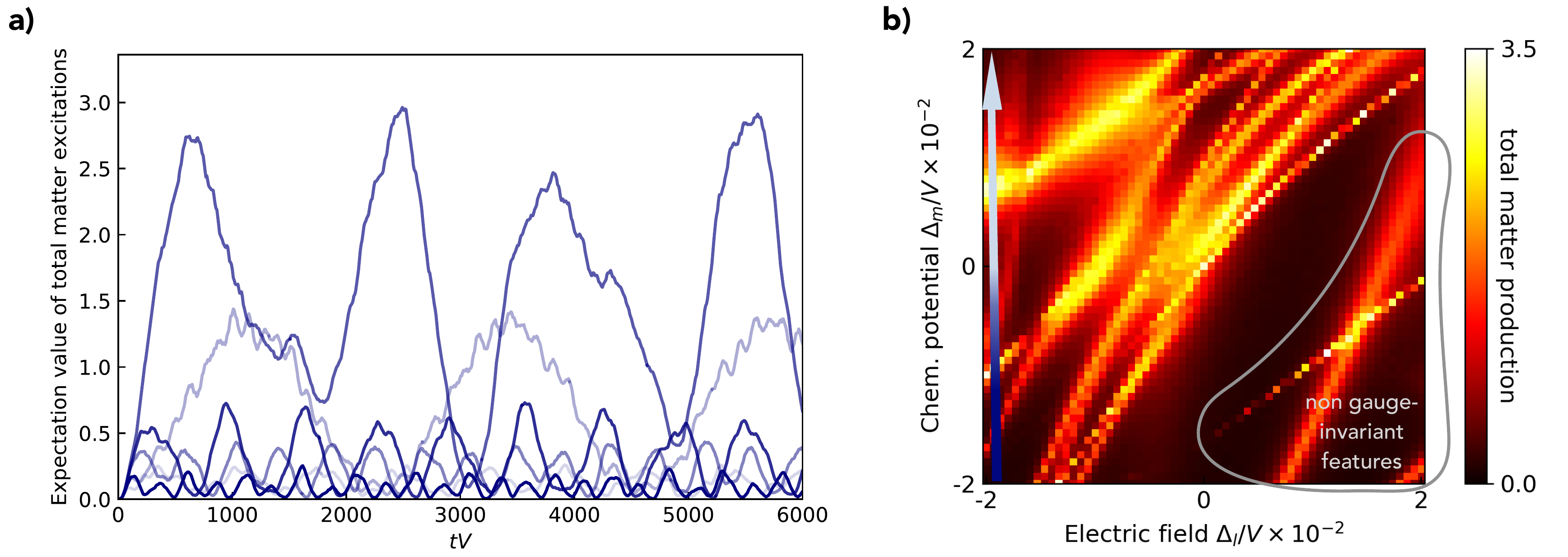}
\caption{\textbf{Schwinger Effect.} We provide complementary plots to the results shown in Fig.~\ref{fig3}b calculated from exact diagonalization studies of the microscopic model~(\ref{eq:micHam}) on the Mercedes star. In panel~\textbf{a)}, we show the number of matter excitations in the target gauge sector, Eq.~(\ref{app:eqSchwinger}), for some exemplary parameters~($\Delta_m,\,\Delta_l$) versus time~$tV$; the parameters are chosen along the arrow on the left hand side of panel~\textbf{b)}. We have checked that the maximal evolution time~$tV=6000$ presumably captures the peak of each time trace. In the main text, Fig.~\ref{fig3}b, we plot the peak value for all parameters~($\Delta_m,\,\Delta_l$). In panel~\textbf{b)}, we show the same result as in Fig.~\ref{fig3}b, but we do not project into the target gauge sector. For large positive detunings~$\Delta_l$, we can see additional, unphysical resonance lines. In particular, the region where we claim gauge-invariant pair creation processes to appear remains unchanged. Therefore, the appearance of resonance lines in these regions are not driven by gauge-symmetry breaking processes.}
\label{app:figSchwinger}
\end{figure*}

The Schwinger effect is a non-perturbative effect from quantum electrodynamics that describes the production of matter excitations from vacuum~\cite{Schwinger1951}.
Due to the weak coupling constant of quantum electrodynamics this effect is only expected to appear at very strong electric fields and has not been observed.
LGTs were originally introduced to study the effects of strong coupling in gauge theories~\cite{Wilson1974} and are therefore a candidate theory to also study the physics of the Schwinger effect.
Recently, digital quantum simulation of the $(1+1)$D Schwinger model on the lattice have examined the Schwinger effect~\cite{Martinez2016}.
Here, we want to present an experimentally measurable quantity by considering the pairwise production of matter excitations.
While the potential between two static charges is of theoretical interest and used as a signature of the Schwinger effect~\cite{Sala2018}, it is numerically challenging to extract in $(2+1)$D and is a topic for future studies.

Our effective model~(\ref{eq:modelH}) allows to explore the Schwinger effect in a \Ztwo{}~mLGT in $(2+1)$D, which has not yet been observed.
The experimental protocol starts by initializing a gauge-invariant state~$\ket{\mathrm{vac}}$ without any matter excitations in one of two vacua, i.e.\ either all links in~$\tau^x_\ij=+1$ or~$\tau^x_\ij=-1$, which is a simply product state.
Then, the system is quenched with the microscopic Hamiltonian~(\ref{eq:micHam}) for time $t$ yielding~$\ket{\psi(t)}=e^{-i\hat{H}_\mathrm{mic}t}\ket{\mathrm{vac}}$.
The effective model~(\ref{eq:modelH}) with quantum-\Ztwo{}~matter, which we expect to correctly capture the physics, contains pairing terms~$\propto (\ad \hat{\tau}^z\ad + \mathrm{H.c.})$ yielding pair creation processes from vacuum.
As soon as the matter excitations are created, they move apart due to the hopping term and interact with the gauge field, which makes the prediction of the dynamics very challenging.

Therefore, an easily accessible quantity to probe the Schwinger effect is the gauge-invariant production rate of matter excitations.
To this end, we let the system time-evolve and calculate
\begin{align} \label{app:eqSchwinger}
    \mathcal{P}(t) = \bra{\psi(t)} \sum_{\j}\nj\prod_{\j}(1+\Gj)/2 \ket{\psi(t)},
\end{align}
which projects onto the gauge sector~$g_{\j}=+1~\forall \j$ and gives the expectation value of the total number of matter excitations in the system.

We calculate~$\mathcal{P}(t)$ for~$t\in [0, 6000/V]$ for different electric fields~$\Delta_m$ and chemical potentials~$\Delta_l$, see Eq.~(\ref{eq:micHam}), for~$\Omega_m=\Omega_l=V/8$.
In Fig.~\ref{app:figSchwinger}a we plot the timetraces for some set of parameters~$(\Delta_m,\Delta_l)$ and we have checked that the maximal time-evolution time of~$6000/V$ captures the peak of each timetrace.

Each timetrace has different amplitude and timescale and therefore only considering $\mathcal{P}(t)$ at a fixed time~$t=t_0$ is not sufficient to extract the productivity of creating matter excitations from vacuum.
To this end, we take the maximum value of each timetrace $\mathrm{max}_t \mathcal{P}(t)$ for each $(\Delta_m,\,\Delta_l)$, which is plotted in Fig.~\ref{fig3}b in the main text.

We find that for some $(\Delta_m,\,\Delta_l)$ the production of matter excitations is significantly higher than for others.
An intuitive picture is that a pair of matter excitations costs an energy $2\mu$ (mass of matter excitations) and due to Gauss's law, the two matter excitations have to be connected by an electric string, which costs~$2h$.
Therefore, if $2\mu +2h = 0 $ the process is on resonance we expect a high production rate of matter pairs.
Note that~$\mu,\,h$ have to be determined from Tab.~\ref{app:TableEffHam} and are not simply given by~$(\Delta_m,\,\Delta_l)$.

Besides the resonances described above, there are several more processes that would have to be taken into account.
E.g. two neighboring matter excitations repel each other with strength~$M$; Eq.~(\ref{app:effHamNoU1}); matter excitations are mobile; plaquette interaction compete with the electric field; finite size effects etc.
Due to these competing interactions, it is very hard to gain a complete picture and thus large-scale numerical simulations as well as experimental observations are needed.
Nevertheless, small-scale numerical calculations of the microscopic model show promising signatures of the Schwinger effect.

An additional feature we monitored in our ED study is the role of gauge-noninvariant dynamics.
In Eq.~(\ref{app:eqSchwinger}) we project out unphysical states.
In contrast we can do the same analysis as before but without projecting on the target gauge sector, which is shown in Fig.~\ref{app:figSchwinger}b.
We find regions with additional resonances, which are caused by gauge-symmetry breaking processes.
However, the resonances we find \textit{with} projecting on the gauge sectors are not altered, which is an evidence that the physics is purely determined by \Ztwo{}-invariant dynamics.

\section{DMRG in the ladder}
\label{app:DMRG}
\begin{figure*}[t]
\includegraphics[width=0.95\textwidth]{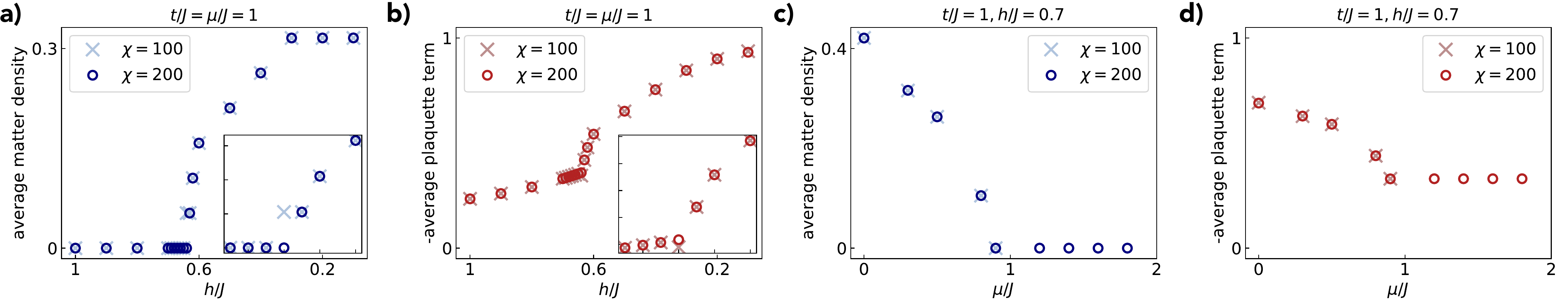}
\caption{\textbf{Convergence of DMRG in the ladder.} We plot results of the DMRG calculations for different bond dimensions~\mbox{$\chi=100$} and~\mbox{$\chi=200$}. Panel~\textbf{a)} and~\textbf{b)} show the expectation value of the matter density and plaquette terms, respectively, as presented in the main text in Fig.~\ref{fig3}c for fixed chemical potential~$\mu/J=1$; note that $J<0$ and hence $\mu<0$. In panel~\textbf{c)} and~\textbf{d)}, we keep the electric field fixed at~$h/J=0.7$ and vary the chemical potential. }
\label{app:figDMRG}
\end{figure*}

In Fig.~\ref{fig2} in the main text, we map out limiting cases of the ground-state phase of the effective model~\eqref{eq:modelH}. 
A numerically more accessible and experimentally realizable model is the \Ztwo{}~mLGT coupled to $U(1)$~matter on a ladder.
While the ladder geometry is not $2$D but mixed-$1$D, it has coordination number~$z=3$ and therefore is applicable for our proposed LPG term~\eqref{eq:LPG}.
Thus, the effective Hamiltonian only has to be modified for the plaquette interaction because the plaquettes on a ladder have four instead six edges; note that we anticipate the plaquette terms to be even stronger in the ladder and we choose~$J<0$ as found in SI~\ref{app:PlaquetteU1} in Eq.~\eqref{app:eqJeffPlaqU1}.

For our numerical simulations, we use Hamiltonian~\eqref{eq:modelH} with~\mbox{$\Delta_1=\Delta_2=0$}, i.e. we have a global $U(1)$~symmetry for the matter, and we tune the electric field~$h$ and chemical potential~$\mu$ for fixed tunneling and plaquette interactions~$t=1$ and~$J=-1$.
Using the density matrix renormalization group (DMRG) technique, we calculate the ground state of the above described Hamiltonian on a ladder with $L=19$ plaquettes.

From the ground state calculated in DMRG, we obtain the average matter excitation density -- an experimentally directly accessible quantity, e.g. by taking snapshots in the atomic ground state and Rydberg basis (see Fig.~\ref{fig1}a).
As shown in Fig.~\ref{fig3}c, the system is in a matter vacuum for large~$h/J$ similar to the $2$D case.
For decreasing $h/J$, we find a sharp increase of matter excitations indicating a phase transition as shown in Fig.~\ref{app:figDMRG}a.
At the same critical electric field value, the plaquette term shows a sharp feature; note that $J<0$ and thus the plaquette expectation value is negative indicated by the reversed sign on the plot label in Fig.~\ref{app:figDMRG}b.

Moreover, we show a similar scan of parameters but now we fix the electric field~$h/J=0.7$ while scanning the chemical potential~$\mu/J$.
We find consistent behaviour with Fig.~\ref{fig2}a, i.e. a sharp transition into the vacuum phase, see Fig.~\ref{app:figDMRG}c and d. 

To characterize the different phases and its phase transitions requires more elaborate studies of our effective model on the ladder and is a topic for future studies.
We emphasize that due to its numerical accessibility and experimental feasibility, the ladder model is a promising playground to probe \Ztwo{}~lattice gauge theories coupled to dynamical matter beyond~$(1+1)$D.

In the following, we discuss numerical details of the DMRG calculation. We use the TeNPy package \cite{Hauschild2018,Hauschild2019} to find the ground state of Hamiltonian~\eqref{eq:modelH} with~\mbox{$\Delta_1=\Delta_2=0$}. Note that while this Hamiltonian conserves the particle number, we do not run the DMRG simulation in a fixed particle number sector. For a given set of Hamiltonian parameters, we find the global ground state, and can therefore use the average matter density as an observable. In the DMRG simulation, we enforce the system to be in the target gauge sector by adding a large energy penalty term proportional to Gauss's law~\eqref{eq:Goperator} to the Hamiltonian. The ground state will therefore always fulfill Gauss's law by construction. We carefully checked our numerical results for convergence, see Fig.~\ref{app:figDMRG}.

\section{Thermal deconfinement}
\label{app:Deconfinement}

In the main text and Methods section, we discuss a thermal deconfinement phase transition in a classical limit of the Hamiltonian~(\ref{eq:modelH}), see Fig.~\ref{fig3}d.
Here we provide numerical details about our Monte Carlo simulations as well as supplementary results.
The Monte Carlo simulations are implemented in C++ using the Boost C++ libraries.
The lattice is represented as a graph which simplifies many operations involving nearest and next-nearest neighbors in a honeycomb lattice.
For I/O operations, multiprocessing and postprocessing, we use Python (NumPy, SciPy, Python multiprocessing, Matplotlib).

\begin{figure}[t]
\includegraphics[width=\textwidth]{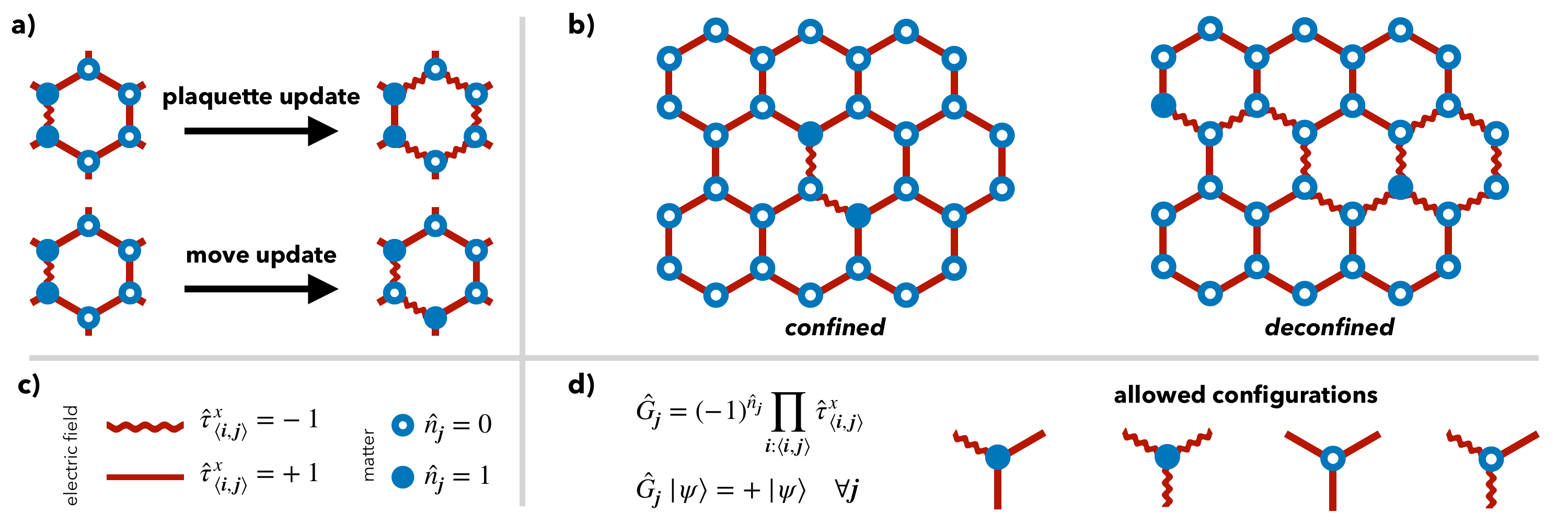}
\caption{\textbf{Monte Carlo sampling.} In panel~\textbf{a)}, we show the update procedures for the Metropolis-Hastings sampling. A plaquette update flips all electric fields in a hexagon. In a move update, a matter excitation moves to an unoccupied lattice site and flips the electric field along its way. All updates conserve the number of matter excitations and Gauss's law, see panel~\textbf{d)}. In panel~\textbf{b)}, we illustrate the connection of percolation and confinement. In the confined phase we have paired matter excitations connected by a short string of electric fields $\tauxij = -1$ while in the deconfined phase a global net of strings spans over the entire lattice. In panel~\textbf{c)}, we introduce the notation for the matter and the electric field. In panel~\textbf{d)}, we illustrate the Gauss's law constraint.}
\label{fig:MC}
\end{figure}

We derive the classical model from the effective Hamiltonian~(\ref{app:eqEffHamZ2mLGTU1matter}) with $U(1)$ matter in the gauge sector $g_{\j} =+1~\forall \j$, see Fig.~\ref{fig:MC}d.
To obtain a purely classical Hamiltonian (energy functional), we remove the matter and gauge field dynamics by setting~$t = J = 0$, which yields (neglecting constant terms)
\begin{align}
\begin{split} \label{app:eqClassicalMC}
    H^{\mathrm{classical}}(\{ n_{\j}, \tau^x_\ij \}_{\j}) &= 
    - h\sum_{\ij} \raisebox{-1ex}{\includegraphics[width=20pt]{TabEffHam/elfield.png}}
    + M\sum_\ij \raisebox{-1ex}{\includegraphics[width=20pt]{TabEffHam/M.png}}
    +\chi_1 \sum_{\ij} \bigg( \raisebox{-1ex}{\includegraphics[width=20pt]{TabEffHam/chi1.png}} +\raisebox{-1ex}{\includegraphics[width=20pt]{TabEffHam/chi1_2.png}} +\raisebox{-1ex}{\includegraphics[width=20pt]{TabEffHam/chi1_3.png}}
    +\raisebox{-1ex}{\includegraphics[width=20pt]{TabEffHam/chi1_4.png}} \bigg)\\
   &+\chi_2 \sum_{\ij} \bigg( \raisebox{-1ex}{\includegraphics[width=20pt]{TabEffHam/chi2.png}} +\raisebox{-1ex}{\includegraphics[width=20pt]{TabEffHam/chi2_2.png}} +\raisebox{-1ex}{\includegraphics[width=20pt]{TabEffHam/chi2_3.png}}
    +\raisebox{-1ex}{\includegraphics[width=20pt]{TabEffHam/chi2_4.png}} \bigg)
    +\chi_3 \sum_{\j} \bigg( \raisebox{-1.25ex}{\includegraphics[width=15pt]{TabEffHam/chi3.png}} +\raisebox{-1.25ex}{\includegraphics[width=15pt]{TabEffHam/chi3_2.png}} +\raisebox{-1.25ex}{\includegraphics[width=15pt]{TabEffHam/chi3_3.png}} \bigg).
\end{split}
\end{align}
The energy~$H^{\mathrm{classical}}(\{ n_{\j}, \tau^x_\ij \})$ is fully determined by the configuration of matter excitations and electric fields~$\{ n_{\j}, \tau^x_\ij \}_{\j}$.
The model should still be understood as an effective theory derived from the microscopic model~Eq.~(\ref{eq:micHam}).
Hence, the drive $\Omega_m/V,\,\Omega_l/V$ and detunings~$\Delta_m, \Delta_l$ determine the coupling strength of $M$, $\chi_1$, $\chi_2$ and $\chi_3$, see Tab.~\ref{app:TableEffHam}.
In our example, we have chosen the experimentally realistic parameters~$\Omega_m=\Omega_l=0.125V$, $\Delta_m=V/2$ and~$\Delta_l=0$, which yields $M/h = 0.2917$, $\chi_1 / h = 0.1483$, $\chi_2 / h = 0.073$, $\chi_3 / h = 0.4347$.
Note that the electric field term~$h$ can be tuned independently by the link detuning~$\Delta_l$.

To probe (de)confinement, we dope the system with exactly two matter excitations in all simulations (see Methods).
The Monte Carlo simulations are based on Metropolis-Hastings sampling using move and plaquette updates presented in Fig.~\ref{fig:MC}a.

\begin{figure}[t]
\includegraphics[width=\textwidth]{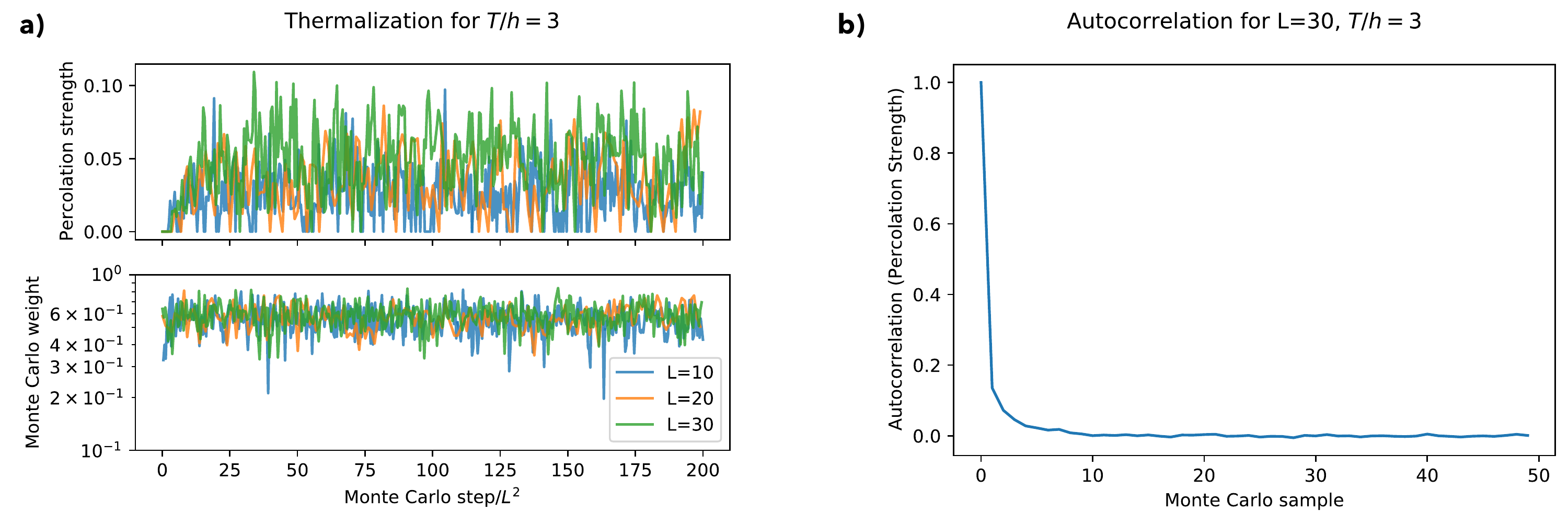}
\caption{\textbf{Monte Carlo thermalization \& autocorrelation.} We show results for Monte Carlo simulations of Hamiltonian~(\ref{app:eqClassicalMC}) at $M/h = 0.2917$, $\chi_1 / h = 0.1483$, $\chi_2 / h = 0.073$, $\chi_3 / h = 0.4347$ and two matter excitations for $T/h=3$. In panel~\textbf{a)} we show the thermalization of the percolation strength (top) and the Monte Carlo weights averaged over 15 runs (bottom). We have confirmed that in our simulations the system thermalizes after $200\!\times\!L^2$~steps for all~$T/h$.  In panel~\textbf{b)}, we show the autocorrelation of the percolation strength. We plot the average over 15 runs with~$10^4$ samples, respectively. Between each sample we perform $2\!\times\!L^2$~steps. We find negligible autocorrelation between samples.}
\label{fig:therm_auto}
\end{figure}

At the beginning of each simulation, the system thermalizes for $200\!\times\!L^2$~steps. Subsequently, we record~$10^4$ samples with $2\!\times\!L^2$~steps between each other.
In Fig.~\ref{fig:therm_auto}, we show exemplary thermalization and autocorrelation plots for $T/h=3$. 
We take the autocorrelation between snapshots into account for every error bar.

We simulate the system at $0.1 \leq T/h \leq 10$ for system size $10\!\times\!10$, $20\!\times\!20$, $30\!\times\!30$ and $35\!\times\!35$ on a honeycomb lattice with open boundaries.
For each sample, we measure 1) the percolation strength, i.e.\ \#(strings in the largest percolating string-cluster)/\#(bonds), 2)  the total number of strings, 3) the size of the largest string-cluster, and 4) the Euclidean distance between the two matter excitations.
We illustrate two snapshots from the deconfined and confined regime in Fig.~\ref{fig:MC}b.

\begin{figure}[t]
\includegraphics[width=\textwidth]{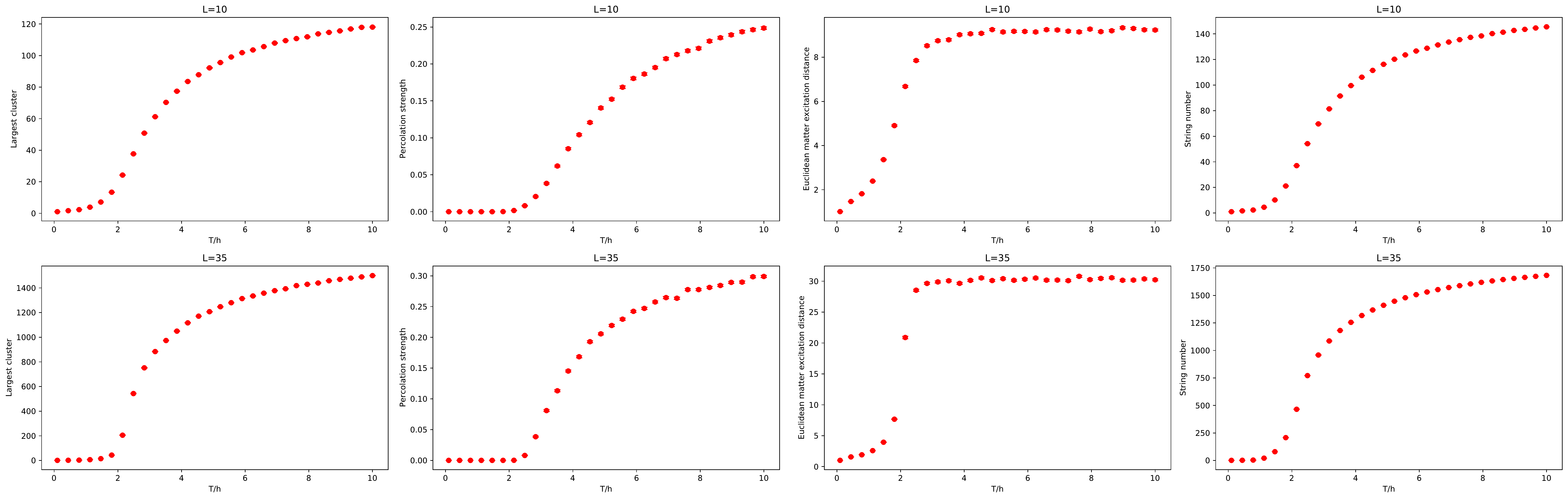}
\caption{\textbf{Monte Carlo results.} We show results for Hamiltonian~(\ref{app:eqClassicalMC}) at $M/h = 0.2917$, $\chi_1 / h = 0.1483$, $\chi_2 / h = 0.073$, $\chi_3 / h = 0.4347$ and two matter excitations. We plot the number of strings in the largest string-cluster, the percolation strength, the Euclidean distance between the two matter excitations and the string number (from left to right). The size of the honeycomb lattice is $10\!\times\!10$ (top) and $35\!\times\!35$ (bottom). We can clearly identify features for thermal deconfinement at $(T/h)_c \approx 2$, which become sharper for increasing system sizes.}
\label{fig:result_cluster}
\end{figure}

In Fig.~\ref{fig:result_cluster} we show Monte Carlo results for system size $10\!\times\!10$ and $35\!\times\!35$.
We observe a clear change of behaviour in all above discussed quantities, which signals a thermal deconfinement phase transition.
For low temperatures, the percolation strength vanishes.
At a critical temperature $(T/h)_c \approx 2$, the percolation strength abruptly increases, i.e. the string-net percolates and the matter excitations are deconfined.
At the same critical temperature $(T/h)_c \approx 2$, the Euclidean distance between the two matter excitations drastically increases to roughly the system size. 
We note that finite-size effects strongly influence $(T/h)_c$ for small system sizes.
However, the transition becomes generally sharper for larger lattice sizes as expected.

\end{document}